\documentclass[fleqn,usenatbib]{mnras}

\usepackage{newtxtext,newtxmath}
\usepackage[T1]{fontenc}

\DeclareRobustCommand{\VAN}[3]{#2}
\let\VANthebibliography\thebibliography
\def\thebibliography{\DeclareRobustCommand{\VAN}[3]{##3}\VANthebibliography}

\usepackage[graphicx]{realboxes}
\usepackage{epsf}
\usepackage{color}
\usepackage{enumitem}
\usepackage{hyperref}
\usepackage{multirow} 
\usepackage{pdflscape}

\usepackage{tabularx} % for 'tabularx' env. and 'X' col. type
\defcitealias{CG2018}{CG18}
\defcitealias{whelan2020}{PW20}
\hypersetup{
    colorlinks=true,
    linkcolor=blue,
    filecolor=magenta,      
    urlcolor=blue,
    pdftitle={Overleaf Example},
    pdfpagemode=FullScreen,
    }

\usepackage{graphicx}	% Including figure files
\usepackage{amsmath}	% Advanced maths commands

\title{The Chemical Homogeneity of Single-Lined Spectroscopic Binaries in Open Clusters}
\author[A. Sinha et al.]{Amaya Sinha,$^{1}$\thanks{E-mail: u1363702@utah.edu} 
Gail Zasowski,$^{1}$
Natalie R. Myers,$^{2,3}$
Catherine Manea,$^{1}$\thanks{NSF Astronomy and Astrophysics Postdoctoral Fellow} 
Peter Frinchaboy,$^{2}$
Katia Cunha,$^{4,5}$
\newauthor
Johanna Müller-Horn,$^{6}$
Yao-Yuan Mao,$^{1}$
Aida Behmard,$^{7,8}$
Joleen Carlberg,$^{9}$
Julio Chanamé,$^{10}$
Polly Frazer$^{11}$
\newauthor
Emily Griffith,$^{12}$\thanks{NASA Hubble Fellow} 
Sarah Loebman,$^{13}$
A. Roman-Lopes,$^{14}$
Jonah Otto,$^{2}$
Diogo Souto$^{15}$
Keivan Staussan,$^{16}$
\newauthor
Guy S. Stringfellow$^{12}$
\\
$^{1}$Department of Physics and Astronomy, University of Utah, Salt Lake City, UT. 84112, USA\\
$^{2}$Department of Physics \& Astronomy, Texas Christian University, Fort Worth, TX 76129, USA \\
$^{3}$Department of Physics and Astronomy, Johns Hopkins University, Baltimore, MD 21218, USA\\
$^{4}$Steward Observatory, University of Arizona, Tucson, AZ 85721, USA\\
$^{5}$Observatório Nacional, Rua General José Cristino, 77, 20921-400 São Cristóvão, Rio de Janeiro, RJ, Brazil\\
$^{6}$Max-Planck-Institut für Astronomie, Königstuhl 17, 69117 Heidelberg, Germany\\
$^{7}$Center for Computational Astrophysics, Flatiron Institute, 162 Fifth Avenue, New York, NY 10010, USA\\
$^{8}$American Museum of Natural History, 200 Central Park West, Manhattan, NY 10024, USA\\
$^{9}$Space Telescope Science Institute, 3700 San Martin Drive, Baltimore, MD 21218, USA\\
$^{10}$Instituto de Astrofísica, Pontificia Universidad Católicade Chile, Av. Vicuña Mackenna 4860, 782-0436 Macul, Santiago, Chile\\
$^{11}$Center for Cosmology and Particle Physics, Department of Physics, New York University, 726~Broadway, New~York,~NY 10003, USA\\
$^{12}$Center for Astrophysics and Space Astronomy, Department of Astrophysical and Planetary Sciences, University  of Colorado, 389~UCB, Boulder,~CO 80309-0389, USA\\
$^{13}$Department of Physics, University of California, Merced, 5200 Lake Road, Merced, CA 95343, USA\\
$^{14}$Department of Astronomy, Universidad de La Serena, Av. Raul Bitran 1302, La Serena, Chile\\
$^{15}$Departamento de Física, Universidade Federal de Sergipe, Av. Marechal Rondon, S/N, 49000-000 São Cristóvão, SE, Brazil\\
$^{16}$Department of Physics and Astronomy, Vanderbilt University, Nashville, TN 37235, USA
}

\date{Accepted XXX. Received YYY; in original form ZZZ}

\pubyear{2025}

\begin{document}
\label{firstpage}
\pagerange{\pageref{firstpage}--\pageref{lastpage}}
\maketitle

\begin{abstract}
Using SDSS-V DR19 Milky Way Mapper APOGEE data, we measure the impact that close binarity has on surface chemistry across the Hertzsprung-Russell diagram in a broad set of abundances by studying single-lined spectroscopic binaries (SB1s) in open clusters. We derive binary membership and orbital parameters for 103 SB1s by analysing APOGEE radial velocities with \textsc{The Joker} and \textsc{UltraNest}. We perform a detailed abundance analysis with \textsc{BACCHUS} to derive abundances in fourteen chemical species:  Si, Fe, C, N, O, Na, Mg, Al, Ca, Ti, Cr, Ni, Ce, and Nd. Leveraging the assumptions of chemical homogeneity in open clusters, we compare the surface abundances of SB1s to non-binary stars at similar evolutionary states. We find that a subset of binaries with significant UV excess have a $\Delta$[C/N] that is 0.2--0.5 dex higher than expected, resulting in overestimated [C/N]-based ages for those stars. This points to pollution from an evolved companion and has implications for [C/N]-based age studies of the broader Milky Way. At the population level, we find that SB1s in our sample can be treated as statistically chemically homogeneous with their single-star counterparts, and we find no connection between orbital separation and chemical enrichment or depletion. We show that at separations up to ~5 pc, co-eval stars can be considered chemically homogeneous with one another within current abundance precisions, regardless of multiplicity.
\end{abstract}


%%%%%%%%%%%%%%%%%%%%%%%%%%%%%%%%%%%%%%%%%%%%%%%%%%

%%%%%%%%%%%%%%%%% BODY OF PAPER %%%%%%%%%%%%%%%%%%
\section{Introduction}
\textit{Two stars, both alike in chemistry?} Understanding the interplay between stellar multiplicity and evolution is key to understanding not just the life cycles of stars, but also the primary chemical enrichment methods that populate galaxies and the broader universe with heavy elements. 

While the Sun lacks a stellar companion, many stars in our Galaxy exist in binaries or higher order systems, and at solar metallicity this fraction is between 10-30$\%$ \citep{offner2023, moe2019,geller2021}. For OB-type stars this fraction approaches unity \citep{offner2023}. With some exceptions, binaries and higher-order systems form co-natally, i.e., from the same molecular gas cloud at the same time \citep{binaries4}. This makes them a useful tracer for many of the same properties as open clusters, such as studying birth chemistry.

Due to their large separations, wide binaries are often resolvable on the sky as independent sources \citep[e.g.,][]{offner,albadry}. It is expected that detached non-interacting binary members at the same evolutionary phase should have identical chemistry \citep{binaries1,binaries2}. This was corroborated by studies such as \citet{binaries} and \citet{andrews2018} in the optical, and \citet{dongwook2024} and \citet{andrews2019} in the near-IR, who showed that wide binary pairs at similar evolutionary stages were chemically homogeneous to between 0.02--0.1~dex.

%KC: THis 10-30 percent is an average number for all spectral types? Might be worth mentioning that for OB stars this fraction is much higher. Almost all of them are binaries.

As the separation between members decreases to tens of AU or even less, the chances of interactions increase. Binaries in this regime---considered ``close binaries"--- are objects of great interest to understanding stellar evolution and chemical enrichment. Mass transfer events can impact the surface abundances of elements like carbon and s-process neutron capture elements (e.g., Barium), which are synthesised in stars on the asymptotic giant branch (AGB), before being dispersed into the interstellar medium through thermal pulses \citep{lugaro2003,cseh2018}. Common examples of post-interaction binaries are blue or yellow stragglers \citep{sandage1953, sun2021} and barium stars \citep{bidelman1951, vanderSwaelmen2017,jorissen2019}. Stars at such small separations can also undergo common envelope evolution, or be found in semi-detached configurations that cause large gradients in temperature and surface gravity. While there is nuance in the label ``close binary", within this paper we only mean single-lined spectroscopic binaries (SB1s), and its conclusions should not be generalized to other types of binary.

However, to our knowledge, no study has directly measured intra-pair chemical homogeneity in close binaries. This arises because resolving both members' spectra at the separations of tens of AU or less is exceptionally challenging. Cases where both objects are similarly luminous result in unresolved spectroscopic binaries (SB2s), where it remains difficult to separate individual contributions from a single spectrum. In other cases, one companion is less luminous and is only detectable through the radial velocity variations (SB1s), or potentially eclipses, it induces in the brighter member \citep{stebbins1911, moe2019}. In either case, detailed elemental abundances are challenging to derive for both members. However, constraining the chemistry of binaries at these close separations is crucial. 

%KC: I think that it may be too strong to say "impossible to derive" chemical abundances. You could say "challenging". I do not know the literature well but I thought there were some studies that attempted to separate the contributions from the stars in SB2. In any case challenging makes it ok.

Studying close binaries in open clusters provides an opportunity to circumvent the limitations of SB2s. Stars in an open cluster form from a homogeneous distribution of gas at the same time, implying that they all have the same age, the same distance \citep{lada}, and most importantly  the same birth abundances. Many studies support this assumption of OC chemical homogeneity \citep{desilva1,bovy, ness,cheng2021,sinha2024}, constraining it within 0.1~dex across the light elements. There are evolutionary effects, such as atomic diffusion, and potential pollution events, such as rocky planet ingestion, that have been shown to cause chemical inhomogeneities. However, these events are either located in predictable parts of evolutionary space or are very short-lived \citep{souto,diffsuion,mack2016,vejar2021}.

%KC: Needs a space between short-lived from the reference

Since close binaries that form in open clusters are also subject to the principle of identical birth chemistry, it is possible to measure the effect of close binarity and stellar interactions without measuring the abundances of both binary members. Indeed, previous works such as \citet{ramos2024} have studied the overall abundances of spectroscopic binaries in clusters. We intend to take this concept further, to compare the chemical abundances of a cluster SB1 primary against that of a nonbinary cluster sibling star at the same evolutionary phase in order to directly measure the chemical similarity between cluster SB1s and single stars.

This work aims to constrain the chemical homogeneity of SB1s, compared to single stars, in open clusters within a large set of abundances. The structure of the paper is as follows: Section~\ref{sec:Data} outlines the survey data, determination of open cluster membership, and characterisation of binarity. Section~\ref{sec:Meth} discusses the calculation of orbital parameters for binaries as well as the derivation of abundances. Section~\ref{sec:results} presents the results of our work, and Section~\ref{sec:discussion} compares our results to previous findings.
\section{Open Cluster Sample Selection}
\label{sec:Data}
\subsection{SDSS-V}
\label{sec:sdss5}

The spectra and radial velocities we use are drawn from the Milky Way Mapper (MWM; J.A.~Johnson, \emph{in prep}), a component of the fifth generation of the Sloan Digital Sky Survey \citep[SDSS, ][]{kollmeier2025,sdssv}. We use data from Data Release 19 \citep[DR19; ][]{dr19}, which builds off of the observing strategies and survey goals outlined in SDSS Data Release 18 \citep{dr18} and includes observations of over one million targets cumulatively observed between SDSS-III and -V \citep{eisenstein2011,blanton2017,dr17}.

SDSS-V/MWM uses two 2.5~m telescopes: the Sloan Foundation Telescope at APO \citep{apo} and the du~Pont Telescope at LCO \citep{lascampanas}. Both are outfitted with nearly identical custom-built 300-fibre APOGEE spectrographs \citep{spectrograph}, with a resolution of $R \sim 22,500$, spanning the range of wavelengths between 1.51--1.70 $\mu$m. Unlike  SDSS-IV, which used a plug-plate system, SDSS-V now uses robotic fibre positioners \citep{fps}, which benefited from the adoption of a three-element corrector for the Sloan telescope at APO \citep{connector}. 

We use stellar metadata, radial velocities, and fundamental stellar parameters from the DR19 \texttt{mwmAllStar}, \texttt{mwmVisit}, and \texttt{astraAllStar} tables\footnote{\url{https://www.sdss.org/dr19/mwm/astra/output-files/}}. We use spectra from the DR19 \texttt{mwmStar} files, which contain reduced, radial velocity corrected, resampled, and pseudo-continuum-normalized data. We also make use of the ASPCAP DR19 linelist, which is described in \citet{linelist}, for our final line selection as detailed in Section~\ref{sec:bacchus}.

\subsection{Stellar Quality Cuts}
\label{sec:quality}
Our goal is to define a sample of open cluster member stars for which we can (i) confidently infer a likelihood of having a close binary companion and (ii) measure reliable elemental abundances. To that end, we apply a series of quality cuts to the stellar spectra at the visit and combined level and to the initial stellar properties inferred from those spectra.

Following the recommendations in \citet{whelan2020}, we enforce the following visit-level quality cuts in the \texttt{spectrum flag} column\footnote{More detailed documentation on the flagging schema and quality cuts can be found here: https://www.sdss.org/dr19/tutorials/} at the \texttt{mwmVisit} level:

\begin{description} \itemsep -2pt
 \item \texttt{BITMASK 3; VERY\_BRIGHT\_NEIGHBOR} == False; Star has neighbor more than 100 times brighter.
 \item \texttt{BITMASK 9; PERSIST\_HIGH} == False; Spectrum has at least 20 percent of pixels in high persistence region.
 \item \texttt{BITMASK 12; PERSIST\_JUMP\_POS} == False; Spectrum has obvious positive jump in blue chip.
 \item \texttt{BITMASK 13; PERSIST\_JUMP\_NEG} == False; Spectrum has obvious negative jump in blue chip.
 \item \texttt{BITMASK 16; SUSPECT\_RV\_COMBINATION} == False; RVs from synthetic template differ significantly ($\sim$2~km/s) from those from combined template.
\end{description}

We enforce the following quality cuts in the \texttt{spectrum flag} column at the \texttt{mwmAllStar} level:

\begin{description} \itemsep -2pt
\item \texttt{BITMASK 2; BRIGHT\_NEIGHBOR} == False; Star has neighbor more than 10 times brighter.
 \item \texttt{BITMASK 3; VERY\_BRIGHT\_NEIGHBOR} == False; Star has neighbor more than 100 times brighter.
 \item \texttt{BITMASK 16; SUSPECT\_RV\_COMBINATION} == False; RVs from synthetic template differ significantly (~2 km/s) from those from combined template.
\end{description}
and in the \texttt{results flag} column at the \texttt{mwmAllStar} level:
\begin{description} \itemsep -2pt
 \item \texttt{BITMASK 6; TEFF\_GRID\_EDGE\_BAD} == False; Teff is within 1/8th of a step from the grid edge
 \item \texttt{BITMASK 8; LOGG\_GRID\_EDGE]\_BAD} == False; logg is within 1/8th of a step from the grid edge
 \item \texttt{BITMASK 10; V\_MICRO\_GRID\_EDGE\_BAD} == False; v\_micro is within 1/8th of a step from the grid edge
 \item \texttt{BITMASK 14; M\_H\_ATM\_GRID\_EDGE\_BAD} == False; [M/H] is within 1/8th of a step from the grid edge
\end{description} 

For this study, we limit our analysis to FGK stars within the range of $4000 < T_{\rm eff} < 6500$~K. To mitigate the risk of contamination from radial velocity jitter, a phenomenon found in upper giant stars \citep{shroyer2020}, we also remove any stars with $\log{g} < 1$. We set a limit of $\rm SNR > 100$ to ensure we can recover blended \ion{Ce}{II} lines. We calculate the number of visits by summing the number of \texttt{mwmVisit} entries that pass the quality cuts outlined above. We also remove any stars identified as SB2s from \citet{kounkel2021}, as studies of double-lined spectroscopic binaries require specialised analysis that is beyond the scope of this project.

%KC: Ce lines > Ce II lines

To ensure the time-series RVs are sufficiently reliable for both the binarity and cluster membership determinations, we enforce that:
\begin{description} \itemsep -2pt
    \item $\texttt{e\_v\_rad} < 1$ km s$^{-1}$. Applied at the \texttt{mwmVisit} and astraStarASPCAP levels. This cut is made to exclude spurious, poorly-constrained, RV measurements
    
    \item $\log{g} \geq 1.0$, to mitigate risk of contamination from RV jitter. Applied at the astraStarASPCAP level.
    \item $\texttt{n\_visits} \geq 5$ Applied at the astraStarASPCAP level.
\end{description}

While it is possible to identify binaries using only 3 RV visits, this is insufficient to constrain parameters such as orbital separation \citep{castro_tapia_2024, badenes2018}. Therefore we opt to only include stars with N $\geq$ 5 visits such that constraints on orbital parameters is plausible. 

Finally, to ensure that we are analysing stars for which we can reliably determine chemical abundances using our existing tools (Section~\ref{sec:bacchus}), we use the entries in \texttt{astraStarASPCAP} to require that there are no failures in the global $[M/H]$ or $[\alpha/M]$:
\begin{description} \itemsep -2pt
    \item $\texttt{m\_h\_atm\_flag}$ == 0
    \item $\texttt{alpha\_m\_atm\_flag}$ == 0
\end{description}

%KC: This is curious that you make this final cut. Do you use the metallicity of alpha over Fe from ASPCAP? It would be interesting to see what metallicity and alpha over Fe you would get these stars that are removed. 

%[ADD ANSWER HERE!!]%

\subsection{Open Cluster Membership}
\label{sec:cluster_membership}
We start with the \citet{cg2020} open cluster catalogue, which includes membership probabilities, ages, and distances for approximately 2000 open clusters based on five inputs ($\alpha$, $\beta$, $\mu _{\alpha *}$, $\mu _{\beta *}$, and $\varpi$) from Gaia~DR2. 
We tested using the Gaia DR3-based catalogue from \citet{hunt} but found it did not expand our sample size. Furthermore, the systematically younger ages in this catalogue compared to other literature values \citep[see, e.g., Section~3 of][]{otto2026} could lead to systematically biased atmospheric parameters (Section~\ref{sec:atmospheric_parameters}).
%We opt to use the Gaia DR2 \citet{cg2020} cluster catalogue as opposed to the Gaia DR3 \citet{hunt} catalogue as \citet{cg2020} was incorporated into the SDSS-V targeting strategy and such there is better overlap. Furthermore the cluster ages from \citet{hunt} are systematically younger than literature ages for clusters older than 500 Myr which will result in incorrectly estimated atmospheric parameters. Finally due to the quality cuts outlined previously and the restrictive cuts required to establish a binary and non-binary sample within each cluster (outlined in Section \ref{sec:Meth}) the final cluster sample is unchanged between the two catalogues \footnote{For a detailed discussion of the choice between \citet{cg2020} and \citet{hunt} please see Section 3.0 of the SDSS-V DR19 OCCAM paper \citep{otto2026}.}.

We match the \texttt{astraStarASPCAP} sample defined in Section~\ref{sec:quality} to all \citet{cg2020} stars with membership probability $P \ge 0.7$ and a position within three cluster radii, defined in \citet{cg2020} as  the radius which contains 50$\%$ of members, from the cluster centre.

%KC: catalogue

Within these crossmatched clusters, we ensure minimal field star contamination by iteratively rejecting stars further than two standard deviations from the cluster median in the following four parameters: radial-velocity (RV), $\delta _{\alpha *}$, $\delta _{\beta *}$, and $\varpi$. Here, the two proper motion parameters and parallax are drawn from Gaia DR3, but radial velocities are drawn from APOGEE spectra. 

After our quality cuts and cluster membership selection, we are left with 267 stars in 20 open clusters. 

\section{Binarity and Binary Properties}
\label{sec:Meth}
%\subsection{Binary Membership}

The methods described in this binary selection are all sensitive to different areas of parameter space, and thus differing classifications of binarity between metrics are to be expected. Due to the 10-year observational baseline of SDSS, we are ideally suited to observe SB1s with periods $\leq$ 4000 days. In particular, both the \textsc{Joker} and \textsc{UltraNest} can leverage this footprint to confidently constrain parameters for edge-on binaries. However, as inclination (sin(i)) increases, the induced RV-variation decreases, reducing these tools' ability to detect binarity in a way that Gaia RUWE is independent of. Therefore, we categorise a star as binary if any one of the above methods classifies it as such. While APOGEE RVs are not sensitive to binaries with periods longer than 4000 days, binaries in this regime are beyond separations at which they can interact. 

Finally, the expected binary fraction in FGK stars at solar metallicity ($[M/H] \ > \ -0.5$) is $\sim$10$\%$, though studies of open clusters measure higher spectroscopic binary fractions ($\sim$30$\%$) \citep{moe2019, geller2021}. Between this binary fraction and our detection sensitivity, it is unlikely that any contaminants in our baseline non-binary sample could skew our results.

\subsection{RV Scatter}
We measure $\Delta RV_{\rm max}$, defined as $RV_{\rm max} - RV_{\rm min}$, for each star in our sample, using the \texttt{mwmVisit} radial velocities. Following the recommendations of \citet{badenes2018}, we classify stars with $\Delta RV_{\rm max} > 3$ km s$^{-1}$ as likely binaries, and stars with $\Delta RV_{\rm max} < 3$ km s$^{-1}$ as potential nonbinaries.

\subsection{The Joker Orbital Estimates}
The \textsc{Joker} is a custom Monte Carlo sampler that uses variability in radial velocity visits to derive posterior solutions in six parameters: period, eccentricity, argument of pericentre, phase, velocity amplitude, and barycentre velocity. To accomplish this, it densely samples from a set of prior distributions in these six parameters that spans the entire parameter space and performs a brute force rejection sampling. While this approach has the advantage of avoiding local minima for which traditional MCMC methods are at risk, it requires a significant upfront investment. Specifically, prior samples must densely sample the entire parameter space. To ensure sufficiently dense prior sampling, we follow the recommendation of \citet{whelan2020} and generate 100,000,000 prior samples. However, in cases where rejection-sampling determines the posterior solution is unimodal, the \textsc{Joker} provides MCMC sampling routines to determine orbital solutions.

We make modifications to the default \textsc{Joker} initialiser to tune it for open cluster studies. In particular, we adopt a normal distribution in barycentric velocity, centred at the median cluster radial velocity ($<RV_{\rm cluster}>$), with a standard deviation equal to the cluster velocity dispersion ($\sigma(RV_{\rm cluster})$); these values are calculated using the median radial velocities for cluster members published in the DR19 astraStarASPCAP file. This choice is driven by the fact that open cluster binaries, being gravitationally bound to the cluster, should have a systemic velocity similar to that of an unpaired star in the same cluster. The prior samples are drawn from the following distributions\footnote{Here $\mathcal{U}$ refers to a uniform distribution and $\mathcal{N}$ refers to a normal distribution.}:

\begin{align*}
\mathrm{Period} \ (P) &= log(\mathcal{U}(2,4000)) \\
\mathrm{eccentricity} \ (e) &= Beta(0.867,3.03) \\
\mathrm{Mean \ anomaly} \ (M_{0}) &= \mathcal{U}(0,2\pi) \\
\mathrm{Pericenter \ arguement}\ (\omega)&= \mathcal{U}(0,2\pi) \\
\mathrm{RV \ jitte} \ (s)&= \mathcal{N}(\mu_{y},\sigma_{y}^{2}) \\
\mathrm{RV \ amplitude} \ (K)&= \mathcal{N}(0,\sigma_{K}^{2}), \\
\mathrm{Systemic \ velocity} \ (v_{0})&= \mathcal{N}(v_{<RV_{\rm cluster}>}, v_{\sigma(RV_{\rm cluster})})
\end{align*}
Where period $P$ is in days; \textit{M$_{0}$} and \textit{$\omega$} are in radians; and \textit{s}, \textit{K}, and \textit{v$_{0}$} are in km~s$^{-1}$. See the \textsc{Joker} documentation\footnote{https://thejoker.readthedocs.io/en/latest/examples/1-Getting-started.html} for more information.

%KC: (See the Joker documentation for more information)

We also correct for any potential RV uncertainty underestimation with Equation 1 from \citet{whelan2020}:
\begin{equation}
\sigma_{\rm RV,total}^{2} = (3.5(\sigma_{\rm RV})^{1.2})^{2} + (0.072)^{2} \ \ \mathrm{km \ s^{-1}}
\end{equation}
to ensure that the visit-level velocity uncertainties are conservatively estimated.

We create a baseline for each star by generating synthetic radial velocity visits drawn from a normal distribution, with the centre at the observed v$_0$, and dispersion equal to the observed star's radial velocity uncertainty. Both synthetic and observed data are passed to the \textsc{Joker} which determines the log-likelihood of an orbital solution. We then compare the log-likelihoods of the best fit solutions, where 
\begin{equation}
\mathcal{L}_{observed} - \mathcal{L}_{synthetic} > 4.6
\end{equation}

indicates that the observed data are likely from a binary star. This threshold was inherited from \citet{whelan2020}, but we tested different thresholds and found no difference in our final results.

In the vast majority of cases, stars that are identified as binaries have highly multimodal posteriors. As a result, no single confident orbital solution can be recovered from them. 

However, in some cases, the posterior sampling is unimodal, and we can converge on a single orbital solution. This typically occurs when the star has sufficient radial velocity visits throughout the entirety of its orbital period. Typically, these stars have $N_{\rm visits} \geq 7$, but we note that many stars with $N_{\rm visits} \geq 7$ do not have unimodal posteriors due to sparse sampling across the entire orbital period. 

\subsection{UltraNest Orbital Estimates}
For stars with $N \geq 7$ visits, we corroborate our \textsc{Joker} orbital solutions using a nested sampling method as implemented in \citet{muellerhorn2025}. While this approach has many similarities to the \textsc{Joker}, it is built on a nested sampler called \textsc{UltraNest}\footnote{\url{https://johannesbuchner.github.io/UltraNest/index.html}} instead of a classic MCMC. As such, it is less prone to being trapped in local minima when sampling the parameter space, which gives it a distinct advantage when constraining potentially multimodal solutions. It also does not require generating a computationally expensive prior cache like the \textsc{Joker}. The prior distributions used are:

%KC: ...to being trapped in local minima

\begin{align*}
e &= \mathcal{U}(0,1) \\
P &= \mathcal{U}(0.2,4000) \\
M_{0} &= \mathcal{U}(0,2\pi) \\
\omega &= \mathcal{U}(0,2\pi) \\
\tau &= \mathcal{U}(0,1) \\
K &= \mathcal{U}(0,200), \\
v_{\mathrm{0}} &= \mathcal{N}(0,100)
\end{align*}

Here $\tau$ is the orbital phase. We apply the same likelihood criteria derived for the \textsc{Joker} here and inspect each orbital solution by eye to ensure it is a good fit to the observed data. From the sample of stars with N $\geq$ 7 radial velocity visits, we determine confident unimodal orbital solutions for 16 of them, hereafter called \textsc{UltraNest} values \citep[for consistency with][]{muellerhorn2025}. These 16 binaries will be used to analyse the relationship between abundance and orbital separation, while our multi-modal binaries will be used to study binary abundances at the population level.

\begin{figure*}
    \centering
    \includegraphics[width=0.8\linewidth]{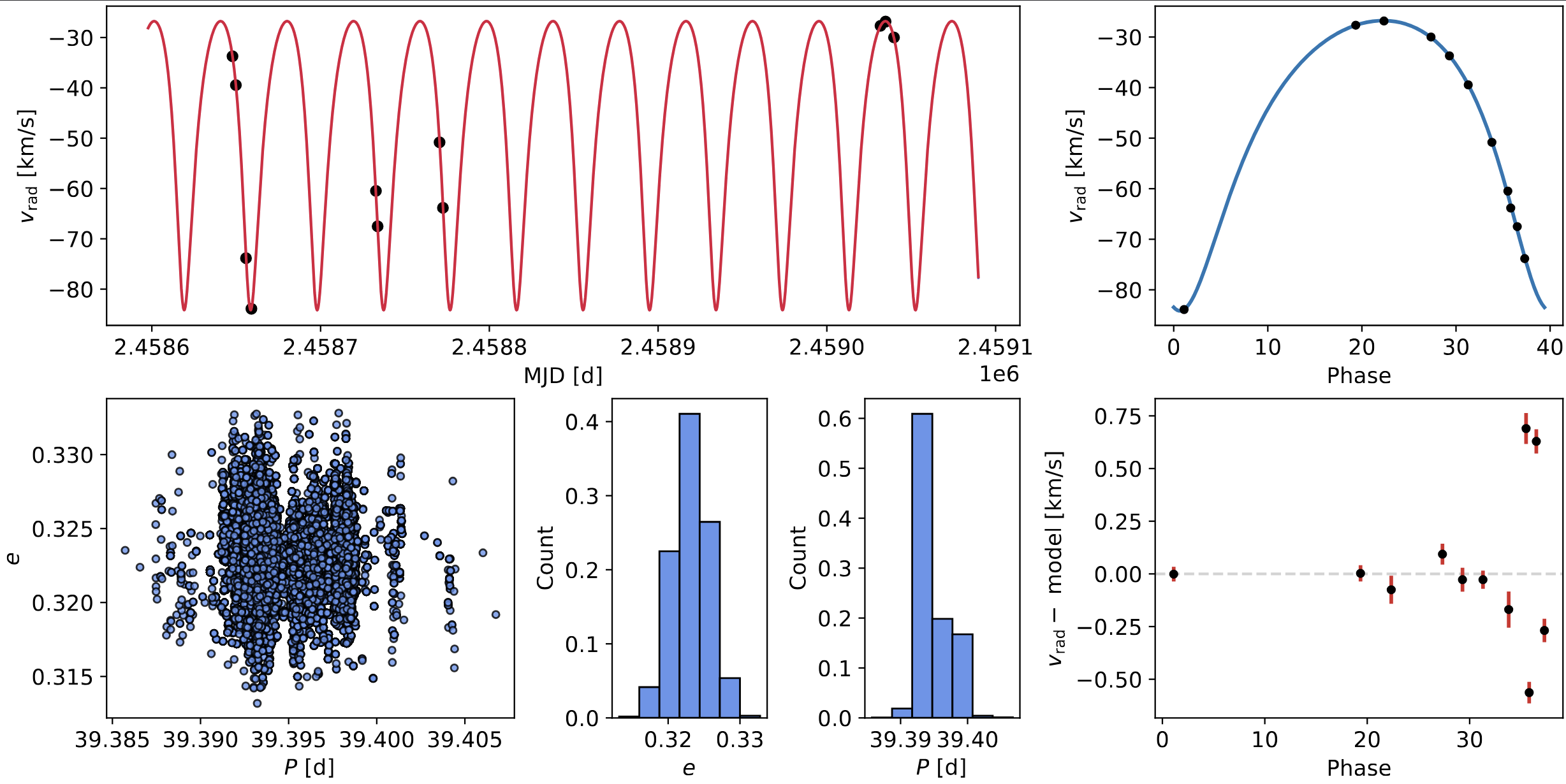}
    \includegraphics[width=0.8\linewidth]{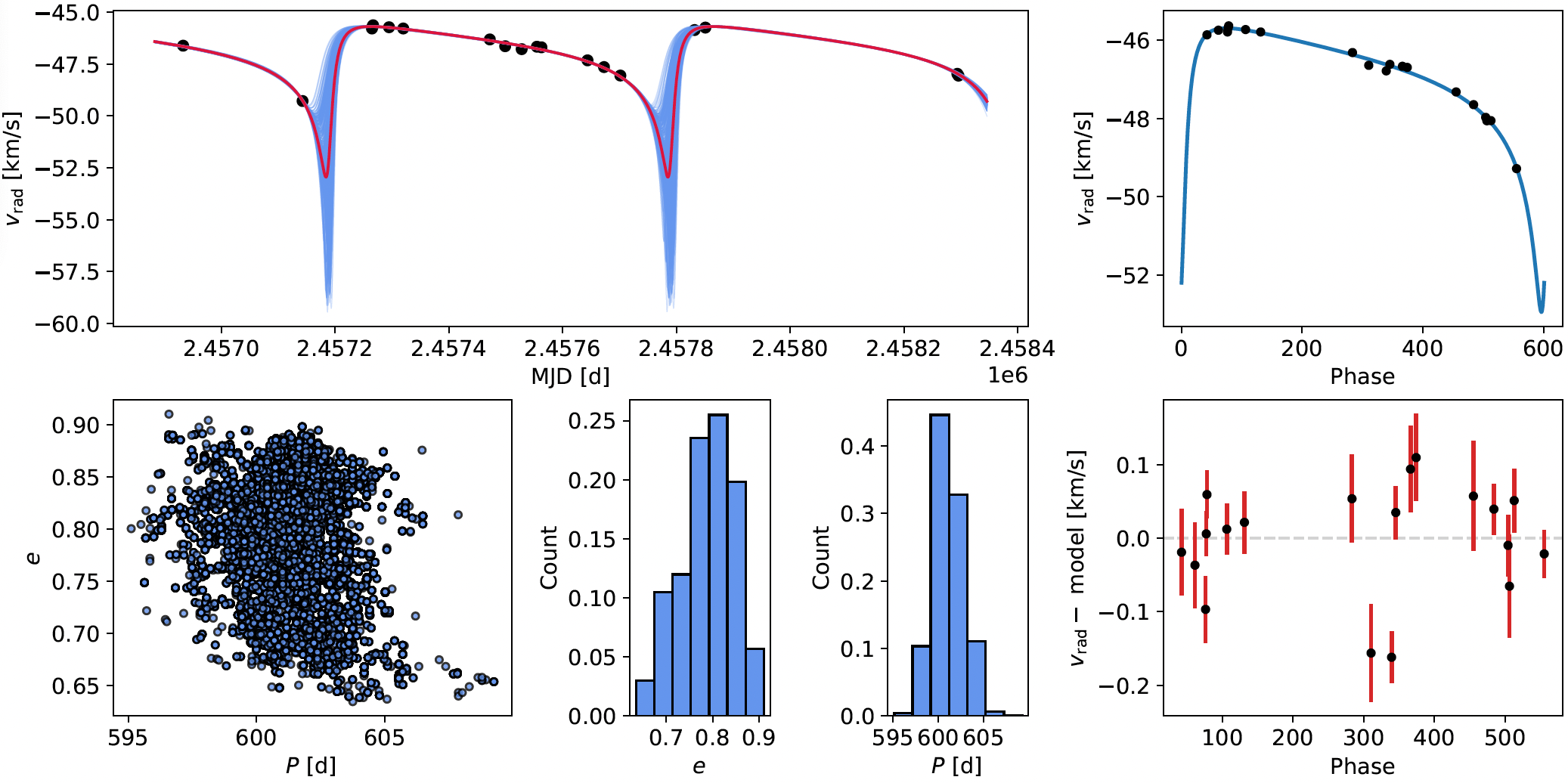}
    \caption{\textit{Top:} The orbital fit for SDSS 66658896, a short-period binary, as determined by \textsc{UltraNest}. We show the best fit orbital solution compared to the observed data points in the top left in MJD and orbital phase across the top row. In the bottom row, we show the \textsc{UltraNest} final live-point sampling for the orbital period, histograms of the sampling counts for period and eccentricity, and the residual between the observed data and model across orbital phase. \textit{Bottom:} The orbital fit for SDSS 66661885, a long-period binary, as determined by \textsc{UltraNest}. The ordering of information in panels is the same as in the top plot.}
    \label{fig:orbital_fits}
\end{figure*}

\begin{figure}
    \centering
    \includegraphics[width=\linewidth]{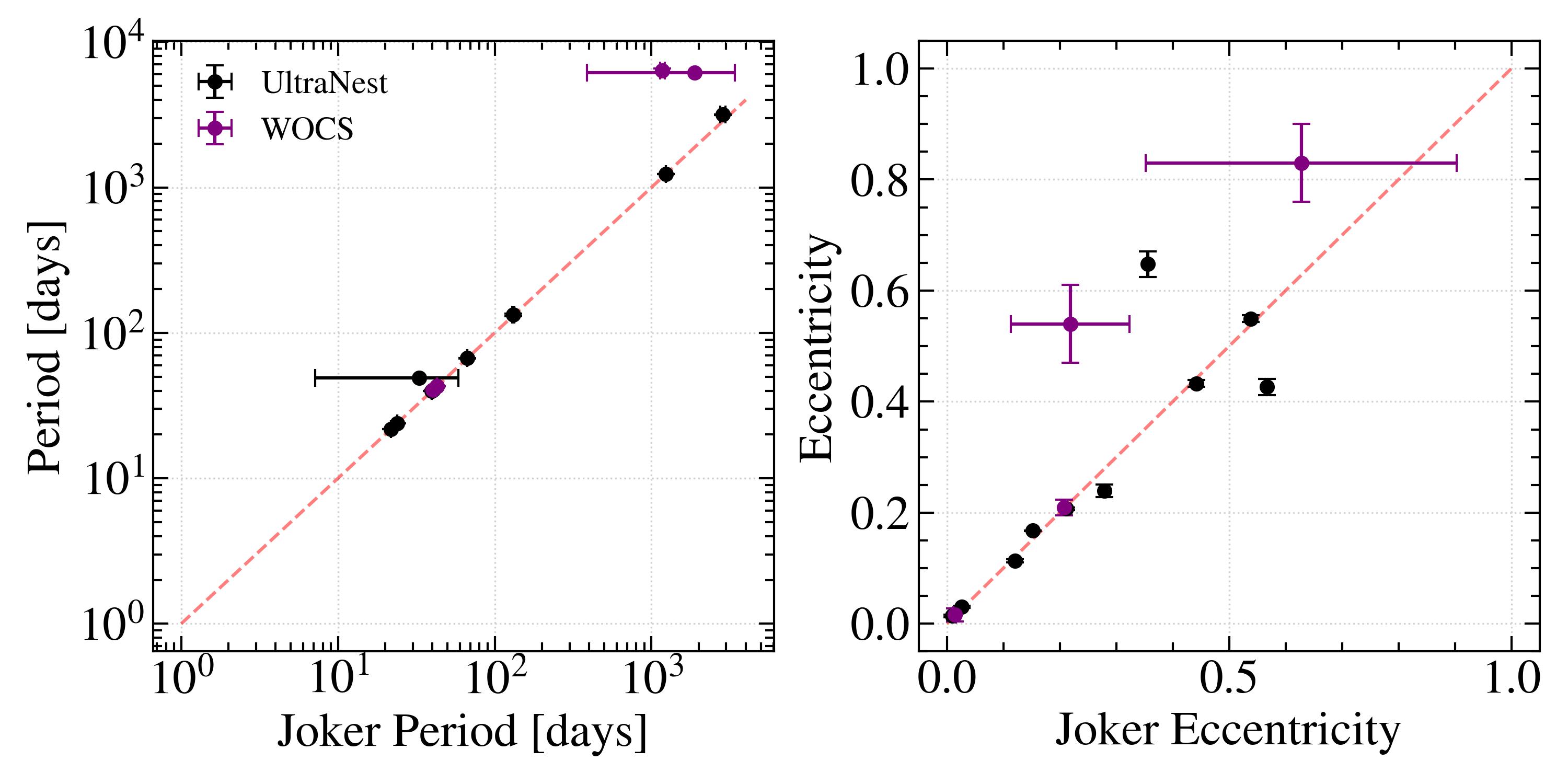}
    \includegraphics[width=\linewidth]{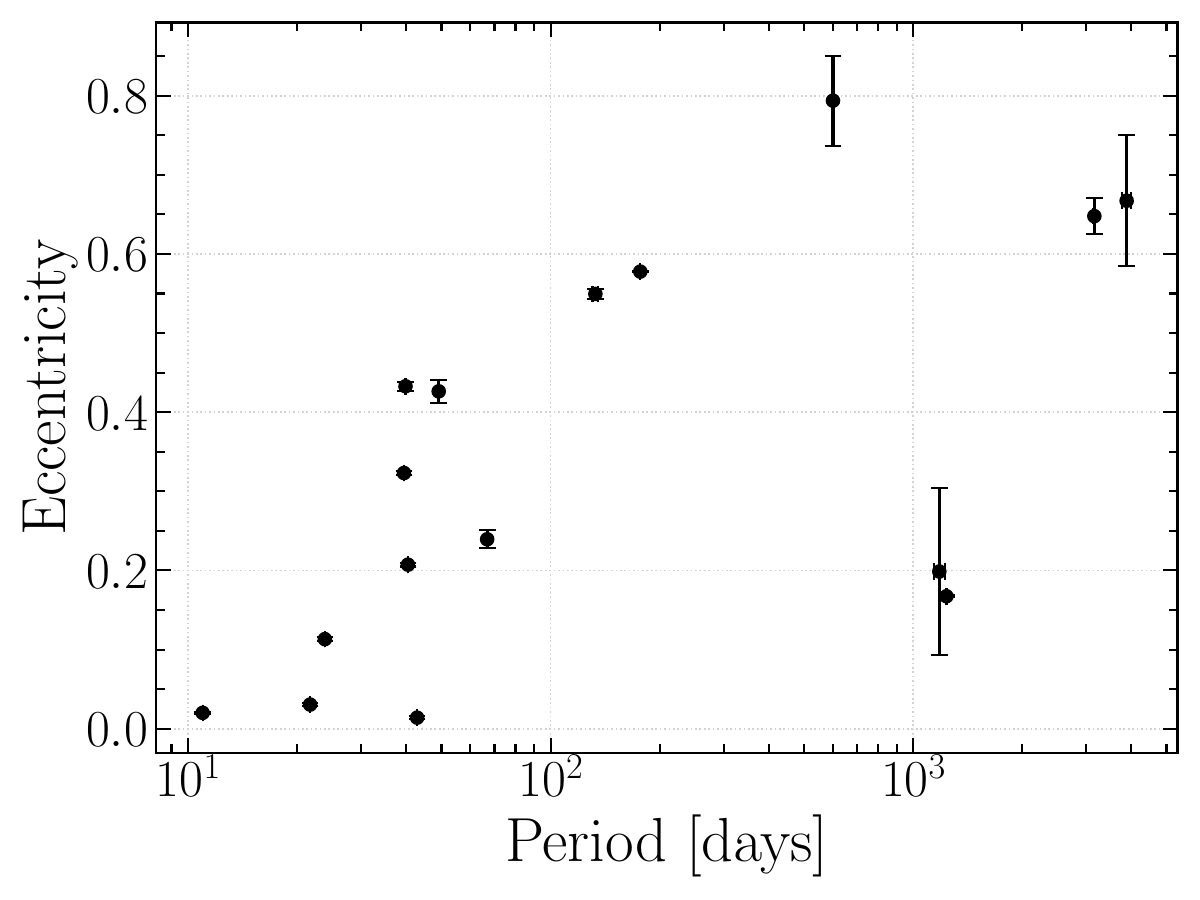}
    \caption{\textit{Top:} A comparison between the orbital solutions for unimodal stars analysed by the \textsc{Joker},  \textsc{UltraNest}, and WOCS. It is clear that while both codes exhibit reasonable agreement in period, there is significantly less agreement in eccentricity. This is likely due to the different priors used in the two codes. It is also clear that the \textsc{Joker/UltraNest} parameters broadly agree with those derived independently by WOCS. \textit{Bottom:} The period-eccentricity distribution of all binaries in our sample with confident orbital solutions, as solved for by  \textsc{UltraNest}.}
    \label{fig:orbital_distribution}
\end{figure}

We provide examples of our orbital fits for unimodal stars in Figure \ref{fig:orbital_fits}, and we compare the orbital solutions from the \textsc{Joker} and \textsc{UltraNest} in Figure \ref{fig:orbital_distribution}. Out of the 16 binaries we find that are confidently unimodal, all but one have consistent periods and eccentricities between the two codes. We also show the distribution of stars in period-eccentricity space in Figure \ref{fig:orbital_distribution}. To further verify our orbital solutions, we cross-match our sample with the WIYN Open Cluster Study \citep[WOCS;][]{mathieu2000}, which has conducted detailed photometric, astrometric, and spectroscopic studies of open clusters,  focusing on open cluster binaries. 

%KC: Is this Figure 2 or Figure 1? The first figure cited should be Figure 1. 

When cross-matched to our sample, we find that only four stars from the WOCS survey are included, all of which are members of M67 and were analysed in \citet{geller2021}. Within these stars, we compare the derived periods and eccentricities as shown in Figure \ref{fig:orbital_distribution}. From this, we see that the \textsc{UltraNest} and the \textsc{Joker} orbital solutions are in good agreement with those derived by \citet{geller2021}. The \textsc{Joker} appears to underestimate eccentricity, likely due to the use of a Beta distribution prior, which favours low eccentricities, instead of a Uniform distribution like \textsc{UltraNest}. 

%KC: The caption of current Figure 1 could be improved for clarity.

\subsection{Gaia RUWE \& NSS}
We supplement our sample with Renormalised Unit Weight Error (RUWE) measurements from Gaia DR3, which measures excess scatter in astrometric position as a potential proxy for binarity \citep{dr3, mueller_horn2025}. Using the \textsc{GaiaUnlimited} package \citep{castro_ginard2024}, we calculate individual RUWE thresholds for each star, which range between 1.19 and 1.3. Stars with measured RUWE greater than their \textsc{GaiaUnlimited} threshold estimate are classified as probable binaries. We choose this method over a flat threshold of 1.4 as \citet{castro_ginard2024} demonstrate that such a restrictive threshold leaves many binaries classified as single stars. Of the 28 stars identified as binaries by their Gaia RUWE, seven were already identified as a binary by our \textsc{Joker/UltraNest} analysis.

%KC: demonstrates > demonstrate
%KC: Any stars that were identified to be binary give just their ruwe that had not been identified as such in the JOker+Ultra nest analyses?

%add comment here!!%

Finally, we cross-reference with the Gaia Non-Single-Star (NSS) catalogue \citep{dr3_nss} to check for any orbital solutions. Only two stars in our sample, Gaia 604917728138508160 and 573967403529044352, have matches in the NSS catalogue, and both were detected as SB1s by our \textsc{Joker/UltraNest} analysis.

We require that each cluster has at least one binary and one non-binary star. After removing clusters that do not pass this threshold, we are left with 222 stars (119 nonbinaries and 103 binaries) across 14 open clusters. A detailed cluster-by-cluster breakdown of our sample can be seen in Table \ref{tab:cluster_pop}. We note that due to the underlying survey selection criteria, exhaustive quality cuts, and restrictive requirements on classification as nonbinary or binary, these are not reflective of the true underlying cluster binary fractions and should not be used as such.

\begin{table}
 \caption{A breakdown of the binary and nonbinary populations in each cluster used in this study. We also provide the ages, distance moduli, and $A_{v}$---drawn from \citet{cg2020}---used to analyse members of each cluster.}
 \label{tab:cluster_pop}
 \centering
    \begin{tabular*}{\columnwidth}{c|c|c|c|c|c|l}
    \hline
       Cluster  & log(Age/yr) & D$_{mod}$ & $A_{v}$ & Nonbinaries  & Binaries \\
    \hline
    \hline
       Collinder 69 & 7.10 & 8.10 & 0.25 & 2 & 1 \\
       Melotte 111 & 8.81 & 4.67 & 0.00 & 2 & 5 \\
       Melotte 20 & 7.71 & 6.17 & 0.30 & 10 & 7 \\
       Melotte 22 & 7.89 & 5.55 & 0.18 & 6 & 2 \\
       NGC 188 & 9.85 & 11.15 & 0.21 & 5 & 6 \\
       NGC 2158 & 9.19 & 13.17 & 1.44 & 4 & 3 \\
       NGC 2168 & 8.17 & 9.79 & 0.46 & 12 & 16 \\
       NGC 2243 & 9.64 & 12.85 & 0.02 & 4 & 3 \\
       NGC 2420 & 9.24 & 12.06 & 0.04 & 9 & 4 \\
       NGC 2632 & 8.83 & 6.31 & 0.00 & 7 & 4 \\
       Messier 67 & 9.63 & 9.75 & 0.07 & 30 & 46 \\
       NGC 6791 & 9.80 & 13.13 & 0.70 & 10 & 10 \\
       NGC 6819 & 9.35 & 12.21 & 0.40 & 13 & 9 \\
       NGC 752 & 9.07 & 8.42 & 0.07 & 10 & 6 \\
       \hline
    \end{tabular*}
\end{table}

%KC: Here, we require...

\section{Abundance Determination}
\label{sec:abundance_determination}
\subsection{Atmospheric Parameters}
\label{sec:atmospheric_parameters}

While large industrial pipelines such as ASPCAP and Astra are well suited for chemical species with clean absorption features---such as Fe, Mg, and Si---they are significantly less equipped to deal with weak or blended features such as those from Ce and Nd \citep{bawlas, cunha2017, hasselquist2017}. As these are $s-$process elements that can identify post-mass transfer objects, accurate determination of them is crucial for this analysis. Therefore we opt to derive our own abundances. To ensure a homogeneous analysis, we derive chemical abundances for every chemical species in our study. This also has the added benefit of allowing us to be judicious with our line selection, and exclude many poorly fit absorption lines or temperature-sensitive features.

%KC: Please reference the Cunha et al. (2017) paper for Ce and the Haselquist et al. (2016) paper for Nd besides the bawlas paper.
%KC:derive abundances > derive chemical abundances
%KC: When we say astra does it mean ASTRA- ASPCAP? Mening they use the ASPCAP methodology? If so, might be good to use ASTRA-ASPCAP. 

In this section we leverage the precise distances and ages from \citet{cg2020}, MIST isochrones, and SDSS-V effective temperatures to derive stellar masses, and empirical atmospheric parameters such as surface gravity and microturbulent velocity.

%KC: SDSS-V temperatures > SDSS-V effective temperatures 
%KC: microturbulence > microturbulent velocity.
%KC: I like that you find better agreement with the incalibrated values! :-)

We make use of the distance moduli (\texttt{DMNN}), extinctions (\texttt{AVNN}), and ages (\texttt{AgeNN}) from \citet{cg2020} for each cluster. To determine the metallicity of each cluster, we use the median [Fe/H] abundance derived by Astra. We then interpolate a MIST isochrone to those values for each cluster using the \textsc{isochrones} package \citep{mist3, isochrones,mist1,mist2}. Observed stars in our sample are matched to corresponding interpolated isochrone Equivalent Evolutionary Points (\texttt{EEP}s) by performing a chi-squared minimization in Gaia $(BP-RP)$ and \textit{G} magnitudes. We also test matching on the 2MASS \citep{shrutskie2006} \textit{J}, \textit{H}, and \textit{K$_{s}$} filters and find that this does not meaningfully change the derived parameters. 

%KC: Might be good to explain how ASPCAP and ASTRA-ASPCAP (if it is the aspcap method) derive effective tempreatures and surface gravities. If you will use them, it might be good to make it more clear that you use them and how they were derived. In case ASRTA uses the ASPCAP method it would be good to reference the paper by Garcia Perez et al. (2015) that is the original paper that explains the methodology.

The APOGEE wavelength range includes many temperature-sensitive lines \citep{meszaros2025}, and the effective temperatures from Astra are broadly trustworthy, the methodology for which is described in \citet{ASPCAP2016}. To verify this, we independently derive effective temperatures for a subset of stars in our sample using their \textit{ugrizJHK} photometry with the InfraRed Flux Method \citep[IRFM; e.g.,][]{casagrande2010}. We follow a similar methodology to \citet{Hawkins2016}, and find strong agreement between the two sets of temperatures ($\Delta$T$_{\mathrm{eff}} < 150\ K$). As we find stronger agreement between IRFM and the uncalibrated temperature (\texttt{raw\_teff}) values, we use them instead of the calibrated effective temperatures.

Unlike effective temperature, the APOGEE region does not have as many surface-gravity dependent lines \citep{meszaros2025}. Therefore, it uses multiple calibration methods, such as comparison to asteroseismic $\log{(g)}$ values, to derive calibrated surface gravities (see Casey et al. in prep). However, these calibrations do not extend to stars with $T_{\rm eff} < 4500$~K due to the lack of asteroseismic calibrations in that part of parameter space. Since many of our stars fall below this limit, rederiving correct surface gravities is imperative for our abundance analysis. To derive surface gravities, we use the Stefan–Boltzmann relation:

%KC: temperature > effective temperature
%KC: APOGEE does not have ... > The Apogee region does not have ...

\begin{multline}
\label{eqn:logg}
log(g)_{\star,empirical} = log(g_{\odot}) + log(\frac{M_{\star}}{M_{\odot}}) \\
+ 4log(\frac{T_{\star}}{T_{\odot}}) + 0.4(M_{bol,\star} - M_{bol,\odot}).
\end{multline}

\begin{figure*}
    \centering
    \includegraphics[width=0.4\linewidth]{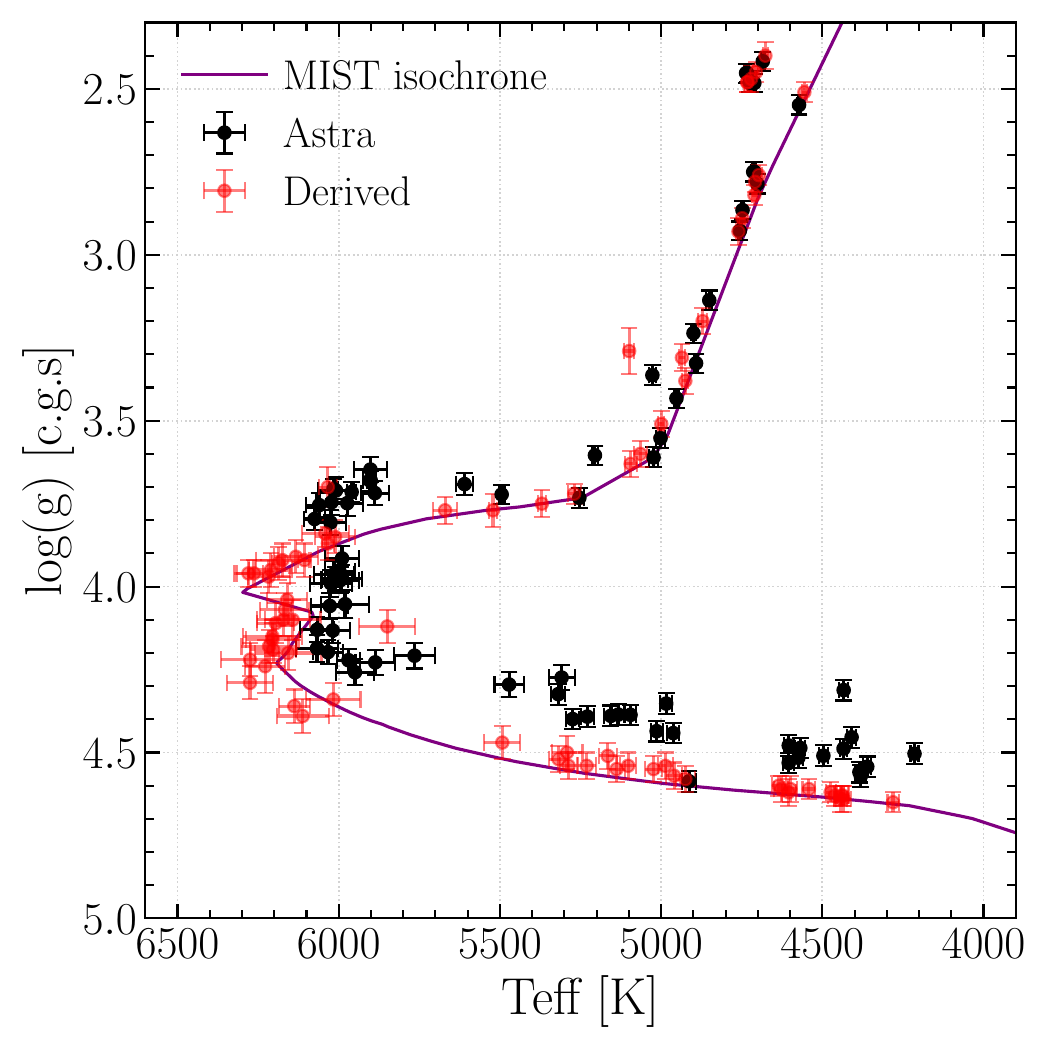}
    \includegraphics[width=0.4\linewidth]{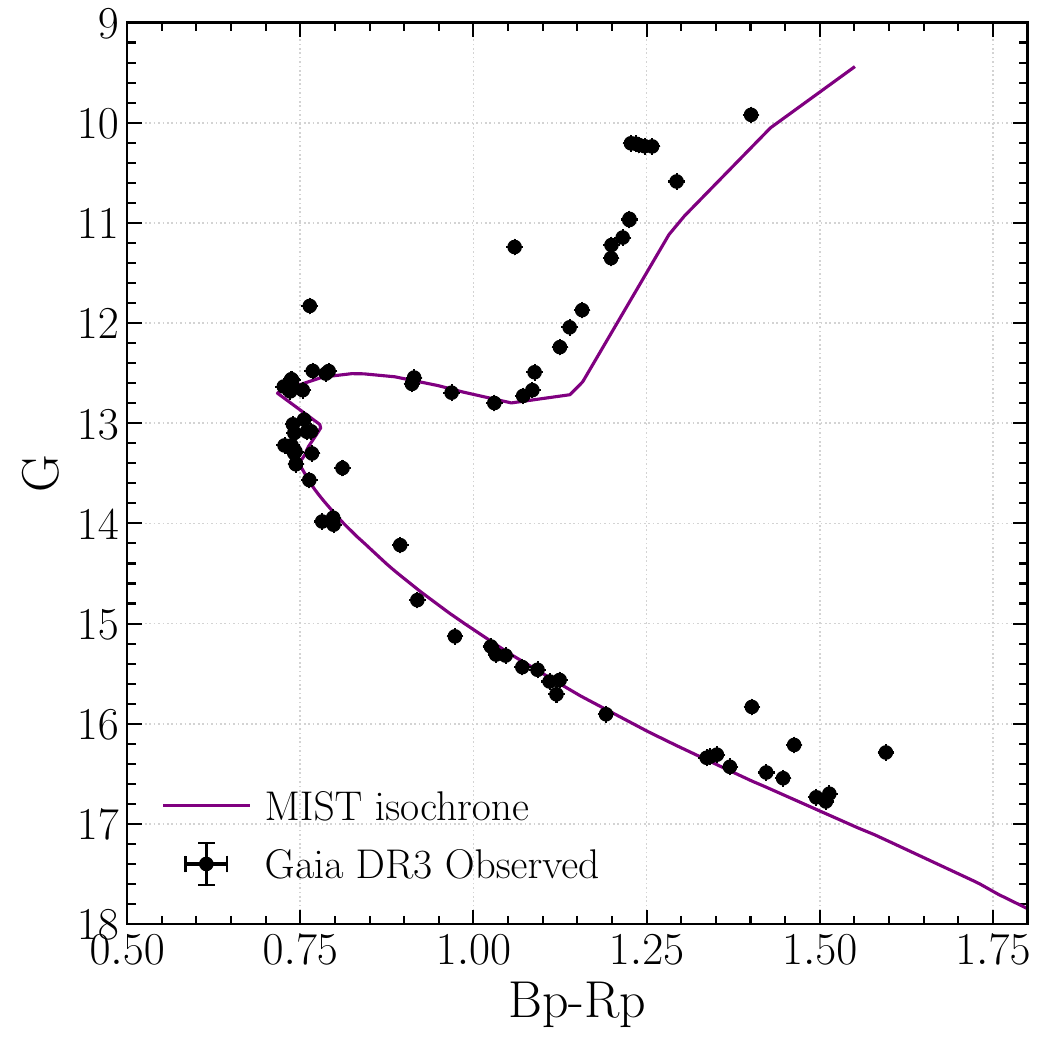}
    \includegraphics[width=1.0\linewidth]{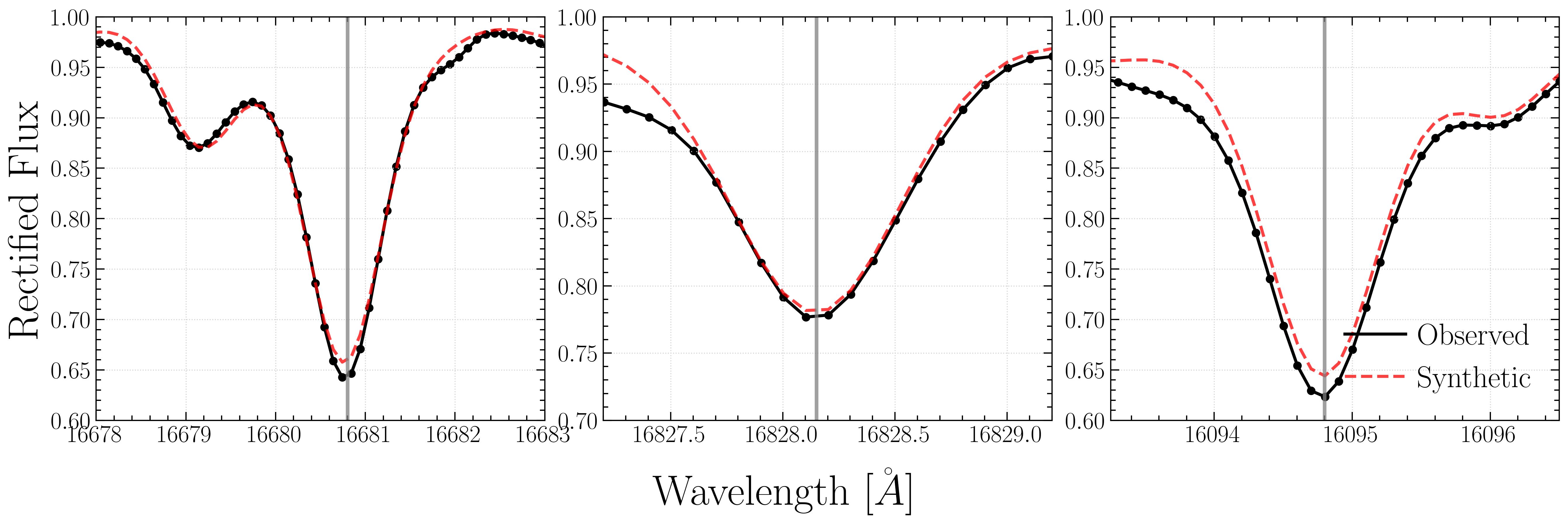}\\
    \caption{\textit{Top Left:} A comparison of the interpolated MIST isochrone for Messier 67 (NGC 2682) and the observed stars in $(BP-RP)$ and $G$. \textit{Top Right:} Here we show the effective temperatures and surface gravities used, as compared to the published values, for M67. \textit{Bottom:} We show the three Si lines used in this study to derive broadening: 16680.8\AA, 16828.2\AA, and 16094.8\AA, with the observed spectrum of a typical star in black. We show a synthetic spectrum generated using an interpolated MARCS stellar atmosphere at that star's derived log(g), \textbf{$T_{\rm eff}$}, [Fe/H], and microturbulence in red. While there are slight differences between the observed and synthetic spectrum on the left wing in the rightmost plot, this does not affect the synthetic fit to the line core---and therefore the derived convolution.}
    \label{fig:iso_fit}
\end{figure*}

Here M$_{\star}$ and M$_{bol,\star}$ are drawn from the star's interpolated \texttt{EEP}, and T$_{\star}$ is the star's \texttt{raw\_teff} value. We use the following solar values: $\log(g)_{\odot} = 4.438$~dex, $M_{\rm bol, \odot} = 4.75$, and $T_{\rm eff,\odot} = 5772$~K. These are the same values recommended by \citet{prsa2016} and are derived from estimates on the solar photospheric radius, radiative luminosity, and measurements of nominal total solar irradiance. We also use these within our abundance analysis to ensure consistency. While stars on the giant branch and red clump have relatively consistent Astra and isochrone-derived values, there are significantly more offsets within the dwarf stars, particularly at lower temperatures, as shown in Figure \ref{fig:iso_fit}. We find that the Astra-derived surface gravities are systematically overestimated for stars below the giant branch, and that for stars near the turn-off, the \texttt{raw\_teff} values are far better aligned with isochrone predictions.

From our empirically derived surface gravity, we determine the microturbulence using the equation from \citet{masseron2019}, where $\mu_{t}$ is in km/s:

\begin{equation}
\label{eqn:vmicrp}
\mu_{t} = 2.488 - 0.866\, \log(g) + 0.1567\, \left(\log(g)\right)^{2}.
\end{equation}

Finally, we measure the $v \sin{i}$ for all the stars in our sample using \textsc{Korg} \citep{korg}, as Astra only measures $v \sin{i}$ for dwarf stars. As the resolution of APOGEE spectra is not sensitive to $v \sin{i} \leq 5$ km s$^{-1}$, and most giant stars are slow rotators, $v \sin{i}$ is not measured for them. This is due to the fact that giant stars are generally slow rotators ($v \sin{i} \leq 5$ km s$^{-1}$) due to magnetized winds from the convective envelope carrying away angular momentum \citep{deMedeiros1996,tayar2015}. However rapid-rotation on the red giant branch (RGB) can be seen as an indicator of unusual history, such through planetary ingestion or binary interaction \citep{carlberg2012,patton2024}. However, to identify potential stars of interest in our sample, we do want to constrain $v \sin{i}$ as much as we can. 

%KC: produces > measures
%KC: paragraph above could be improved... to say from the onset that maeasures v sin i only for stars having v sin i larger than 5km.s. This is only said at the end.

%KC: Caption of Figure 3 does not seem to be completely OK. please double check.

\begin{figure}
    \centering
    \includegraphics[width=\linewidth]{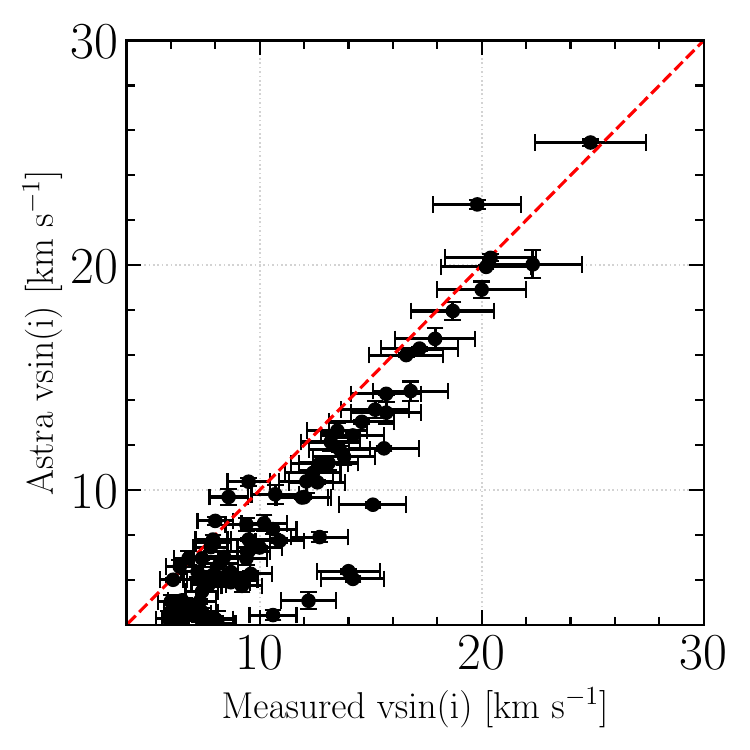}
    \caption{Here we compare the derived values of $v \sin{i}$ to the published Astra values. We find strong agreement between the two samples at values below $\sim$30~km~s$^{-1}$ ($\sim$85\% of the sample).}
    \label{fig:vsini}
\end{figure}

Using the effective temperatures and now-determined surface gravities, as well as the star's [Fe/H], [Mg/H], and [Si /H] abundance from  BACCHUS, we synthesize a set of synthetic spectra between $0 \le v \sin{i} \le 80$~km~s$^{-1}$, with a stepsize of 5~km~s$^{-1}$. We then perform a chi-squared minimization across the entire spectrum and use a \texttt{CubicSpline}, from the \textsc{SciPy} package, to determine the best-fit $v \sin{i}$. For dwarf stars with Astra $v \sin{i}$, we compare them to our derived values and find strong agreement, as shown in Figure \ref{fig:vsini}. We note that the difference between the Astra and our $v \sin{i}$ estimates increases at large values, but it is almost always less than  $\Delta v \sin{i} \sim 5$~km s$^{-1}$. We find strong agreement between the two samples at values below $\sim$30~km~s$^{-1}$ ($\sim$94\% of the sample).

%KC: Using the now determined effective temperature? You derived the Teff? KC got confused... Surface gravities you derived your own.
%KC: Also I think you did not make it clear earlier that you will use the iron, Mg and Si abundances from ASTRa-aspcap. you said they were reliable but was it clear that you would adopt these?

For the giant stars in our sample, we find that $\sim85\%$ have $v \sin{i} \leq 5$~km~s$^{-1}$. We only find one giant star (Gaia 865402924795303040) that could be considered a rapid-rotator, with $v \sin{i} =12.7 \pm 1.7$~km~s$^{-1}$, though it was not classified as a binary by our analysis. 

%KC: Figure 4 is for dwarfs. Best to make this clear in the caption. Do you show the comparison for the giants somewhere? Might be a good second panel to this figure. A histogram with the distributon of V sin i of redgiants would be of interest. But I understand this is too late.

% e    =````````````````````````````````````````````````````````````````````````````````````````````````````````````````````````````````````````````````````````````````````````````````````````````````````````````````````````````````````````````````````````````````````````````````````````````````````````````````````````````````````````````````````````````````````````````````````````````````````````````````````````````````````````````````````````````````````````````````````````````````````````````````````````````````````````````````````````````````````````````````````````````````````````````````````````````````````````````````````````````````````````````````````````````````````````````````````````````````````````````````````````````````````````````````````````````````````````````````````````````      - Kuiper

\subsection{Derivation of  Chemical Abundances with BACCHUS}
\label{sec:bacchus}

%KC: Abudnances > Chemical Abundances

We use the Brussels Automatic Code for Characterizing High accUracy Spectra \citep[BACCHUS;][]{bacchus}, an ``on the fly" spectrum synthesis code, to derive elemental abundances from stacked mwmStar spectra. \textsc{BACCHUS} relies on the grid of MARCS model atmospheres \citep{MARCS}, Masseron's model atmosphere thermodynamic structure interpolator, and the radiative transfer code Turbospectrum \citep{TURBOSPECTRUM}. One of the primary benefits of \textsc{BACCHUS} over the Astra abundances is that it is better suited to determine abundances for weak or strongly blended lines, such as Ce and Nd. To model a stellar atmosphere, \textsc{BACCHUS} requires five parameters: T$_{\mathrm{eff}}$, log(g), $\mu_{t}$, [M/H], and convolution. Given the lack of usable \ion{Fe}{ii} lines within the APOGEE wavelength range required for a traditional excitation-ionization balance, we forego the derivation of temperature and surface gravity within \textsc{BACCHUS}. Instead, we fix each star's T$_{\mathrm{eff}}$, log(g), and $\mu_{t}$ at the values derived in the previous section. We set the initial guess for [M/H] at the same median cluster [Fe/H] used to interpolate its isochrone. We then solve for each star's individual metallicity using \ion{Fe}{i}. 

Unlike other spectral synthesis codes like \textsc{MOOG} \citep{moog} or \textsc{Korg}, \textsc{BACCHUS} does not measure $v \sin{i}$; instead, it uses a single convolution parameter that acts as a catch-all for rotation and instrumental broadening. To solve for this parameter, we follow a similar methodology to \citet{bawlas} and use three \ion{Si}{i} lines: 16680.8\AA, 16828.2\AA, and 16094.8\AA, as shown in Figure~\ref{fig:iso_fit}. We choose \ion{Si}{i} as the lines are strong and clean, but we also experiment with OH and \ion{Ti}{i} lines. We find that there is no significant difference in the derived convolutions. Finally, due to carbon's impact on model atmospheres, especially in cooler stars where molecules are present (e.g., CO and CN), we solve for the carbon abundance of each star to establish a proper chemical equilibrium. 

Once all five parameters and C have been established, we derive abundances for the following chemical species: Si, Fe, C, N, O, Na, Mg, Al, Ca, Ti, Cr, Ni, Ce, and Nd. While we initially measure abundances for all the observed absorption features measured in the APOGEE DR19 linelist, our final results only include information from a select set of confident lines, discussed in Section \ref{sec:line_selection}. We iterate this entire analysis twice for each star to ensure self-consistency in the atmospheric parameter and abundance determination. We run \textsc{BACCHUS} in ``manual" mode, using the packages \textsc{bacchus$\_$tools}\footnote{https://github.com/catherine-manea/BACCHUS$\_$Tools} and \textsc{PyBACCHUS}\footnote{https://github.com/rocketxturtle/PyBACCHUS} to interface with \textsc{BACCHUS} from within Python. Finally, we solar-scale our abundances using \citet{grevesse2007} recommended values.

\begin{figure*}
    \centering
    \includegraphics[width=\linewidth]{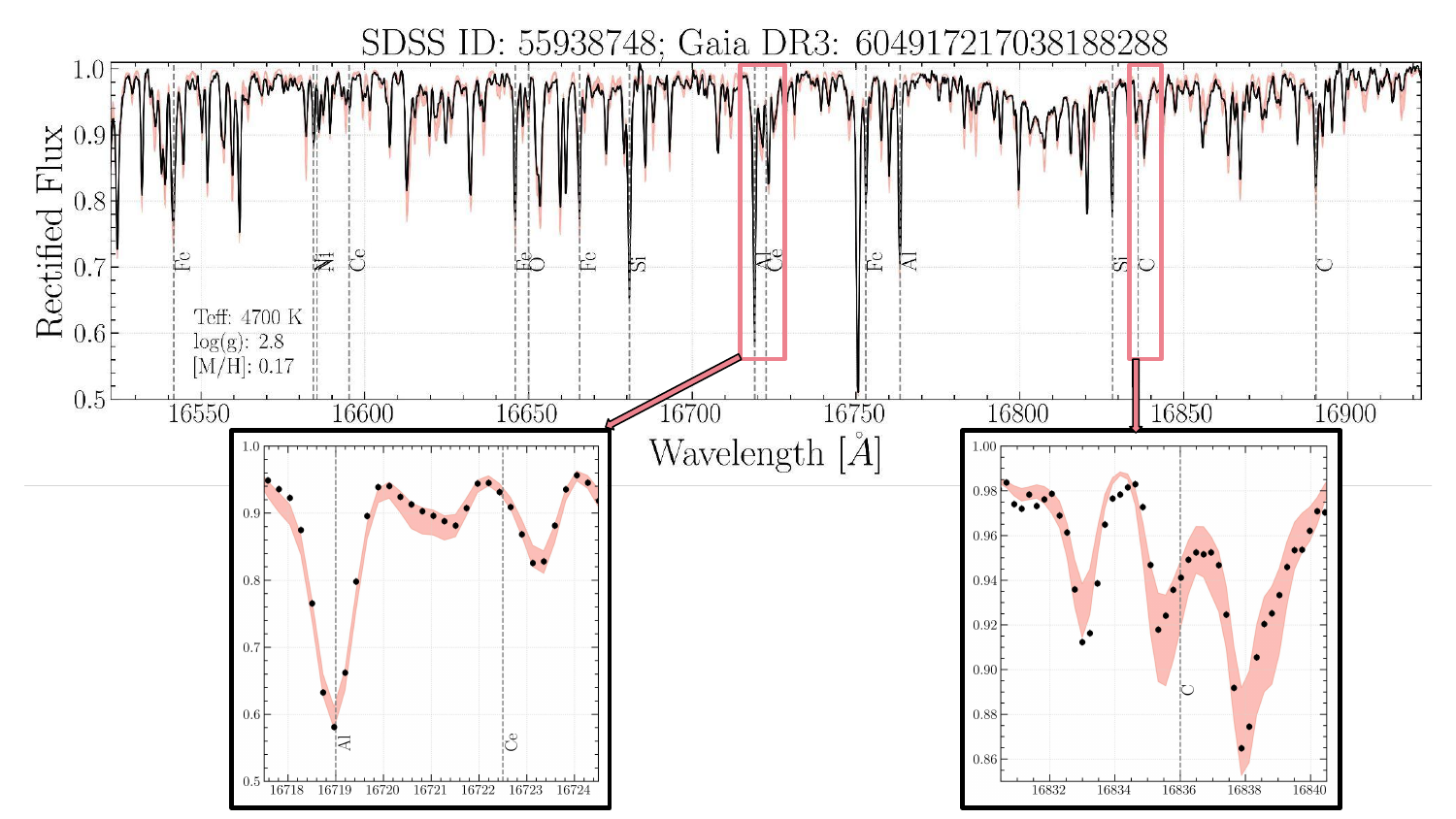}
    \caption{Here we show comparisons between the observed spectrum (black) and synthetic spectrum (salmon) for one of the stars in our sample, with our chosen absorption features shown as dashed lines (\S\ref{sec:line_selection}). We also show cutouts of two regions that contain C and Ce features to demonstrate the quality of our spectral synthesis. The shaded band shows a typical uncertainty in synthetic fit when considering the uncertainty in all the atmospheric parameters.}
    \label{fig:example_spectra}
\end{figure*}

Our final abundances for each star are taken as the median of all lines flagged as good by the \texttt{chi2} method \footnote{For a more detailed discussion of the strengths and weaknesses of BACCHUS's abundance determination methods, as well as its flagging metrics, please see Section 3.0 in \citet{bawlas}.}.  We also only select lines where the difference in line abundance derived through \texttt{chi2} and \texttt{eqw} is less than 0.3~dex. Finally, for species with more than two lines, we remove any individual line with abundance measurements more than 3$\sigma$ from the median. Because every chemical species depends on a correctly estimated microturbulence and convolution, we limit our sample to stars with confidently measured Si and Fe abundances. For these two elements, we require all four abundance determinations to have \texttt{flag} = 1 for a line to be used. Our derived model atmospheres closely agree with the observed spectra, as shown in Figure \ref{fig:example_spectra}.

\subsubsection{Line Selection $\&$ Temperature Trends}
\label{sec:line_selection}

%KC: ATTENTION: Table 1 has N I, CI, OI many lines! These are not neutral atomic lines. these are CN, CO, OH 
%KC: ATTENTION

To ensure that we have no trends in temperature that could affect our results, we measure the $T_{\rm eff}$ vs. A(X) relation for every elemental line in the APOGEE wavelength range, using M67 and Melotte 22 (the Pleiades). We choose the Melotte 22 as it allows us to compare with the line selection done by \citet{grilo2024}. Furthermore, these two clusters provide excellent coverage across the entire HR diagram, with robust coverage on the main sequence. 

We quantify each temperature-line abundance relationship using \textsc{scipy}'s \texttt{linregress} function. We remove any lines with significant systematic offsets ($\Delta \mathrm{A}(x) \geq 0.2$~dex), or trends in temperature, which we classify as a slope greater than $4 \times 10^{-5}$~dex~K$^{-1}$. Lines with slopes greater than this limit would induce 0.1~dex of chemical scatter (if they were the only lines being used) across the temperature range spanned by our sample. We also remove lines with $R^2 < 0.5$ as this implies the line has a large star-to-star scatter. Our final line selection can be found in Table~\ref{tab:lines}.

\begin{table}
 \caption{Atomic and molecular lines used for abundance determination. All wavelengths are given in air and in Angstroms (\AA).}
 \label{tab:lines}
 \centering
 \begin{tabular*}{0.8\columnwidth}{lll}
  \hline
  Element & Solar Value & Wavelengths (\AA)\\
  \hline
  \hline
  CO & 8.39 & 15783.9, 16004.9,16021.7, \\
    & & 16836.0, 16890.4 \\
  CN & 7.78 &  15222.0, 15228.8, 15242.5, \\
   & & 15708.5\\
  OH & 8.66 &  15719.7, 16650.0 \\
  \ion{Na}{i} & 6.17 &  16373.9, 16388.9 \\
  \ion{Mg}{i} & 7.53 &  15879.5, 15886.2, 15954.5 \\
  \ion{Al}{i} & 6.37 &  16763.4, 16719.0 \\
  \ion{Si}{i} & 7.51 &  16680.8, 16828.2, 16094.8 \\
  \ion{Ca}{i} & 6.31 &  16136.8, 16157.4 \\
  \ion{Ti}{i} & 4.90 &  15715.6 \\
  \ion{Cr}{i} & 5.64 &  15680.1 \\
  \ion{Fe}{i} & 7.45 &  15207.5, 15662.0, 15686.3, 15774.1,\\
    & & 15904.4, 15911.3, 15920.7, 16006.8, \\
    & & 16042.7, 16125.9, 16180.9, 16207.7,\\ 
    & & 16213.5, 16225.6, 16522.1, 16541.6, \\
    & & 16645.9, 16665.5, 16753.0\\
\ion{Ni}{i} & 6.23 & 15605.6, 15632.6, 16584.4, 16585.4 \\
\ion{Ce}{ii} & 1.58 & 15784.8, 16376.5, 16595.2, 16722.5 \\
\ion{Nd}{ii} & 1.45 &  15368.1, 16053.6, 16262.0 \\
  \hline
 \end{tabular*}
\end{table}

\subsubsection{Abundance Uncertainties}

\textsc{BACCHUS} does not propagate the uncertainties on the atmospheric parameters through the abundance determination. As a result, the abundance uncertainties it measures are strictly from the line-to-line dispersion. To ensure that the abundance uncertainties are correctly estimated, we create a grid in temperature, surface gravity, and [M/H] that spans the entire parameter space of our sample, with steps of 200 K, 0.5 dex, and 0.2 dex, respectively. Given that we limit our sample to stars with $\rm SNR > 100$, we assume there is no dependency on that parameter. We experimented with finer step sizes in these three dimensions and found no significant difference in uncertainties; therefore, these steps were chosen for the sake of computational efficiency. Within each bin, we arbitrarily choose ten stars and Monte-Carlo their temperatures, isochrone-derived surface gravities, and derived microturbulences within their uncertainties. Given that the derivation of [M/H] and convolution is determined in part by these three parameters, this effectively Monte-Carlos all five stellar parameters. For bins with fewer than ten stars, stars were reused as independent Monte-Carlo instances. 

From this, we repeat the abundance derivation methodology described above to get abundances for each star. Within each bin of temperature, surface gravity, and metallicity, we measure the [X/H] dispersion to estimate an empirical abundance uncertainty for stars in our sample at that effective temperature, surface gravity, and metallicity. 

The cluster membership, orbital and atmospheric parameters, and derived abundances from this study will be included in a machine-readable table. The columns of this table are shown in Table \ref{tab:results_table}.

\begin{table*}
 \caption{Summary Table of Parameters: List of columns in the stellar parameters table. [X/H] abundances are measured for the following elements: Fe, Si, C, N, O, Mg, Ca, Ni, Cr, Ti, Na, Al, Ce, and Nd. This table will be made available in machine readable format.}
 \label{tab:results_table}
 \begin{tabular*}{0.8\paperwidth}{lcl}
  \hline
  Column \ Name & Units & Description (\AA)\\
  \hline
  \hline
 sdss$\_$id & & SDSS DR19 identifier\\
 gaia$\_$dr3$\_$source$\_$id & & Gaia Data Release 3 identifier\\
 RA & deg & Right ascension from Gaia DR3\\
 DEC & deg & Declination from Gaia DR3 \\
 parallax & mas & Parallax from Gaia DR3 \\
 pmRA & mas~yr$^{-1}$ & Right ascension proper motion from Gaia DR3\\
 pmDEC & mas~yr$^{-1}$ & Declination proper motion from Gaia DR3 \\
 X$\_$mag & & Apparent magnitude in the \textit{B$_{\rm p}$R$_{\rm p}$G, JHK, W1W2, ugriz, y} filters  \\
 snr & & Signal-to-noise of the stacked mwmStar spectrum used. \\
 v$\_$rad & km s$^{-1}$ & Median radial-velocity across all APOGEE visits \\
 e$\_$v$\_$rad & km s$^{-1}$ & Uncertainty in median radial-velocity  \\
 std$\_$v$\_$rad & km s$^{-1}$ & Radial-velocity scatter around the median. \\
 cluster & & Name of the star's host open cluster, as listed in \citet{cg2020}.\\
 min$\_$mjd & days & First Modified Julian Date of observation by SDSS. \\
 max$\_$mjd & days & Last Modified Julian Date of observation by SDSS. \\
 n$\_$total$\_$visits & & Total number of visits across all fields in SDSS III-V. \\
 ruwe$\_$threshold & & Renormalized Unit Weight Error threshold for binarity, calculated using \textsc{GaiaUnlimited}. \\
 ruwe & & Gaia DR3 RUWE measurement \\
 ruwe$\_$binary & bool & Binary classification using only RUWE \\
 delta$\_$rv$\_$max & km s$^{-1}$ & Maximum RV difference across all visits \\
 delta$\_$rv$\_$max$\_$binary & bool & Binary classification using only delta$\_$rv$\_$max.\\
 joker$\_$ln$\_$total & & \textsc{Joker} log-likelihood from observed RV visits. \\
 joker$\_$ln$\_$total$\_$control & & \textsc{Joker} log-likelihood from synthetic normally distributed RV visits. \\
 N$\_$posterior$\_$samples & & Number of posterior samples returned by the \textsc{Joker}. \\
 joker$\_$P & days & Best-fit period from the \textsc{Joker}\\
 joker$\_$e & & Best-fit eccentricity from the \textsc{Joker} \\
 joker$\_$K & km s$^{-1}$ & Best-fit RV amplitude from the \textsc{Joker} \\
 ultranest$\_$P & days & Best-fit period from \textsc{UltraNest}\\
 ultranest$\_$e & & Best-fit eccentricity from \textsc{UltraNest} \\
 ultranest$\_$K & km s$^{-1}$ & Best-fit RV amplitude from \textsc{UltraNest} \\
 joker$\_$binary & bool & Binarity from ratio of joker$\_$ln$\_$total$\_$control and joker$\_$ln$\_$total \\
 binary & bool & Final binarity classification across all methods\\
 unimodal & bool & Whether the star has a confident orbital solution \\
 fm & M$_{\odot}^{2}$ & Mass function calculated from stellar mass and orbital parameters. \\
 orbital$\_$m2 & M$_{\odot}$  & Minimum secondary mass assuming no inclination.  \\
 raw$\_$teff & K  & Uncalibrated effective temperature measured by ASPCAP. \\
 iso$\_$logg & & Empirically derived surface gravity from raw$\_$teff and interpolated \texttt{EEP}. \\
 iso$\_$vmicro & km s$^{-1}$ & Microturbulence calculated using iso$\_$logg. \\
 iso$\_$mass & M$_{\odot}$ & Present-day stellar mass from interpolated isochrone \texttt{EEP}.  \\
 iso$\_$fe$\_$h & dex & [Fe/H] from interpolated isochrone \texttt{EEP}.\\
 iso$\_$fe$\_$chi2 & & $\chi^{2}$ fit between iso$\_$fe$\_$h and observed spectrum.  \\
 bacchus$\_$x$\_$h & dex & [X/H] abundance from \textsc{BACCHUS}.\\
 sp$\_$e$\_$x$\_$h & dex & Uncertainty in [X/H] from atmospheric parameters. \\
 line$\_$e$\_$x$\_$h & dex & Line-to-line uncertainty in [X/H].\\
 bacchus$\_$e$\_$x$\_$h & dex & Total uncertainty in [X/H]. \\
  \hline
 \end{tabular*}
\end{table*}

\section{Results}
\label{sec:results}

  % Cluster & Alias & Age & $A_{v}$ & $D_{mod}$ & $\#$\ of\ Nonbinaries & $\#$\ of\ Binaries\\

% \begin{table}
%  \caption{A breakdown of the binary and nonbinary populations in each cluster used in this study.}
%  \label{tab:cluster_pop}
%  \centering
%  \begin{tabular*}{0.8\columnwidth}{lcr}
%   \hline
%   Cluster & $\#$\ of\ Nonbinaries & $\#$\ of\ Binaries\\
%   \hline
%   \hline
%  Collinder 69 & 2 & 1\\
%  Melotte 111 & 2 & 4\\
%  Melotte 20 & 9 & 4\\
%  Melotte 22 & 5& 1\\
%  NGC 188 & 5& 5\\
%  NGC 2158 & 4& 3\\
%  NGC 2168 & 11& 8\\
%  NGC 2243 & 4& 3\\
%  NGC 2420 & 9& 4\\
%  NGC 2632 & 7& 3\\
%  Messier 67 & 29& 46\\
%  NGC 6791 & 10& 10\\
%  NGC 6819 & 13& 9\\
%  NGC 752 &9 & 2\\
%   \hline
%  \end{tabular*}
% \end{table}

%KC: Figure 5 labels of diagnostic lines should refer to the molecule whose line is measured or ionization stage of the measured line. For example, Ce II, Fe I... or OH etc... 
%KC: Y axes have to go to at least bottom=0.5 so that you can show the bottom of all lines, for example 

\subsection{Population Abundance Comparison}
\label{sec:pop_comp}
\begin{figure*}
  \includegraphics[width=\textwidth]{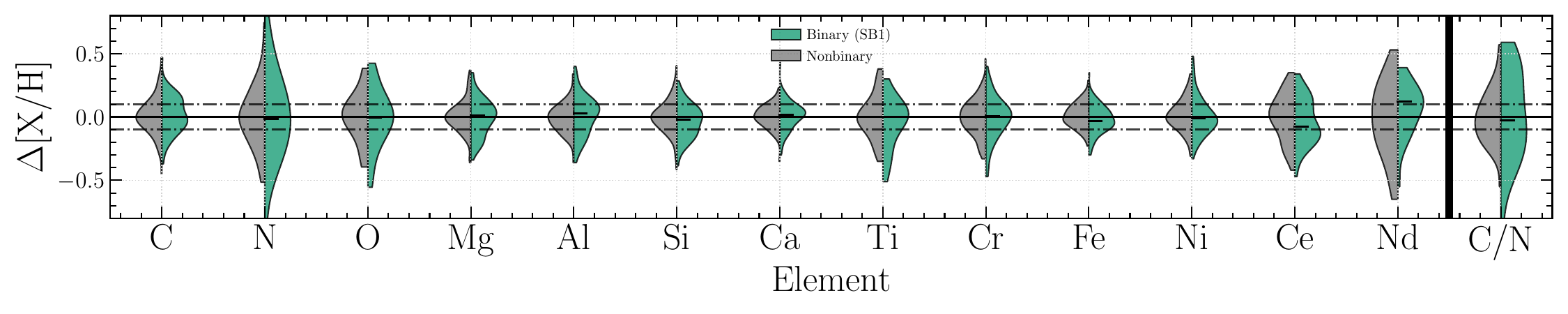}
  \includegraphics[width=\textwidth]{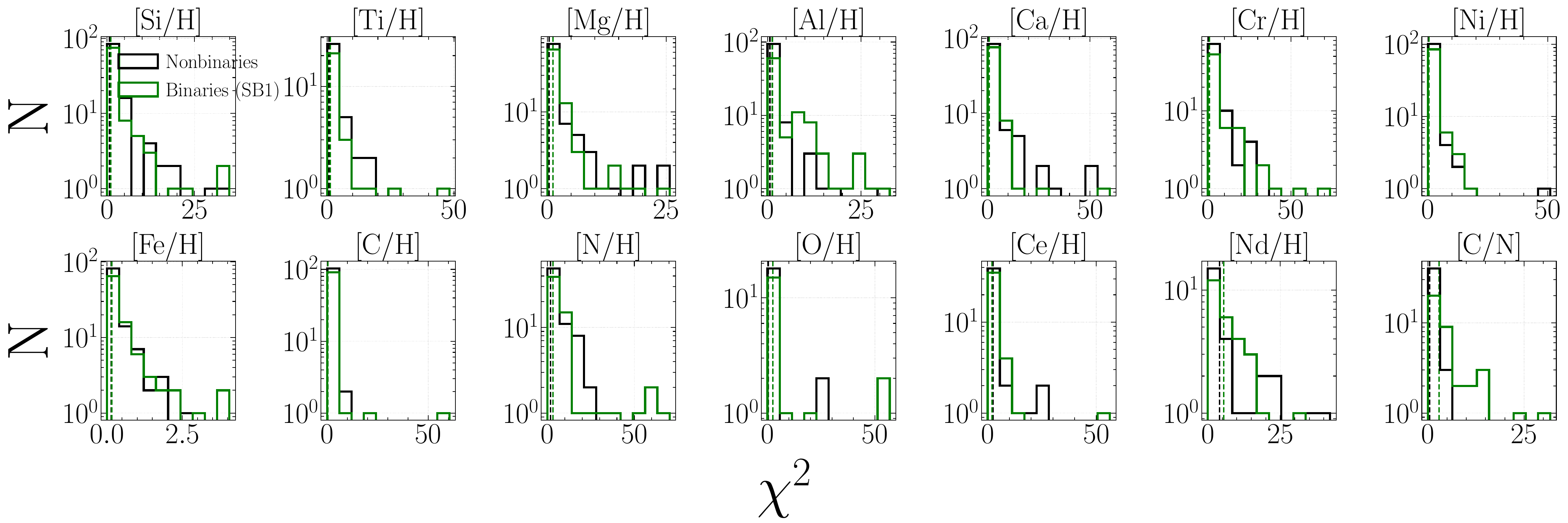}
  \caption{\textit{Top}: Here we compare the $\Delta$[X/H] and $\Delta$[C/N] distributions for the SB1-nonbinary (\textit{green}) and nonbinary-nonbinary (\textit{grey}) pairs in all the elements we measure. The black lines indicate a $\Delta$[X/H] of $\pm$0.1 dex. \textit{Bottom}: Here we compare the elemental $\chi^2$ distributions for the SB1-nonbinary (\textit{green}) and nonbinary-nonbinary (\textit{grey}) pairs in all the elements we measure. The black lines and green lines indicate the median nonbinary-nonbinary and median SB1-nonbinary $\chi^2$, respectively.}
  \label{fig:apo_abund_comparison}
\end{figure*}

Here we compare the individual abundances of stars in our binary sample to nonbinary stars within the same cluster. To remove the influence of stellar evolution and observational noise on any part of the analysis, we compare binary-nonbinary pairs with $\rm \Delta T_{eff} \le 250$~K and $\Delta \log{g} \le 0.25$~dex. This is particularly important for [C/N], which is mass dependent and therefore sensitive to large differences in $\rm T_{eff}/\log{g}$, even within a given cluster. Within each element, we measure the $\Delta$[X/H], defined as the intra-pair difference in [X/H] ([X/H]$_{\rm binary}$ - [X/H]$_{\rm nonbinary}$), and compare it with the limits on open cluster homogeneity derived in \citet{sinha2024}. As a control, we compare the $\Delta$[X/H] from our binary-nonbinary pairs to a set of $\Delta$[X/H] measured with nonbinary-nonbinary pairs. 

The results of this analysis can be seen in Figure \ref{fig:apo_abund_comparison}.
We find no systematic, universal enrichment or depletion in any of the elements we study -- \textit{that is, the stars in the binary-nonbinary pairs are as similar to each other as the stars in the nonbinary-nonbinary pairs are.} To corroborate this, we measure the similarity between the binary-nonbinary and nonbinary-nonbinary datasets using a two-sample Kolmogorov–Smirnov test. We measure $p>0.5$ for all elements, implying that, as a population, the SB1 star abundances and single star abundances were drawn from the same underlying sample in a given cluster.

To ensure that the impact of abundance uncertainties is also considered in our analysis, we compute a binary-binary and nonbinary-nonbinary $\chi^2$ value for each element, using Equation~3 from \citet{ness}.
We find that the $\chi^2$ distributions, also shown in Figure \ref{fig:apo_abund_comparison}, for the binary-nonbinary and nonbinary-nonbinary samples are drawn from the same population across all the chemical species we measure. This implies that, as a population, close binarity has not had a statistical impact on the surface abundances of our SB1 stars.

\subsection{Extended Analysis with GALAH}

\begin{figure*}
  \includegraphics[width=\textwidth]{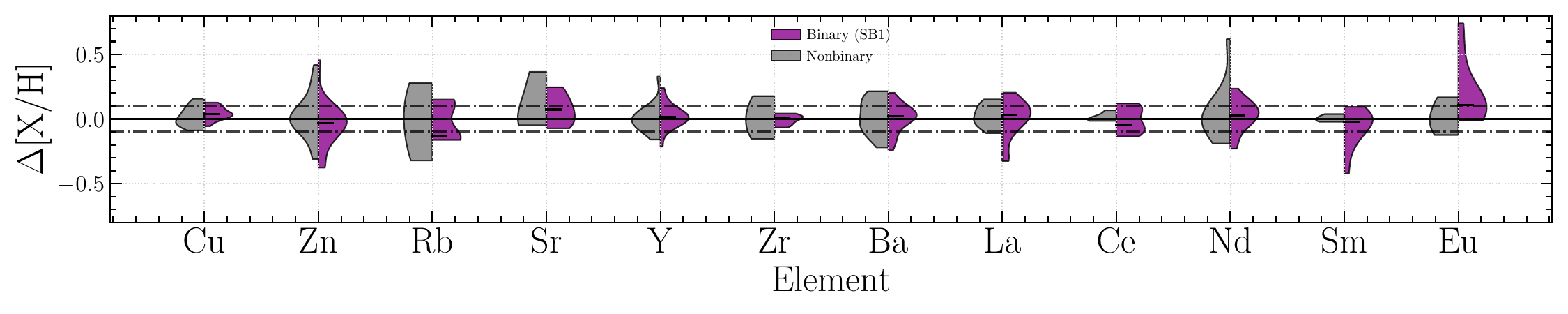}
  \caption{Similar to Figure~\ref{fig:apo_abund_comparison} but for the neutron-capture elements for GALAH targets. The black lines indicate a $\Delta$[X/H] of $\pm$0.1 dex.}
  \label{fig:galah_abund_comparison}
\end{figure*}

%KC: What does the [nbt!] label above the figure mean?
%KC: This seems a little bit strange to me that the distributions have somewhat different shapes. Eu Nd, but probably becuse of small numer statistics or some sort of sysntematics in their abundances. Not sure... 

Due to the wavelength range of the APOGEE spectra, only Ce and Nd are available to trace s-process neutron capture abundances. While Ce is primarily an s-process element, Nd receives approximately 50$\%$ contribution from the r-process \citep{prantzos2020}, making it less suitable to study potential post-mass-transfer objects. We supplement our results with abundance information from the Galactic Archaeology with HERMES (GALAH) Survey DR4 \citep{galah1, galahdr4}, which publishes abundances for 12 neutron capture elements: Rb, Sr, Y, Zr, Mo, Ba, La, Ce, Nd, Ru, Sm, and Eu. Given that Ba is an almost pure s-process element \citep{Lugaro2023}, it is the ideal tracer for identifying post-AGB mass-transfer objects. Using the GALAH DR4 \texttt{allstar} file, we apply the following quality cuts\footnote{The quality cuts for SNR, stellar parameters, and [Fe/H] are all drawn from https://www.galah-survey.org/dr4/overview/. The GALAH DR4 \texttt{allstar\_240705.fits} file can also be found here.} to ensure trustworthy abundance measurements:

%KC: NOt clear what is meant by Ce is a s process element in the solar neighborhood. Is it solar composition?

\begin{description} \itemsep -2pt
    \item $\texttt{snr\_px\_ccd3} >$   30, 
    
    \item $\texttt{flag\_sp}$ == 0, 

    \item $\texttt{flag\_fe\_h}$ == 0, and
    
    \item $\texttt{flag\_X\_fe}$ == 0, 
\end{description}
where $\texttt{flag\_X\_fe}$ includes the elements listed above.

When cross-matched to our MWM open cluster sample, we find only 38 common stars, all within M67 and NGC 2632. None of our unimodal binary stars appear in the GALAH cross-match. 
We repeat our analysis from Section \ref{sec:pop_comp}, identifying pairs of binary-nonbinary and nonbinary-nonbinary stars with similar temperatures and surface gravities. We then measure the $\Delta$[X/H] within the neutron-capture elements for these pairs.

Within our cross-matched GALAH sample, we find that none of the likely binaries are enhanced in the s-process neutron-capture elements as compared to their nonbinary counterparts, as shown in Figure \ref{fig:galah_abund_comparison}. In particular, we find no evidence that any star is a post-AGB mass-transfer object, as the 
binary-nonbinary and nonbinary-nonbinary $\Delta$[Ba/H] are consistent with one another. While it visually appears that Eu is potentially enhanced in binaries compared to nonbinaries, we note that this sample of Eu abundances is limited in size ($N=9$) and strong conclusions cannot be drawn from it. 

\subsection{Chemical Variations versus Orbital Parameters}

\begin{figure*}
    \centering
    \includegraphics[width=\linewidth]{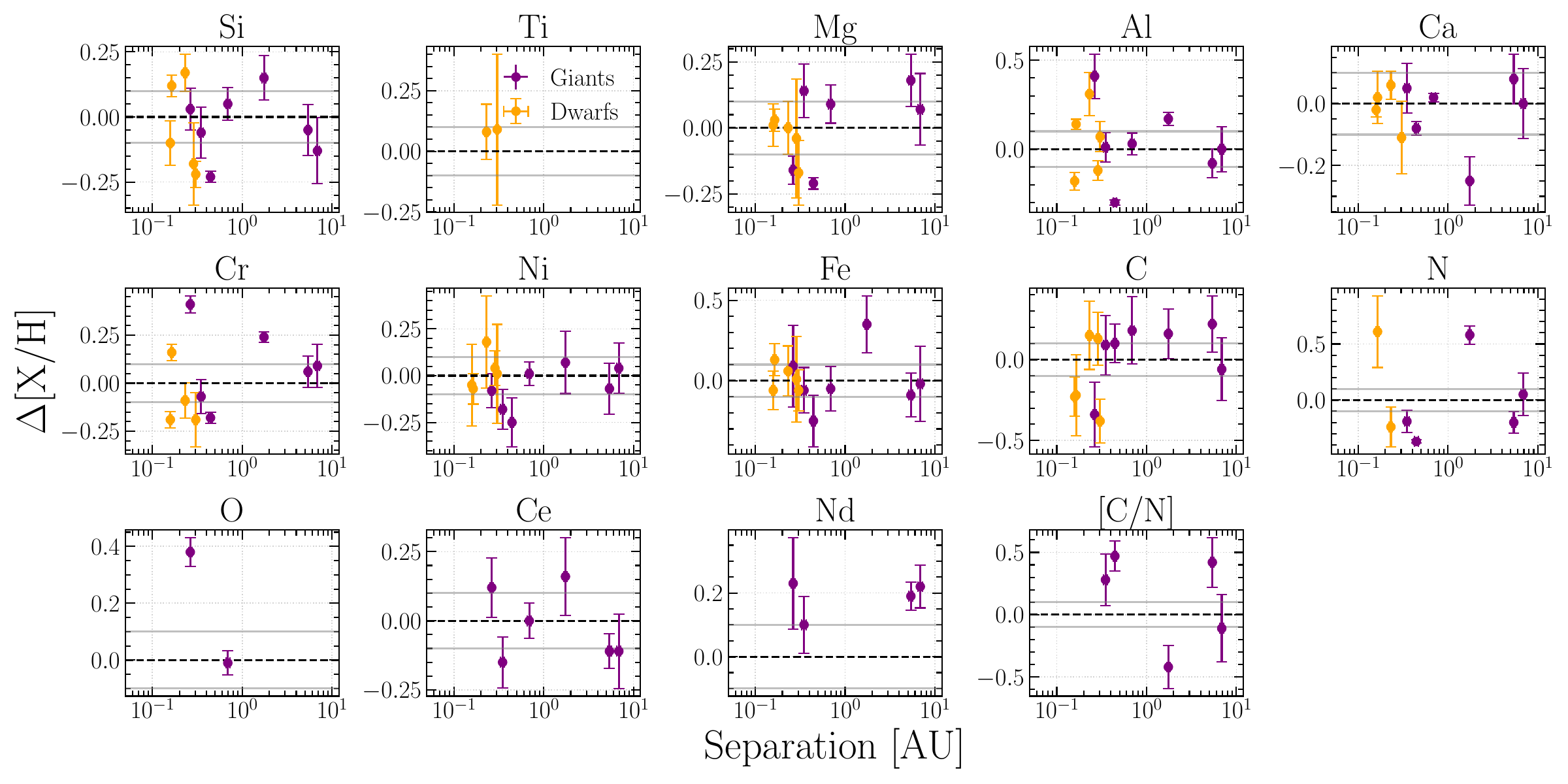}
    \caption{Here we show the measured semi-major axis as orbital separation, compared to $\Delta$[X/H], for unimodal binaries with a nonbinary counterpart at similar temperature and surface gravity (Section~\ref{sec:pop_comp}). We do not find any trend between orbital parameter and surface abundance difference. Giant stars are shown in purple and dwarf stars are shown in orange.}
    \label{fig:orbital_comparison}
\end{figure*}

Previous studies in wide binaries, such as \citet{ramirez2019} have shown a potential connection between abundance difference and orbital separation. We attempt to isolate a similar phenomenon in the close binary regime by checking for any correlation between chemical differences and orbital properties for the subset of binary stars with unimodal orbital solutions. 
We use each star's \texttt{EEP}, derived in Section \ref{sec:atmospheric_parameters}, to estimate the present-day stellar mass of the primary. Since none of our binaries have directly measured secondary masses or eclipses, we cannot fully solve the orbital solution. However, assuming an edge-on inclination angle ($\sin{i}=1$) in Equation \ref{eqn:kepler} below \citep{kepler1609}:
\begin{equation}
\label{eqn:kepler}
\frac{M_{2}^{3}sin^{3}i}{(M_{1}+M_{2})^{2}} = \frac{PK^{3}}{2\pi G} {(1-e^{2})^{3/2}},
\end{equation}
we can compute a minimum secondary mass (\textit{M$_{2, min}$}) by solving  Equation \ref{eqn:kepler} as a straightforward cubic polynomial. We can also compute a minimum semi-major axis (\textit{a$_{min}$}) 
\begin{equation}
\label{eqn:semimajor}
a_{min} = \frac{P}{1 - e^{2}},
\end{equation}
and periastron ($r_{min}$), or point of closest approach: 
\begin{equation}
\label{eqn:periastron}
r_{min} = \frac{P}{1 + e}.
\end{equation}
As shown in Figure \ref{fig:orbital_comparison}, we find that the orbital separations for unimodal binaries in our sample range between 10$^{-1}$ and 10$^{1}$ astronomical units (AU). We measure the Spearman correlation between $\Delta$[X/H], a$_{min}$, and $r_{min}$, and we find no relationship between any abundance difference and orbital separation. However, due to the strict selection criteria imposed by this analysis, we are very limited in sample size.

We also measure that the minimum periastrons are all larger than the Roche Lobe radii for the primaries in our binary sample, calculated using Equation \ref{eqn:roche_lobe}, originally derived in \citet{eggleton1983}: 

\begin{equation}
\label{eqn:roche_lobe}
\frac{r}{a_{min}}=\frac{0.49q^{2/3}}{0.6q^{2/3} + ln(1 + q^{1/3})} .
\end{equation}

Here $q$ is the mass ratio $M_{2,min}/M_{1}$ and $r$ is the Roche Lobe radius. While this result does not preclude the possibility of past mass transfer from the less luminous companion, in addition to orbital evolution, we cannot make predictions without independent constraints on the secondary masses. 

\section{Discussion}
\label{sec:discussion}
\subsection{[C/N] Comparison}

Determining accurate stellar ages is a key area of interest in galactic archaeology. Knowing the ages of large sets of stars is crucial for understanding the Milky Way's chemical composition and enrichment throughout its lifetime. The current gold standard for field star ages is via asteroseismology, through missions such as Kepler/K2 \citep{kepler, K2}, TESS \citep{tess}, CoRoT \citep{corot}, and the upcoming PLATO and Roman missions \citep{plato, rotac_2025}. However, there has also been significant work in so-called ``chemical clocks", or surface abundance ratios corresponding to stellar age \citep{espinoza2021, sales_silva_2022, casali2025}. 

%KC: I think we should cite some works from Casali. A recent one is Casali et al. (2025). Also please cite Sales Silva et al. (2022) (using APOGEE for open clusters). 

One commonly used chemical clock for stars on the red giant branch is the [C/N] ratio. This is because, as stars ascend the red giant branch, their convective envelopes expand and pull CNO-cycled material from above the core to the surface. As the amount of dredge-up is linked to stellar mass, the total surface [C/N] measured after a star begins its ascent on the RGB can be used as an age indicator \citep{masseron2015, martig2016, spoo2022}. Recent work has demonstrated that this method is viable for developing stellar age catalogues, and it can be informative for dating stars in the absence of other metrics \citep{stonemartinez2025}.

%KC: Below We need to cite the work of Nadege Lagarde. Please check out her papers...

However, it has also been shown through works like \citet{bufanda2023} and \citet{frazer2025} that not all stars follow the expected [C/N]--mass relationship. While many potential causes have been proposed, including extra mixing and rotation, binary interactions leading to mass transfer is of particular interest to this paper \citep{lagarde2022,carlberg2012, tayar2015, patton2024}. In a close binary, if one member ascends the AGB before the other, it is possible that the binary would undergo a mass-transfer event, causing the less-evolved companion to become more massive. In addition, it is expected that both carbon and s-process elements like Ba or Ce would become enhanced in the less evolved star due to this mass transfer \citep{foster2024}, with $s-$process enhancement larger than $+$0.3 dex, as noted in \citet{deCastro2016}. The combination of these two effects would lead to unphysically ancient age measurements for observed stars \citep{spoo2022}. One proposed method to study this has been to look for UV excess. Binaries that are brighter in the ultraviolet than expected have been shown to host white dwarfs \citep{anguinao2022}; therefore, elevated [C/N] in UV-excess binaries is physically motivated and a potential indicator of past mass transfer. In fact, in \citet{frazer2025}, the authors find APOKASC stars with GALEX UV excess as potential mass accretors. In a similar vein, \citet{dixon2020} measure a relationship between RGB rapid rotators and NUV excess.

%KC: Dixon et al measures > measure

\begin{figure}
    \centering
    \includegraphics[width=\linewidth]{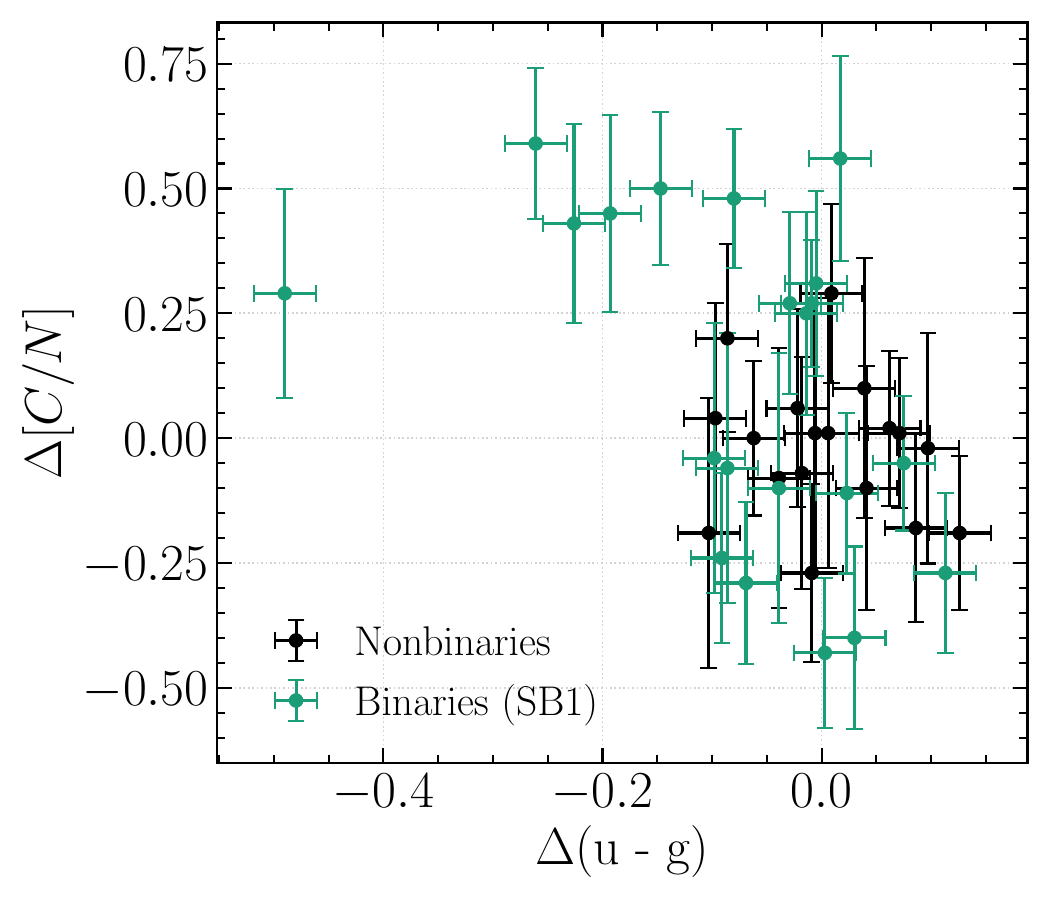}
    \caption{Here we show $\Delta$[C/N] as a function of $\Delta(u-g)$. SB1 primaries with large UV excess have significant increases in [C/N], and thus increases in the age that would be measured with that abundance.}
    \label{fig:cn_uv}
\end{figure}

To examine this, we compare the $\Delta$[C/N] measured between our binary-nonbinary and nonbinary-nonbinary sample as a function of UV excess, $\Delta(u-g)$, measured as the difference in $(u-g)$ colour between a binary and its matched single star. 
We find a potential relationship between binaries with UV excess and $\Delta$[C/N], as shown in Figure \ref{fig:cn_uv}. In particular, we find that binary members with significant UV excess exhibit a 0.2--0.5~dex increase in [C/N]. However, none of these UV excess binaries show enhancement in Ce or Nd ($\rm \Delta[Ce/H] < 0.2$), and all have $v \sin i \leq 7$~km s$^{-1}$, so further study is needed to confirm whether this relationship is truly caused by close proximity to a white dwarf. 

%KC: It looks like there is a threshold in delta (u-g) below which delta C/N is always enhanced. Below delta (u-g) ~-0.1 or a little bit below that value. TRy doing a mean/median of the delta C/N for two populations: one having delta (u-g) below -0.1 and one above. None of the black points are beloew Delta u-g ~-0.1. Above delta g-u -0.1 you have the full range in delta C/N. When you do the averages we will see the differences in C/N clearly.
%KC: Did you do a plot for C/N versus Ce color coding with delta (u-g)?

\subsection{Chemical Homogeneity of Close Binaries in a Galactic Context}
\label{sec:all_together}

Understanding how elements are dispersed from their production sites throughout the Milky Way, and the different mixing lengths for each nucleosynthetic pathway, is central to galactic archaeology. One of the ways we study these processes is through the variation in chemical abundances at different spatial scales. Previous work has shown that Milky Way field stars at similar ages, galactocentric radii, [M/H], and [$\alpha$/M] have a $\sim$0.2 dex dispersion in chemical abundances \citep{hayden, ness}. On the other hand, co-natal stellar populations, such as open clusters, have dispersions of $\leq$ 0.1 dex \citep[e.g.,][]{bovy, poovelil2020, ness2021, sinha2024}. As the basic building blocks of the Milky Way, the dissolution of these clusters populates the disc and bar/bulge with stars. Understanding the spatial correlation between co-eval stars and chemical homogeneity can inform us how elements are mixed throughout the Galaxy.

%KC: THey have the same [alpha over metallicity] and metallicity but different in what elements? What do you mean by light elements? noting that alpha elements are light elements. [HAs Hayden et al. (2015) really shown that?]

\begin{figure}
  \includegraphics[width=0.45\textwidth]{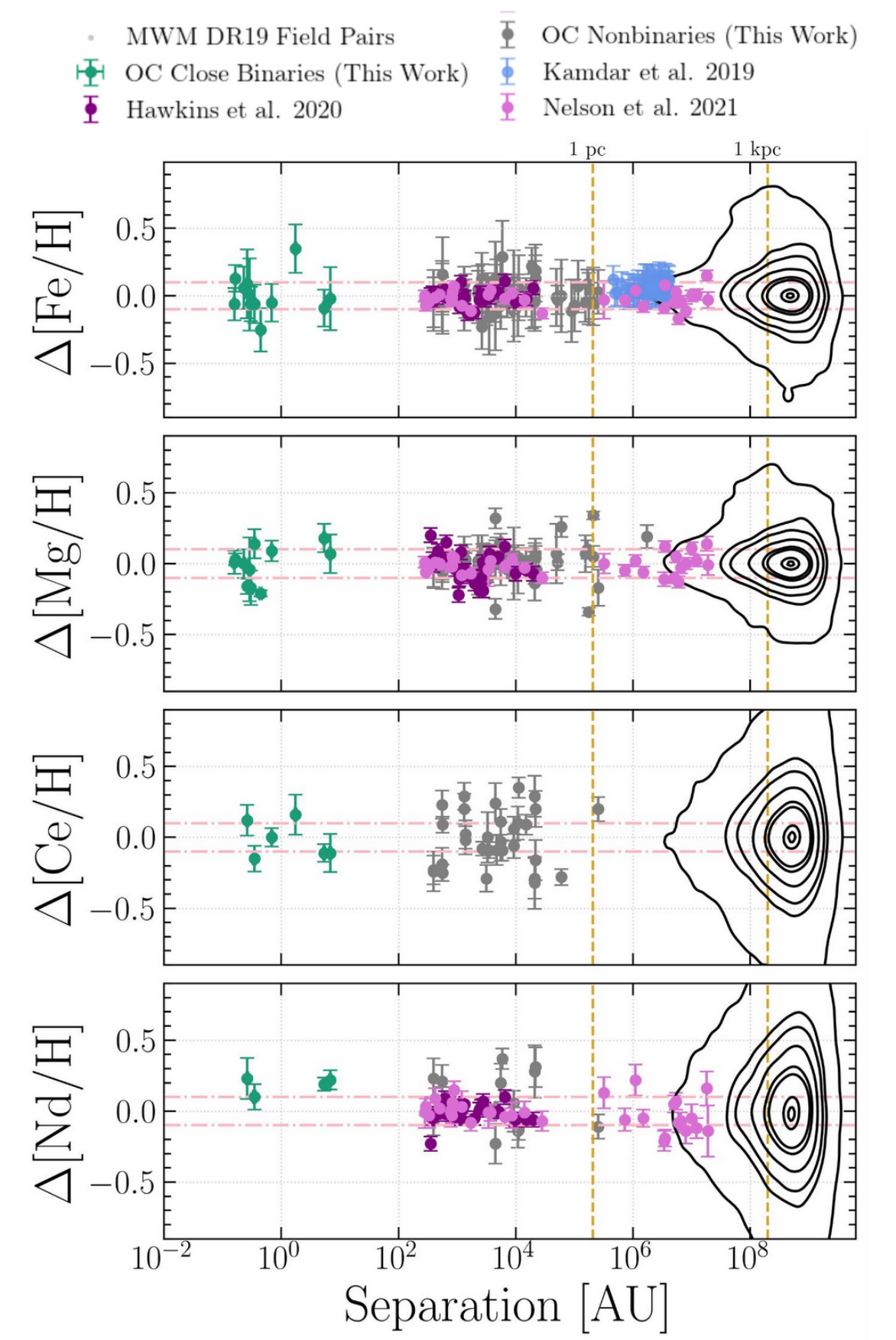}
  \includegraphics[width=0.45\textwidth]{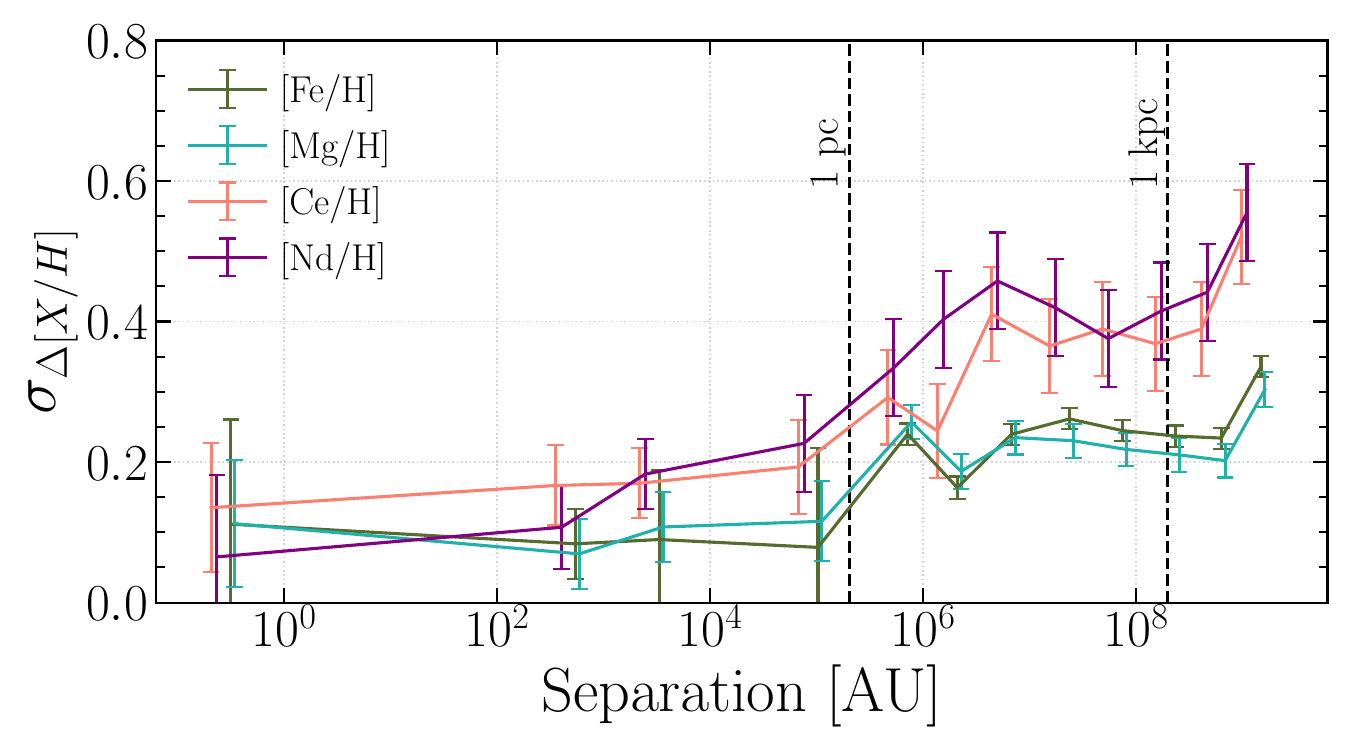}
  \caption{\textit{Top:} Here we compare the $\Delta$[X/H] distributions for the binary-nonbinary co-eval stars at increasing separations across four elements: Fe, Mg, Ce, and Nd. We note that the plot for $\Delta$[Ce/H] does not include any wide binaries or co-moving pairs as \citet{binaries} and \citet{nelson2021} did not measure Ce in their analyses. \textit{Bottom:} We find that below separations of 1 pc, stars can be considered chemically homogeneous across all nucleosynthetic channels. Beyond one pc but within one kpc, there is a sudden increase in $\sigma_{\Delta\mathrm{[X/H]}}$, with the light and heavy elements separating into two different tracks. Lastly, at separations beyond one kpc, $\sigma_{\Delta\mathrm{[X/H]}}$ steadily increases. We show contours for the 5th, 25th, 34th, 50th, 68th, 75th, and 95th percentiles for the top four panels.}
  \label{fig:stellar_pop_comparison}
\end{figure}

In this section and Figure~\ref{fig:stellar_pop_comparison}, we collect results from previous literature to investigate the relationship between chemical homogeneity and physical scales in the Milky Way. To accomplish this, we compare the binary-nonbinary homogeneity measured in Section \ref{sec:results} to the chemical homogeneity between pairs of open cluster stars, the intra-pair homogeneity of wide binaries and co-moving stars, and finally co-eval field stars at increasing separations within the disc. For separations smaller than 10$^{6}$ AU, we require a minimum of three pairs per bin, resulting in unequal spacing. At separations larger than 10$^{6}$ AU, we use bins with separations of 0.35 in log(AU) space, and all bins in this regime contain N $\gg$ 3 pairs.

%KC: "we collect estimates of chemical homogeneity on physical scales in the Milky Way from 10−1 AU to 109 AU (or ∼4.8 kpc)." Collect estimates of chemical homegeneity of an ackward sentence to me. Collect results from the literature to investigate y versus x?

%KC:temperature > effective temperature

%KC:"should look like'? Or rather something like benchmark?

With separations between 10$^{3}$ and 10$^{5}$ astronomical units, wide binaries are the next step up from close binaries on the distance ladder of co-eval and co-natal stars. Previous studies of wide binary chemistry, such as \citet{andrews2018,andrews2019, binaries, dongwook2024}, have found that they are homogeneous within 0.05~dex across a broad suite of elements. We compare the $\Delta$[X/H] distributions of our unimodal binaries to the results of \citet{binaries}, and find that our SB1s exhibit a similar level of chemical similarity to previous studies of wide binaries. We note, though, that the metrics for $\Delta$[X/H] are slightly different between the wide and close binaries. Our SB1s are being compared to nonbinaries, in the same cluster at the same effective temperature and surface gravity, with the idea that the nonbinary stars are a benchmark for the SB1 primary. In \citet{binaries}, since both binary members are resolvable on the sky, they can measure the intra-pair $\Delta$[X/H] directly. Figure \ref{fig:stellar_pop_comparison} does not include any wide binary data for $\Delta$[Ce/H] as \citet{binaries} did not measure Ce in their analysis.

With separations between 10$^{4}$ and 10$^{5}$ astronomical units, nonbinaries in open clusters are the largest sample of co-eval/co-natal stars that are not in a binary. We measure the spatial separation for all the nonbinary-nonbinary pairs assembled in Section \ref{sec:pop_comp}. We find that their dispersion in [Fe/H] is $\sim$0.1 dex, which is in agreement with previous literature on open cluster abundance dispersion \citep{desilva1,bovy,sinha2024,grilo2024}. 

Lastly, we use co-moving pairs from \citet{kamdar2019} and \citet{nelson2021} to study the chemical similarity of co-natal pairs between 10$^{5}$ and 10$^{7}$ AU. Both studies found that co-moving pairs of stars were chemically homogeneous in [Fe/H] within 0.1 dex. While \citet{kamdar2019} did not measure the intra-pair abundance difference for any other element, \citet{nelson2021} measured $\Delta$[X/H] for two dozen elements, including Mg and Nd.

We compare our sample of binaries to a set of co-eval, but not co-natal, field stars in the Milky Way. Using the MWM DR19 StarFlow Value Added catalogue\footnote{\url{https://www.sdss.org/dr19/data_access/value-added-catalogues/?vac_id=10003}} of stellar masses and ages \citep{stonemartinez2025}, we apply the same quality cuts outlined in Section \ref{sec:Data}, resulting in a sample of $\sim$31,000 stars. Using the same binary selection methods outlined in \ref{sec:Meth}, as well as a threshold of seven radial velocity visits, we identify $\sim$12,000 binaries and $\sim$19,000 non-binaries in this field star sample. Finally, for the reasons outlined in \ref{sec:bacchus}, we use Ce and Nd measurements for stars on the giant branch. 

Using \textsc{Galpy} \citep{galpy1}, we integrate the orbits of the StarFlow sample using the \texttt{MWPotential2014} gravitational potential \citep{galpy2} in order to determine each star's guiding radius ($R_{\rm guide}$). Finally, we pair field stars in the StarFlow sample with their closest match in age, temperature, and surface gravity, where we use the same temperature and surface gravity thresholds as in Section \ref{sec:pop_comp}. Stars without a match within the temperature and surface gravity thresholds were removed from the sample.

We measure the difference in $R_{\rm guide}$ within each pair and take that as the separation between the two stars, finding field star separations span 0.6 and 9.5 kpc, with an average uncertainty of $\sim$ 0.1 kpc.

Finally, we measure the intra-pair $\Delta$[X/H] for all our field star pairs within four chemical species: [Fe/H], [Mg/H], [Ce/H], and [Nd/H]. These four were chosen as they all trace different nucleosynthetic pathways, informing us about the iron-peak (Fe), alpha- (Mg), and neutron-capture (Nd and Ce) pathways. 

Between our sample of open cluster SB1s and nonbinaries, wide binaries from \citet{binaries}, and matched field stars from StarFlow, we have a large sample of ostensibly co-eval pairs that are increasingly less co-natal. From this dataset, we measure the $\sigma_{\Delta\mathrm{[X/H]}}$ as a function of separation in order to quantify the relationship between intra-pair homogeneity and spatial separation. The results of this are shown in Figure \ref{fig:stellar_pop_comparison}. With the potential exception of Nd, at all separations smaller than one parsec, $\sigma_{\Delta\mathrm{[X/H]}}$ is $\leq 0.1$~dex across all the different nucleosynthetic tracers. 

While the dispersion in $\sigma_{\Delta\mathrm{[Nd/H]}}$ appears to increase to $\sim$ 0.3 dex at 10$^{4}$ AU, this is less likely to be driven by astrophysical mechanisms, and more due to the limited sample of stars with $\Delta$[Nd/H] measurements at that separation. This hypothesis is strengthened by the fact that Ce follows similar enrichment pathways to Nd, but has more measurements at 10$^{4}$ AU, and $\sigma_{\Delta\mathrm{[Ce/H]}}$ is consistent with homogeneity at 10$^{4}$ AU. Furthermore, when using only results from  \citet{binaries}, we find $\sigma_{\Delta\mathrm{[Nd/H]}}$ $\leq$0.1 dex at similar separations.

The fact that within one pc we measure a $\sigma_{\Delta\mathrm{[X/H]}}$ consistent with homogeneity implies that at distances smaller than one pc, gas is being mixed homogeneously enough that there is no statistical difference in surface abundance, which is consistent with results from simulations \citep{Armillotta2018, kolberg2022,krumholz2025}. Equally interesting, though, is the fact that beyond one pc of separation, even when controlling for factors such as age, temperature, and surface gravity, we find an immediate increase in $\sigma_{\Delta\mathrm{[X/H]}}$. Within separations of 1~kpc, we find that the dispersion in light element and neutron-capture channels are $\sim$ 0.2 dex and $\sim$ 0.35 dex respectively, which is in agreement with previous findings \citep{ness2021,manea2023}.

Important throughout this analysis is the result that at all scales between 10$^{-1}$ and 10$^{7}$ AU, there is no statistical difference in the chemical homogeneity of co-natal and co-eval pairs of stars across any nucleosynthetic channel. Given that at all scales below one pc, n-capture elements appear as homogeneous as the light elements, larger dispersions than the light elements at larger separations imply they are potentially being mixed differently into the ISM. Therefore, as also noted by \citet{manea2025}, n-capture elements are far more suited to distinguish co-natal populations dissolved in the field than the lighter elements. 

%KC: many captions (all?) in the paper have : Here we show... Here is not necessary. You can start directly

\section{Conclusions}
\label{sec:conc}
%KC: Is it conclusion or Conclusions? I think plural

The purpose of this study was to measure the chemistry of SB1 primaries in order to determine how close binarity can affect surface abundances. To accomplish this, we determine the SB1 membership of 14 open clusters in SDSS-V Milky Way Mapper DR19 using a combination of radial velocities and Gaia DR3 RUWE. Within our open cluster sample, we identify 119 likely-single stars and 103 likely-SB1s with FGK primaries. Only 16 targets in our sample of likely binaries have well-constrained orbital parameters. 

To study the difference between binaries and nonbinaries as a population, we measure the $\Delta$[X/H] between binaries and nonbinaries at similar effective temperatures and surface gravities. We verify, and re-derive where necessary, the atmospheric parameters for all the stars in our sample. Using MWM DR19 spectra and the \textsc{BACCHUS} spectral synthesis code, we re-derive abundances in the following 13 elements, as well as the [C/N] ratio:  Si, Fe, C, N, O, Na, Mg, Al, Ca, Ti, Cr, Ni, Ce, and Nd.

%KC: temperature > effective temperature

From this analysis, we find the following:
\begin{enumerate}
    \item Compared to a non-binary baseline using a K-S test, we find no statistical difference between the two populations across all the elements we studied. 
    
    \item When crossmatched to GALAH DR4 to study potential neutron-capture variations, we still find no statistical difference in $\Delta$[X/H] between the binary-nonbinary and nonbinary-nonbinary populations.

    \item We measure the relationship between $\Delta$[X/H] and orbital parameter for the sample of SB1s for which we have constrained orbital solutions. We find no trend in abundance difference between 10$^{-1}$ and 10$^{1}$ AU.

    \item Though there is no statistical difference between the binary-nonbinary and nonbinary-nonbinary $\Delta$[C/N] as a whole population, we do find a trend when measuring the relationship between UV excess and $\Delta$[C/N]. We find that binary stars with UV excess have an enhanced [C/N] ratio compared to single stars of comparable mass. We find that [C/N] can be enhanced between 0.2--0.5 dex, which corresponds to an increase in the inferred log(age)$_{\rm [C/N]}$ by up to $\sim$1 dex. This overestimation in age has implications for the efficacy of [C/N] as a chemical clock for red giant stars, underscoring the importance of considering binarity when measuring stellar ages.

    \item We find that at separations of 1 pc or less, co-eval stars can be considered chemically homogeneous across all nucleosynthetic channels. However, at separations beyond one pc, we see the dispersion in $\rm \Delta[X/H]$ increases at different rates for the light and neutron-capture elements. This difference in dispersion implies that elements from these nucleosynthetic channels are being mixed from their production sites into the broader Galaxy through different means. This also strongly supports the idea that neutron-capture elements are far more useful at identifying co-natal populations than the lighter elements.
\end{enumerate}

\section*{Acknowledgements}
We would like to thank the referee for their helpful comments, which have improved this analysis. A.S. would also like to acknowledge Shelly S. for her helpful conversations and who sadly passed away prior to this work's publication. This material is based upon work supported by the National Science Foundation under Grant No. AST-2206542. K.C. acknowledges support from NSF Grant No. AST-2206543. D.S. acknowledges support from the Foundation for Research and Technological Innovation Support of the State of Sergipe (FAPITEC/SE) and the National Council for Scientific and Technological Development (CNPq), under grant numbers 404056/2021-0, 794017/2013, and 444372/2024-5. J.C. acknowledges support from the Agencia Nacional de Investigación y 
Desarrollo (ANID) via Proyecto Fondecyt Regular 1231345, and by ANID BASAL 
project FB210003.

Funding for the Sloan Digital Sky Survey V has been provided by the Alfred P. Sloan Foundation, the Heising-Simons Foundation, the National Science Foundation, and the Participating Institutions. SDSS acknowledges support and resources from the Center for High-Performance Computing at the University of Utah. SDSS telescopes are located at Apache Point Observatory, funded by the Astrophysical Research Consortium and operated by New Mexico State University, and at Las Campanas Observatory, operated by the Carnegie Institution for Science. The SDSS web site is \url{www.sdss.org}. 

SDSS is managed by the Astrophysical Research Consortium for the Participating Institutions of the SDSS Collaboration, including the Carnegie Institution for Science, Chilean National Time Allocation Committee (CNTAC) ratified researchers, Caltech, the Gotham Participation Group, Harvard University, Heidelberg University, The Flatiron Institute, The Johns Hopkins University, L'Ecole polytechnique f\'{e}d\'{e}rale de Lausanne (EPFL), Leibniz-Institut f\"{u}r Astrophysik Potsdam (AIP), Max-Planck-Institut f\"{u}r Astronomie (MPIA Heidelberg), Max-Planck-Institut f\"{u}r Extraterrestrische Physik (MPE), Nanjing University, National Astronomical Observatories of China (NAOC), New Mexico State University, The Ohio State University, Pennsylvania State University, Smithsonian Astrophysical Observatory, Space Telescope Science Institute (STScI), the Stellar Astrophysics Participation Group, Universidad Nacional Aut\'{o}noma de M\'{e}xico, University of Arizona, University of Colorado Boulder, University of Illinois at Urbana-Champaign, University of Toronto, University of Utah, University of Virginia, Yale University, and Yunnan University.

%%%%%%%%%%%%%%%%%%%%%%%%%%%%%%%%%%%%%%%%%%%%%%%%%%
\section*{Software}
\textsc{astropy}\citep{astropy:2013,astropy:2018,astropy:2022}, \textsc{scipy} \citep{scipy2020}, \textsc{matplotlib} \citep{matplotlib}, \textsc{numpy} \citep{numpy}, \textsc{pandas} \citep{pandas_1, pandas_2}, \textsc{isochrones} \citep{isochrones}, \textsc{galpy} \citep{galpy}, \textsc{bacchus\_tools}, and \textsc{PyBACCHUS} \citep{PyBACCHUS}.

\section*{Data Availability}
The data for this project is publicly available through SDSS-V and the Gaia Collaboration. SDSS-V spectra and radial velocities can be found here (\hyperlink{https://www.sdss.org/dr19/data_access/}{https://www.sdss.org/dr19/data$\_$access/}), and instructions on how to query the Gaia main$\_$table as well as NSS catalogue can be found here (\hyperlink{https://www.cosmos.esa.int/web/gaia/dr3}{https://www.cosmos.esa.int/web/gaia/dr3}). The cluster membership, atmospheric parameters, binary membership, and elemental abundances derived from this analysis will be made available as a machine-readable table.



\bibliographystyle{mnras}
\bibliography{references} % if your bibtex file is called example.bib

@ARTICLE{ramirez2019,
       author = {{Ram{\'\i}rez}, I. and {Khanal}, S. and {Lichon}, S.~J. and {Chanam{\'e}}, J. and {Endl}, M. and {Mel{\'e}ndez}, J. and {Lambert}, D.~L.},
        title = "{The chemical composition of HIP 34407/HIP 34426 and other twin-star comoving pairs}",
      journal = {\mnras},
         year = 2019,
        month = dec,
       volume = {490},
       number = {2},
        pages = {2448-2457},
          doi = {10.1093/mnras/stz2709},
archivePrefix = {arXiv},
       eprint = {1909.07460},
 primaryClass = {astro-ph.SR},
       adsurl = {https://ui.adsabs.harvard.edu/abs/2019MNRAS.490.2448R}
}

@ARTICLE{ramos2024,
       author = {{Ramos}, A.~A. and {Holanda}, N. and {Drake}, N.~A. and {Rain}, M.~J. and {Maia}, F.~F.~S. and {Daflon}, S. and {Pereira}, C.~B.},
        title = "{A study of chemical abundances, rotational velocities, and orbital elements in single-lined spectroscopic binary stars in open clusters}",
      journal = {\mnras},
         year = 2024,
        month = jan,
       volume = {527},
       number = {3},
        pages = {6211-6226},
          doi = {10.1093/mnras/stad3632},
       adsurl = {https://ui.adsabs.harvard.edu/abs/2024MNRAS.527.6211R}
}

@ARTICLE{vanderSwaelmen2017,
       author = {{Van der Swaelmen}, M. and {Boffin}, H.~M.~J. and {Jorissen}, A. and {Van Eck}, S.},
        title = "{The mass-ratio and eccentricity distributions of barium and S stars, and red giants in open clusters}",
      journal = {\aap},
         year = 2017,
        month = jan,
       volume = {597},
          eid = {A68},
        pages = {A68},
          doi = {10.1051/0004-6361/201628867},
archivePrefix = {arXiv},
       eprint = {1608.04949},
 primaryClass = {astro-ph.SR},
       adsurl = {https://ui.adsabs.harvard.edu/abs/2017A&A...597A..68V}
}

@ARTICLE{deCastro2016,
       author = {{de Castro}, D.~B. and {Pereira}, C.~B. and {Roig}, F. and {Jilinski}, E. and {Drake}, N.~A. and {Chavero}, C. and {Sales Silva}, J.~V.},
        title = "{Chemical abundances and kinematics of barium stars}",
      journal = {\mnras},
         year = 2016,
        month = jul,
       volume = {459},
       number = {4},
        pages = {4299-4324},
          doi = {10.1093/mnras/stw815},
archivePrefix = {arXiv},
       eprint = {1604.03031},
 primaryClass = {astro-ph.SR},
       adsurl = {https://ui.adsabs.harvard.edu/abs/2016MNRAS.459.4299D}
}

@ARTICLE{grevesse2007,
       author = {{Grevesse}, N. and {Asplund}, M. and {Sauval}, A.~J.},
        title = "{The Solar Chemical Composition}",
      journal = {\ssr},
         year = 2007,
        month = jun,
       volume = {130},
       number = {1-4},
        pages = {105-114},
          doi = {10.1007/s11214-007-9173-7},
       adsurl = {https://ui.adsabs.harvard.edu/abs/2007SSRv..130..105G}
}

@ARTICLE{otto2026,
       author = {{Otto}, Jonah M. and {Frinchaboy}, Peter M. and {Myers}, Natalie R. and {Johnson}, James W. and {Donor}, John and {Hossain}, Ahabar and {M{\'e}sz{\'a}ros}, Szabolcs and {Wallace}, Hailey and {Cunha}, Katia and {Bhattarai}, Binod and {Sinha}, Amaya and {Zasowski}, Gail and {Loebman}, Sarah R. and {Wiggins}, Alessa I. and {Price-Whelan}, Adrian M. and {Spoo}, Taylor and {Souto}, Diogo and {Bizyaev}, Dmitry and {Pan}, Kaike and {Saydjari}, Andrew K. and {Song}, Ying-Yi},
        title = "{The Open Cluster Chemical Abundances and Mapping Survey. VIII. Galactic Chemical Gradient and Azimuthal Analysis from SDSS/MWM DR19}",
      journal = {\aj},
         year = 2026,
        month = feb,
       volume = {171},
       number = {2},
          eid = {91},
        pages = {91},
          doi = {10.3847/1538-3881/ae28d8},
archivePrefix = {arXiv},
       eprint = {2507.07264},
 primaryClass = {astro-ph.GA},
       adsurl = {https://ui.adsabs.harvard.edu/abs/2026AJ....171...91O}
}

@ARTICLE{lagarde2022,
       author = {{Lagarde}, N. and {Minkevi{\v{c}}i{\={u}}t{\.{e}}}, R. and {Drazdauskas}, A. and {Tautvai{\v{s}}ien{\.{e}}}, G. and {Charbonnel}, C. and {Reyl{\'e}}, C. and {Miglio}, A. and {Kushwahaa}, T. and {Bale}, B.},
        title = "{$^{12}$C/$^{13}$C of Kepler giant stars: The missing piece of the mixing puzzle}",
      journal = {\aap},
         year = 2024,
        month = apr,
       volume = {684},
          eid = {A70},
        pages = {A70},
          doi = {10.1051/0004-6361/202348336},
archivePrefix = {arXiv},
       eprint = {2312.08197},
 primaryClass = {astro-ph.SR},
       adsurl = {https://ui.adsabs.harvard.edu/abs/2024A&A...684A..70L}
}

@ARTICLE{sales_silva_2022,
       author = {{Sales-Silva}, J.~V. and {Daflon}, S. and {Cunha}, K. and {Souto}, D. and {Smith}, V.~V. and {Chiappini}, C. and {Donor}, J. and {Frinchaboy}, P.~M. and {Garc{\'\i}a-Hern{\'a}ndez}, D.~A. and {Hayes}, C. and {Majewski}, S.~R. and {Masseron}, T. and {Schiavon}, R.~P. and {Weinberg}, D.~H. and {Beaton}, R.~L. and {Fern{\'a}ndez-Trincado}, J.~G. and {J{\"o}nsson}, H. and {Lane}, R.~R. and {Minniti}, D. and {Manchado}, A. and {Moni Bidin}, C. and {Nitschelm}, C. and {O'Connell}, J. and {Villanova}, S.},
        title = "{Exploring the S-process History in the Galactic Disk: Cerium Abundances and Gradients in Open Clusters from the OCCAM/APOGEE Sample}",
      journal = {\apj},
         year = 2022,
        month = feb,
       volume = {926},
       number = {2},
          eid = {154},
        pages = {154},
          doi = {10.3847/1538-4357/ac4254},
archivePrefix = {arXiv},
       eprint = {2112.02196},
 primaryClass = {astro-ph.GA},
       adsurl = {https://ui.adsabs.harvard.edu/abs/2022ApJ...926..154S}
}

@ARTICLE{casali2025,
       author = {{Casali}, G. and {Montalb{\'a}n}, J. and {Miglio}, A. and {Casagrande}, L. and {Magrini}, L. and {Chiappini}, C. and {Bragaglia}, A. and {Matteuzzi}, M. and {Brogaard}, K. and {Stokholm}, A. and {Grisoni}, V. and {Tailo}, M. and {Willett}, E.},
        title = "{Tracing the Milky Way: calibrating chemical ages with high-precision Kepler data}",
      journal = {\mnras},
         year = 2025,
        month = aug,
       volume = {541},
       number = {3},
        pages = {2631-2650},
          doi = {10.1093/mnras/staf1047},
archivePrefix = {arXiv},
       eprint = {2506.15546},
 primaryClass = {astro-ph.GA},
       adsurl = {https://ui.adsabs.harvard.edu/abs/2025MNRAS.541.2631C}
}

@ARTICLE{hasselquist2017,
       author = {{Hasselquist}, Sten and {Shetrone}, Matthew and {Cunha}, Katia and {Smith}, Verne V. and {Holtzman}, Jon and {Lawler}, J.~E. and {Allende Prieto}, Carlos and {Beers}, Timothy C. and {Chojnowski}, Drew and {Fern{\'a}ndez-Trincado}, J.~G. and {Garc{\'\i}a-Hern{\'a}ndez}, D.~A. and {Hearty}, Fred R. and {Majewski}, Steven R. and {Pereira}, C.~B. and {Placco}, Vinicius M. and {Villanova}, Sandro and {Zamora}, Olga},
        title = "{Identification of Neodymium in the Apogee H-Band Spectra}",
      journal = {\apj},
         year = 2016,
        month = dec,
       volume = {833},
       number = {1},
          eid = {81},
        pages = {81},
          doi = {10.3847/1538-4357/833/1/81},
       adsurl = {https://ui.adsabs.harvard.edu/abs/2016ApJ...833...81H}
}

@ARTICLE{cunha2017,
       author = {{Cunha}, Katia and {Smith}, Verne V. and {Hasselquist}, Sten and {Souto}, Diogo and {Shetrone}, Matthew D. and {Allende Prieto}, Carlos and {Bizyaev}, Dmitry and {Frinchaboy}, Peter and {Garc{\'\i}a-Hern{\'a}ndez}, D. Anibal and {Holtzman}, Jon and {Johnson}, Jennifer A. and {J{\H{o}}nsson}, Henrik and {Majewski}, Steven R. and {M{\'e}sz{\'a}ros}, Szabolcs and {Nidever}, David and {Pinsonneault}, Mark and {Schiavon}, Ricardo P. and {Sobeck}, Jennifer and {Skrutskie}, Michael F. and {Zamora}, Olga and {Zasowski}, Gail and {Fern{\'a}ndez-Trincado}, J.~G.},
        title = "{Adding the s-Process Element Cerium to the APOGEE Survey: Identification and Characterization of Ce II Lines in the H-band Spectral Window}",
      journal = {\apj},
         year = 2017,
        month = aug,
       volume = {844},
       number = {2},
          eid = {145},
        pages = {145},
          doi = {10.3847/1538-4357/aa7beb},
       adsurl = {https://ui.adsabs.harvard.edu/abs/2017ApJ...844..145C}
}

@ARTICLE{jorissen2019,
       author = {{Jorissen}, A. and {Boffin}, H.~M.~J. and {Karinkuzhi}, D. and {Van Eck}, S. and {Escorza}, A. and {Shetye}, S. and {Van Winckel}, H.},
        title = "{Barium and related stars, and their white-dwarf companions. I. Giant stars}",
      journal = {\aap},
         year = 2019,
        month = jun,
       volume = {626},
          eid = {A127},
        pages = {A127},
          doi = {10.1051/0004-6361/201834630},
archivePrefix = {arXiv},
       eprint = {1904.03975},
 primaryClass = {astro-ph.SR},
       adsurl = {https://ui.adsabs.harvard.edu/abs/2019A&A...626A.127J}
}

@ARTICLE{sun2021,
       author = {{Sun}, M. and {Mathieu}, Robert D. and {Leiner}, Emily M. and {Townsend}, R.~H.~D.},
        title = "{WOCS 5379: Detailed Analysis of the Evolution of a Post-mass-transfer Blue Straggler}",
      journal = {\apj},
         year = 2021,
        month = feb,
       volume = {908},
       number = {1},
          eid = {7},
        pages = {7},
          doi = {10.3847/1538-4357/abd402},
archivePrefix = {arXiv},
       eprint = {2012.08502},
 primaryClass = {astro-ph.SR},
       adsurl = {https://ui.adsabs.harvard.edu/abs/2021ApJ...908....7S}
}

@ARTICLE{frazer2025,
       author = {{Frazer}, Polly and {Griffith}, Emily J. and {Hogg}, David W. and {Sinha}, Amaya and {Tayar}, Jamie},
        title = "{Evolved stars with inconsistent age estimates: Abundance outliers or mass transfer products?}",
      journal = {arXiv e-prints},
         year = 2025,
        month = oct,
          eid = {arXiv:2510.26927},
        pages = {arXiv:2510.26927},
          doi = {10.48550/arXiv.2510.26927},
archivePrefix = {arXiv},
       eprint = {2510.26927},
 primaryClass = {astro-ph.SR},
       adsurl = {https://ui.adsabs.harvard.edu/abs/2025arXiv251026927F}
}

@misc{PyBACCHUS,
  author = {{Sinha}, Amaya},
  title = {PyBACCHUS: A Python Wrapper for running the Brussels Automatic Code for Characterizing High accUracy Spectra (BACCHUS)},
  year = {2025},
  month = sep,
  publisher = {Zenodo},
  doi = {10.5281/zenodo.17070806},
  url = {https://github.com/rocketxturtle/PyBACCHUS}
}

@ARTICLE{nelson2021,
       author = {{Nelson}, Tyler and {Ting}, Yuan-Sen and {Hawkins}, Keith and {Ji}, Alexander and {Kamdar}, Harshil and {El-Badry}, Kareem},
        title = "{Distant Relatives: The Chemical Homogeneity of Comoving Pairs Identified in Gaia}",
      journal = {\apj},
         year = 2021,
        month = nov,
       volume = {921},
       number = {2},
          eid = {118},
        pages = {118},
          doi = {10.3847/1538-4357/ac14be},
archivePrefix = {arXiv},
       eprint = {2104.12883},
 primaryClass = {astro-ph.SR},
       adsurl = {https://ui.adsabs.harvard.edu/abs/2021ApJ...921..118N}
}

@ARTICLE{kamdar2019,
       author = {{Kamdar}, Harshil and {Conroy}, Charlie and {Ting}, Yuan-Sen and {Bonaca}, Ana and {Smith}, Martin C. and {Brown}, Anthony G.~A.},
        title = "{Stars that Move Together Were Born Together}",
      journal = {\apjl},
         year = 2019,
        month = oct,
       volume = {884},
       number = {2},
          eid = {L42},
        pages = {L42},
          doi = {10.3847/2041-8213/ab4997},
archivePrefix = {arXiv},
       eprint = {1904.02159},
 primaryClass = {astro-ph.GA},
       adsurl = {https://ui.adsabs.harvard.edu/abs/2019ApJ...884L..42K}
}

@ARTICLE{andrews2019,
       author = {{Andrews}, Jeff J. and {Anguiano}, Borja and {Chanam{\'e}}, Julio and {Ag{\"u}eros}, Marcel A. and {Lewis}, Hannah M. and {Hayes}, Christian R. and {Majewski}, Steven R.},
        title = "{Using APOGEE Wide Binaries to Test Chemical Tagging with Dwarf Stars}",
      journal = {\apj},
         year = 2019,
        month = jan,
       volume = {871},
       number = {1},
          eid = {42},
        pages = {42},
          doi = {10.3847/1538-4357/aaf502},
archivePrefix = {arXiv},
       eprint = {1811.12032},
 primaryClass = {astro-ph.SR},
       adsurl = {https://ui.adsabs.harvard.edu/abs/2019ApJ...871...42A}
}

@ARTICLE{andrews2018,
       author = {{Andrews}, Jeff J. and {Chanam{\'e}}, Julio and {Ag{\"u}eros}, Marcel A.},
        title = "{Wide binaries in Tycho-Gaia II: metallicities, abundances and prospects for chemical tagging}",
      journal = {\mnras},
         year = 2018,
        month = feb,
       volume = {473},
       number = {4},
        pages = {5393-5406},
          doi = {10.1093/mnras/stx2685},
archivePrefix = {arXiv},
       eprint = {1710.04678},
 primaryClass = {astro-ph.SR},
       adsurl = {https://ui.adsabs.harvard.edu/abs/2018MNRAS.473.5393A}
}

@ARTICLE{dixon2020,
       author = {{Dixon}, Don and {Tayar}, Jamie and {Stassun}, Keivan G.},
        title = "{Rotationally Driven Ultraviolet Emission of Red Giant Stars}",
      journal = {\aj},
         year = 2020,
        month = jul,
       volume = {160},
       number = {1},
          eid = {12},
        pages = {12},
          doi = {10.3847/1538-3881/ab9080},
archivePrefix = {arXiv},
       eprint = {2005.00577},
 primaryClass = {astro-ph.SR},
       adsurl = {https://ui.adsabs.harvard.edu/abs/2020AJ....160...12D}
}

@ARTICLE{carlberg2012,
       author = {{Carlberg}, Joleen K. and {Cunha}, Katia and {Smith}, Verne V. and {Majewski}, Steven R.},
        title = "{Observable Signatures of Planet Accretion in Red Giant Stars. I. Rapid Rotation and Light Element Replenishment}",
      journal = {\apj},
         year = 2012,
        month = oct,
       volume = {757},
       number = {2},
          eid = {109},
        pages = {109},
          doi = {10.1088/0004-637X/757/2/109},
archivePrefix = {arXiv},
       eprint = {1208.1775},
 primaryClass = {astro-ph.SR},
       adsurl = {https://ui.adsabs.harvard.edu/abs/2012ApJ...757..109C}
}

@ARTICLE{patton2024,
       author = {{Patton}, Rachel A. and {Pinsonneault}, Marc H. and {Cao}, Lyra and {Vrard}, Mathieu and {Mathur}, Savita and {Garc{\'\i}a}, Rafael A. and {Tayar}, Jamie and {Daher}, Christine Mazzola and {Beck}, Paul G.},
        title = "{Spectroscopic identification of rapidly rotating red giant stars in APOKASC-3 and APOGEE DR16}",
      journal = {\mnras},
         year = 2024,
        month = feb,
       volume = {528},
       number = {2},
        pages = {3232-3248},
          doi = {10.1093/mnras/stae074},
archivePrefix = {arXiv},
       eprint = {2303.08151},
 primaryClass = {astro-ph.SR},
       adsurl = {https://ui.adsabs.harvard.edu/abs/2024MNRAS.528.3232P}
}

@ARTICLE{tayar2015,
       author = {{Tayar}, Jamie and {Ceillier}, Tugdual and {Garc{\'\i}a-Hern{\'a}ndez}, D.~A. and {Troup}, Nicholas W. and {Mathur}, Savita and {Garc{\'\i}a}, Rafael A. and {Zamora}, O. and {Johnson}, Jennifer A. and {Pinsonneault}, Marc H. and {M{\'e}sz{\'a}ros}, Szabolcs and {Allende Prieto}, Carlos and {Chaplin}, William J. and {Elsworth}, Yvonne and {Hekker}, Saskia and {Nidever}, David L. and {Salabert}, David and {Schneider}, Donald P. and {Serenelli}, Aldo and {Shetrone}, Matthew and {Stello}, Dennis},
        title = "{Rapid Rotation of Low-mass Red Giants Using APOKASC: A Measure of Interaction Rates on the Post-main-sequence}",
      journal = {\apj},
         year = 2015,
        month = jul,
       volume = {807},
       number = {1},
          eid = {82},
        pages = {82},
          doi = {10.1088/0004-637X/807/1/82},
archivePrefix = {arXiv},
       eprint = {1505.03536},
 primaryClass = {astro-ph.SR},
       adsurl = {https://ui.adsabs.harvard.edu/abs/2015ApJ...807...82T}
}

@ARTICLE{deMedeiros1996,
       author = {{de Medeiros}, J.~R. and {Da Rocha}, C. and {Mayor}, M.},
        title = "{The distribution of rotational velocity for evolved stars.}",
      journal = {\aap},
         year = 1996,
        month = oct,
       volume = {314},
        pages = {499-502},
       adsurl = {https://ui.adsabs.harvard.edu/abs/1996A&A...314..499D}
}

@BOOK{kepler1609,
       author = {{Kepler}, Johann},
        title = "{Astronomia nova.}",
         year = 1609,
       adsurl = {https://ui.adsabs.harvard.edu/abs/1609anov.book.....K}
}

@ARTICLE{mueller_horn2025,
       author = {{M{\"u}ller-Horn}, Johanna and {Rix}, Hans-Walter and {El-Badry}, Kareem and {Pennell}, Ben and {Green}, Matthew and {Li}, Jiadong and {Seeburger}, Rhys},
        title = "{Dormant BH candidates from Gaia DR3 summary diagnostics}",
      journal = {arXiv e-prints},
         year = 2025,
        month = oct,
          eid = {arXiv:2510.05982},
        pages = {arXiv:2510.05982},
          doi = {10.48550/arXiv.2510.05982},
archivePrefix = {arXiv},
       eprint = {2510.05982},
 primaryClass = {astro-ph.SR},
       adsurl = {https://ui.adsabs.harvard.edu/abs/2025arXiv251005982M}
}

@ARTICLE{castro_tapia_2024,
       author = {{Castro-Tapia}, Matias and {Aguilera-G{\'o}mez}, Claudia and {Chanam{\'e}}, Julio},
        title = "{Are lithium-rich giants binaries? A radial velocity variability analysis of 1400 giants}",
      journal = {\aap},
         year = 2024,
        month = oct,
       volume = {690},
          eid = {A367},
        pages = {A367},
          doi = {10.1051/0004-6361/202349106},
archivePrefix = {arXiv},
       eprint = {2401.00049},
 primaryClass = {astro-ph.SR},
       adsurl = {https://ui.adsabs.harvard.edu/abs/2024A&A...690A.367C}
}

@ARTICLE{vejar2021,
       author = {{Vejar}, George and {Schuler}, Simon C. and {Stassun}, Keivan G.},
        title = "{Detailed Abundances of Planet-hosting Open Clusters. The Praesepe (Beehive) Cluster}",
      journal = {\apj},
         year = 2021,
        month = oct,
       volume = {919},
       number = {2},
          eid = {100},
        pages = {100},
          doi = {10.3847/1538-4357/ac10c3},
archivePrefix = {arXiv},
       eprint = {2107.01472},
 primaryClass = {astro-ph.EP},
       adsurl = {https://ui.adsabs.harvard.edu/abs/2021ApJ...919..100V}
}

@ARTICLE{mack2016,
       author = {{Mack}, III, Claude E. and {Stassun}, Keivan G. and {Schuler}, Simon C. and {Hebb}, Leslie and {Pepper}, Joshua A.},
        title = "{Detailed Abundances of Planet-hosting Wide Binaries. II. HD80606+HD80607}",
      journal = {\apj},
         year = 2016,
        month = feb,
       volume = {818},
       number = {1},
          eid = {54},
        pages = {54},
          doi = {10.3847/0004-637X/818/1/54},
archivePrefix = {arXiv},
       eprint = {1601.00018},
 primaryClass = {astro-ph.SR},
       adsurl = {https://ui.adsabs.harvard.edu/abs/2016ApJ...818...54M}
}

@ARTICLE{cseh2018,
       author = {{Cseh}, B. and {Lugaro}, M. and {D'Orazi}, V. and {de Castro}, D.~B. and {Pereira}, C.~B. and {Karakas}, A.~I. and {Moln{\'a}r}, L. and {Plachy}, E. and {Szab{\'o}}, R. and {Pignatari}, M. and {Cristallo}, S.},
        title = "{The s process in AGB stars as constrained by a large sample of barium stars}",
      journal = {\aap},
         year = 2018,
        month = dec,
       volume = {620},
          eid = {A146},
        pages = {A146},
          doi = {10.1051/0004-6361/201834079},
archivePrefix = {arXiv},
       eprint = {1810.01788},
 primaryClass = {astro-ph.SR},
       adsurl = {https://ui.adsabs.harvard.edu/abs/2018A&A...620A.146C}
}

@ARTICLE{lugaro2003,
       author = {{Lugaro}, Maria and {Herwig}, Falk and {Lattanzio}, John C. and {Gallino}, Roberto and {Straniero}, Oscar},
        title = "{s-Process Nucleosynthesis in Asymptotic Giant Branch Stars: A Test for Stellar Evolution}",
      journal = {\apj},
         year = 2003,
        month = apr,
       volume = {586},
       number = {2},
        pages = {1305-1319},
          doi = {10.1086/367887},
archivePrefix = {arXiv},
       eprint = {astro-ph/0212364},
 primaryClass = {astro-ph},
       adsurl = {https://ui.adsabs.harvard.edu/abs/2003ApJ...586.1305L}
}

@ARTICLE{stebbins1911,
       author = {{Stebbins}, Joel},
        title = "{The Discovery of Eclipsing Variable Stars}",
      journal = {\apj},
         year = 1911,
        month = sep,
       volume = {34},
        pages = {105},
          doi = {10.1086/141874},
       adsurl = {https://ui.adsabs.harvard.edu/abs/1911ApJ....34..105S}
}

@ARTICLE{eggleton1983,
       author = {{Eggleton}, P.~P.},
        title = "{Aproximations to the radii of Roche lobes.}",
      journal = {\apj},
         year = 1983,
        month = may,
       volume = {268},
        pages = {368-369},
          doi = {10.1086/160960},
       adsurl = {https://ui.adsabs.harvard.edu/abs/1983ApJ...268..368E}
}

@ARTICLE{krumholz2025,
       author = {{Krumholz}, Mark R. and {Ting}, Yuan-Sen and {Li}, Zefeng and {Zhang}, Chuhan and {Mead}, Jennifer and {Ness}, Melissa K.},
        title = "{Metallicity fluctuation statistics in the interstellar medium and young stars -- II. Elemental cross-correlations and the structure of chemical abundance space}",
      journal = {arXiv e-prints},
         year = 2025,
        month = jul,
          eid = {arXiv:2507.14572},
        pages = {arXiv:2507.14572},
          doi = {10.48550/arXiv.2507.14572},
archivePrefix = {arXiv},
       eprint = {2507.14572},
 primaryClass = {astro-ph.GA},
       adsurl = {https://ui.adsabs.harvard.edu/abs/2025arXiv250714572K}
}

@ARTICLE{kolberg2022,
       author = {{Kolborg}, Anne Noer and {Martizzi}, Davide and {Ramirez-Ruiz}, Enrico and {Pfister}, Hugo and {Sakari}, Charli and {Wechsler}, Risa H. and {Soares-Furtado}, Melinda},
        title = "{Supernova-driven Turbulent Metal Mixing in High-redshift Galactic Disks: Metallicity Fluctuations in the Interstellar Medium and its Imprints on Metal-poor Stars in the Milky Way}",
      journal = {\apjl},
         year = 2022,
        month = sep,
       volume = {936},
       number = {2},
          eid = {L26},
        pages = {L26},
          doi = {10.3847/2041-8213/ac8c98},
archivePrefix = {arXiv},
       eprint = {2111.02619},
 primaryClass = {astro-ph.GA},
       adsurl = {https://ui.adsabs.harvard.edu/abs/2022ApJ...936L..26K}
}

@ARTICLE{casagrande2010,
       author = {{Casagrande}, L. and {Ram{\'\i}rez}, I. and {Mel{\'e}ndez}, J. and {Bessell}, M. and {Asplund}, M.},
        title = "{An absolutely calibrated T$_{eff}$ scale from the infrared flux method. Dwarfs and subgiants}",
      journal = {\aap},
         year = 2010,
        month = mar,
       volume = {512},
          eid = {A54},
        pages = {A54},
          doi = {10.1051/0004-6361/200913204},
archivePrefix = {arXiv},
       eprint = {1001.3142},
 primaryClass = {astro-ph.SR},
       adsurl = {https://ui.adsabs.harvard.edu/abs/2010A&A...512A..54C}
}

@ARTICLE{shrutskie2006,
       author = {{Skrutskie}, M.~F. and {Cutri}, R.~M. and {Stiening}, R. and {Weinberg}, M.~D. and {Schneider}, S. and {Carpenter}, J.~M. and {Beichman}, C. and {Capps}, R. and {Chester}, T. and {Elias}, J. and {Huchra}, J. and {Liebert}, J. and {Lonsdale}, C. and {Monet}, D.~G. and {Price}, S. and {Seitzer}, P. and {Jarrett}, T. and {Kirkpatrick}, J.~D. and {Gizis}, J.~E. and {Howard}, E. and {Evans}, T. and {Fowler}, J. and {Fullmer}, L. and {Hurt}, R. and {Light}, R. and {Kopan}, E.~L. and {Marsh}, K.~A. and {McCallon}, H.~L. and {Tam}, R. and {Van Dyk}, S. and {Wheelock}, S.},
        title = "{The Two Micron All Sky Survey (2MASS)}",
      journal = {\aj},
         year = 2006,
        month = feb,
       volume = {131},
       number = {2},
        pages = {1163-1183},
          doi = {10.1086/498708},
       adsurl = {https://ui.adsabs.harvard.edu/abs/2006AJ....131.1163S}
}

@software{isochrones,
       author = {{Morton}, Timothy D.},
        title = "{isochrones: Stellar model grid package}",
 howpublished = {Astrophysics Source Code Library, record ascl:1503.010},
         year = 2015,
        month = mar,
          eid = {ascl:1503.010},
archivePrefix = {ascl},
       eprint = {1503.010},
       adsurl = {https://ui.adsabs.harvard.edu/abs/2015ascl.soft03010M}
}

@ARTICLE{anguinao2022,
       author = {{Anguiano}, Borja and {Majewski}, Steven R. and {Stassun}, Keivan G. and {Badenes}, Carles and {Daher}, Christine Mazzola and {Dixon}, Don and {Allende Prieto}, Carlos and {Schneider}, Donald P. and {Price-Whelan}, Adrian M. and {Beaton}, Rachael L.},
        title = "{White Dwarf Binaries across the H-R Diagram}",
      journal = {\aj},
         year = 2022,
        month = oct,
       volume = {164},
       number = {4},
          eid = {126},
        pages = {126},
          doi = {10.3847/1538-3881/ac8357},
archivePrefix = {arXiv},
       eprint = {2207.13992},
 primaryClass = {astro-ph.SR},
       adsurl = {https://ui.adsabs.harvard.edu/abs/2022AJ....164..126A}
}

@ARTICLE{grilo2024,
       author = {{Grilo}, Vinicius and {Souto}, Diogo and {Cunha}, Katia and {Guer{\c{c}}o}, Rafael and {Vieira}, Rodrigo and {Smith}, Verne and {Vilar}, Deusalete and {Silva-Andrade}, Anderson and {Wanderley}, F{\'a}bio and {Daflon}, Simone and {Silva}, Jo{\~a}o Victor Sales},
        title = "{Chemical abundances for a sample of FGK dwarfs in the Pleiades open cluster from APOGEE}",
      journal = {\mnras},
         year = 2024,
        month = nov,
       volume = {534},
       number = {4},
        pages = {3005-3021},
          doi = {10.1093/mnras/stae2209},
archivePrefix = {arXiv},
       eprint = {2409.15207},
 primaryClass = {astro-ph.SR},
       adsurl = {https://ui.adsabs.harvard.edu/abs/2024MNRAS.534.3005G}
}

@ARTICLE{moog,
       author = {{Sneden}, C.},
        title = "{The nitrogen abundance of the very metal-poor star HD 122563.}",
      journal = {\apj},
         year = 1973,
        month = sep,
       volume = {184},
        pages = {839},
          doi = {10.1086/152374},
       adsurl = {https://ui.adsabs.harvard.edu/abs/1973ApJ...184..839S}
}

@ARTICLE{korg,
       author = {{Wheeler}, Adam J. and {Abruzzo}, Matthew W. and {Casey}, Andrew R. and {Ness}, Melissa K.},
        title = "{KORG: A Modern 1D LTE Spectral Synthesis Package}",
      journal = {\aj},
         year = 2023,
        month = feb,
       volume = {165},
       number = {2},
          eid = {68},
        pages = {68},
          doi = {10.3847/1538-3881/acaaad},
archivePrefix = {arXiv},
       eprint = {2211.00029},
 primaryClass = {astro-ph.SR},
       adsurl = {https://ui.adsabs.harvard.edu/abs/2023AJ....165...68W}
}

@ARTICLE{masseron2019,
       author = {{Masseron}, T. and {Garc{\'\i}a-Hern{\'a}ndez}, D.~A. and {M{\'e}sz{\'a}ros}, Sz. and {Zamora}, O. and {Dell'Agli}, F. and {Allende Prieto}, C. and {Edvardsson}, B. and {Shetrone}, M. and {Plez}, B. and {Fern{\'a}ndez-Trincado}, J.~G. and {Cunha}, K. and {J{\"o}nsson}, H. and {Geisler}, D. and {Beers}, T.~C. and {Cohen}, R.~E.},
        title = "{Homogeneous analysis of globular clusters from the APOGEE survey with the BACCHUS code. I. The northern clusters}",
      journal = {\aap},
         year = 2019,
        month = feb,
       volume = {622},
          eid = {A191},
        pages = {A191},
          doi = {10.1051/0004-6361/201834550},
archivePrefix = {arXiv},
       eprint = {1812.08817},
 primaryClass = {astro-ph.SR},
       adsurl = {https://ui.adsabs.harvard.edu/abs/2019A&A...622A.191M}
}

@ARTICLE{Lugaro2023,
       author = {{Lugaro}, Maria and {Pignatari}, Marco and {Reifarth}, Ren{\'e} and {Wiescher}, Michael},
        title = "{The s Process and Beyond}",
      journal = {Annual Review of Nuclear and Particle Science},
         year = 2023,
        month = sep,
       volume = {73},
        pages = {315-340},
          doi = {10.1146/annurev-nucl-102422-080857},
       adsurl = {https://ui.adsabs.harvard.edu/abs/2023ARNPS..73..315L}
}

@ARTICLE{prantzos2020,
       author = {{Prantzos}, N. and {Abia}, C. and {Cristallo}, S. and {Limongi}, M. and {Chieffi}, A.},
        title = "{Chemical evolution with rotating massive star yields II. A new assessment of the solar s- and r-process components}",
      journal = {\mnras},
         year = 2020,
        month = jan,
       volume = {491},
       number = {2},
        pages = {1832-1850},
          doi = {10.1093/mnras/stz3154},
archivePrefix = {arXiv},
       eprint = {1911.02545},
 primaryClass = {astro-ph.GA},
       adsurl = {https://ui.adsabs.harvard.edu/abs/2020MNRAS.491.1832P}
}

@ARTICLE{dongwook2024,
       author = {{Lim}, Dongwook and {Koch-Hansen}, Andreas J. and {Hong}, Seungsoo and {Chun}, Sang-Hyun and {Lee}, Young-Wook},
        title = "{Chemical Homogeneity of Wide Binary Systems: An Approach from Near-Infrared Spectroscopy}",
      journal = {\aj},
         year = 2024,
        month = jan,
       volume = {167},
       number = {1},
          eid = {3},
        pages = {3},
          doi = {10.3847/1538-3881/ad0a62},
archivePrefix = {arXiv},
       eprint = {2311.08461},
 primaryClass = {astro-ph.GA},
       adsurl = {https://ui.adsabs.harvard.edu/abs/2024AJ....167....3L}
}

@ARTICLE{manea2025,
       author = {{Manea}, Catherine and {Ness}, Melissa and {Hawkins}, Keith and {Zeimann}, Greg and {Hogg}, David W. and {Filion}, Carrie and {Griffith}, Emily J. and {Johnston}, Kathryn and {Casey}, Andrew and {Hackshaw}, Zoe and {Nelson}, Tyler and {Marks}, Micah},
        title = "{Optical Spectroscopy Reveals Hidden Neutron-capture Elemental Abundance Differences among APOGEE-identified Chemical Doppelg{\"a}ngers}",
      journal = {arXiv e-prints},
         year = 2025,
        month = aug,
          eid = {arXiv:2508.16717},
        pages = {arXiv:2508.16717},
          doi = {10.48550/arXiv.2508.16717},
archivePrefix = {arXiv},
       eprint = {2508.16717},
 primaryClass = {astro-ph.SR},
       adsurl = {https://ui.adsabs.harvard.edu/abs/2025arXiv250816717M}
}

@ARTICLE{stonemartinez2025,
       author = {{Stone-Martinez}, Alexander and {Holtzman}, Jon A. and {Lu}, Yuxi(Lucy) and {Hasselquist}, Sten and {Imig}, Julie and {Griffith}, Emily J. and {Bellinger}, Earl P. and {Saydjari}, Andrew K.},
        title = "{StarFlow: Leveraging Normalizing Flows for Stellar Age Estimation in SDSS-V DR19}",
      journal = {\aj},
         year = 2025,
        month = aug,
       volume = {170},
       number = {2},
          eid = {66},
        pages = {66},
          doi = {10.3847/1538-3881/addd18},
archivePrefix = {arXiv},
       eprint = {2503.03138},
 primaryClass = {astro-ph.SR},
       adsurl = {https://ui.adsabs.harvard.edu/abs/2025AJ....170...66S}
}

@ARTICLE{foster2024,
       author = {{Foster}, Steve and {Schiavon}, Ricardo P. and {de Castro}, Denise B. and {Lucatello}, Sara and {Daher}, Christine and {Penoyre}, Zephyr and {Price-Whelan}, Adrian and {Badenes}, Carles and {Fern{\'a}ndez-Trincado}, Jos{\'e} G. and {Garc{\'\i}a-Hern{\'a}ndez}, Domingo An{\'\i}bal and {Holtzman}, Jon and {J{\"o}nsson}, Henrik and {Shetrone}, Matthew},
        title = "{Carbon enrichment in APOGEE disk stars as evidence of mass transfer in binaries}",
      journal = {\aap},
         year = 2024,
        month = sep,
       volume = {689},
          eid = {A230},
        pages = {A230},
          doi = {10.1051/0004-6361/202450014},
archivePrefix = {arXiv},
       eprint = {2407.18130},
 primaryClass = {astro-ph.GA},
       adsurl = {https://ui.adsabs.harvard.edu/abs/2024A&A...689A.230F}
}

@ARTICLE{bufanda2023,
       author = {{Bufanda}, Erica and {Tayar}, Jamie and {Huber}, Daniel and {Hasselquist}, Sten and {Lane}, Richard R.},
        title = "{Investigating APOKASC Red Giant Stars with Abnormal Carbon-to-nitrogen Ratios}",
      journal = {\apj},
         year = 2023,
        month = dec,
       volume = {959},
       number = {2},
          eid = {123},
        pages = {123},
          doi = {10.3847/1538-4357/acf9a5},
archivePrefix = {arXiv},
       eprint = {2310.19872},
 primaryClass = {astro-ph.SR},
       adsurl = {https://ui.adsabs.harvard.edu/abs/2023ApJ...959..123B}
}

@ARTICLE{masseron2015,
       author = {{Masseron}, T. and {Gilmore}, G.},
        title = "{Carbon, nitrogen and {\ensuremath{\alpha}}-element abundances determine the formation sequence of the Galactic thick and thin discs}",
      journal = {\mnras},
         year = 2015,
        month = oct,
       volume = {453},
       number = {2},
        pages = {1855-1866},
          doi = {10.1093/mnras/stv1731},
archivePrefix = {arXiv},
       eprint = {1503.00537},
 primaryClass = {astro-ph.SR},
       adsurl = {https://ui.adsabs.harvard.edu/abs/2015MNRAS.453.1855M}
}

@ARTICLE{martig2016,
       author = {{Martig}, Marie and {Fouesneau}, Morgan and {Rix}, Hans-Walter and {Ness}, Melissa and {M{\'e}sz{\'a}ros}, Szabolcs and {Garc{\'\i}a-Hern{\'a}ndez}, D.~A. and {Pinsonneault}, Marc and {Serenelli}, Aldo and {Silva Aguirre}, Victor and {Zamora}, Olga},
        title = "{Red giant masses and ages derived from carbon and nitrogen abundances}",
      journal = {\mnras},
         year = 2016,
        month = mar,
       volume = {456},
       number = {4},
        pages = {3655-3670},
          doi = {10.1093/mnras/stv2830},
archivePrefix = {arXiv},
       eprint = {1511.08203},
 primaryClass = {astro-ph.SR},
       adsurl = {https://ui.adsabs.harvard.edu/abs/2016MNRAS.456.3655M}
}

@ARTICLE{spoo2022,
       author = {{Spoo}, Taylor and {Tayar}, Jamie and {Frinchaboy}, Peter M. and {Cunha}, Katia and {Myers}, Natalie and {Donor}, John and {Majewski}, Steven R. and {Bizyaev}, Dmitry and {Garc{\'\i}a-Hern{\'a}ndez}, D.~A. and {J{\"o}nsson}, Henrik and {Lane}, Richard R. and {Pan}, Kaike and {Longa-Pe{\~n}a}, Pen{\'e}lope and {Roman-Lopes}, A.},
        title = "{The Open Cluster Chemical Abundances and Mapping Survey. VII. APOGEE DR17 [C/N]-Age Calibration}",
      journal = {\aj},
         year = 2022,
        month = may,
       volume = {163},
       number = {5},
          eid = {229},
        pages = {229},
          doi = {10.3847/1538-3881/ac5d53},
archivePrefix = {arXiv},
       eprint = {2203.05463},
 primaryClass = {astro-ph.GA},
       adsurl = {https://ui.adsabs.harvard.edu/abs/2022AJ....163..229S}
}

@ARTICLE{espinoza2021,
       author = {{Espinoza-Rojas}, Francisca and {Chanam{\'e}}, Julio and {Jofr{\'e}}, Paula and {Casamiquela}, Laia},
        title = "{The Consistency of Chemical Clocks among Coeval Stars}",
      journal = {\apj},
         year = 2021,
        month = oct,
       volume = {920},
       number = {2},
          eid = {94},
        pages = {94},
          doi = {10.3847/1538-4357/ac15fd},
archivePrefix = {arXiv},
       eprint = {2105.01096},
 primaryClass = {astro-ph.SR},
       adsurl = {https://ui.adsabs.harvard.edu/abs/2021ApJ...920...94E}
}

@INPROCEEDINGS{plato,
       author = {{Baglin}, Annie and {Auvergne}, Michel and {Barge}, Pierre and {Deleuil}, Magali and {Michel}, Eric and {CoRoT Exoplanet Science Team}},
        title = "{CoRoT: Description of the Mission and Early Results}",
    booktitle = {Transiting Planets},
         year = 2009,
       editor = {{Pont}, Fr{\'e}d{\'e}ric and {Sasselov}, Dimitar and {Holman}, Matthew J.},
       series = {IAU Symposium},
       volume = {253},
        month = feb,
        pages = {71-81},
          doi = {10.1017/S1743921308026252},
       adsurl = {https://ui.adsabs.harvard.edu/abs/2009IAUS..253...71B}
}

@INPROCEEDINGS{corot,
       author = {{Baglin}, Annie and {Auvergne}, Michel and {Barge}, Pierre and {Deleuil}, Magali and {Michel}, Eric and {CoRoT Exoplanet Science Team}},
        title = "{CoRoT: Description of the Mission and Early Results}",
    booktitle = {Transiting Planets},
         year = 2009,
       editor = {{Pont}, Fr{\'e}d{\'e}ric and {Sasselov}, Dimitar and {Holman}, Matthew J.},
       series = {IAU Symposium},
       volume = {253},
        month = feb,
        pages = {71-81},
          doi = {10.1017/S1743921308026252},
       adsurl = {https://ui.adsabs.harvard.edu/abs/2009IAUS..253...71B}
}

@ARTICLE{tess,
       author = {{Ricker}, George R. and {Winn}, Joshua N. and {Vanderspek}, Roland and {Latham}, David W. and {Bakos}, G{\'a}sp{\'a}r {\'A}. and {Bean}, Jacob L. and {Berta-Thompson}, Zachory K. and {Brown}, Timothy M. and {Buchhave}, Lars and {Butler}, Nathaniel R. and {Butler}, R. Paul and {Chaplin}, William J. and {Charbonneau}, David and {Christensen-Dalsgaard}, J{\o}rgen and {Clampin}, Mark and {Deming}, Drake and {Doty}, John and {De Lee}, Nathan and {Dressing}, Courtney and {Dunham}, Edward W. and {Endl}, Michael and {Fressin}, Francois and {Ge}, Jian and {Henning}, Thomas and {Holman}, Matthew J. and {Howard}, Andrew W. and {Ida}, Shigeru and {Jenkins}, Jon M. and {Jernigan}, Garrett and {Johnson}, John Asher and {Kaltenegger}, Lisa and {Kawai}, Nobuyuki and {Kjeldsen}, Hans and {Laughlin}, Gregory and {Levine}, Alan M. and {Lin}, Douglas and {Lissauer}, Jack J. and {MacQueen}, Phillip and {Marcy}, Geoffrey and {McCullough}, Peter R. and {Morton}, Timothy D. and {Narita}, Norio and {Paegert}, Martin and {Palle}, Enric and {Pepe}, Francesco and {Pepper}, Joshua and {Quirrenbach}, Andreas and {Rinehart}, Stephen A. and {Sasselov}, Dimitar and {Sato}, Bun'ei and {Seager}, Sara and {Sozzetti}, Alessandro and {Stassun}, Keivan G. and {Sullivan}, Peter and {Szentgyorgyi}, Andrew and {Torres}, Guillermo and {Udry}, Stephane and {Villasenor}, Joel},
        title = "{Transiting Exoplanet Survey Satellite (TESS)}",
      journal = {Journal of Astronomical Telescopes, Instruments, and Systems},
         year = 2015,
        month = jan,
       volume = {1},
          eid = {014003},
        pages = {014003},
          doi = {10.1117/1.JATIS.1.1.014003},
       adsurl = {https://ui.adsabs.harvard.edu/abs/2015JATIS...1a4003R}
}

@ARTICLE{Armillotta2018,
       author = {{Armillotta}, Lucia and {Krumholz}, Mark R. and {Fujimoto}, Yusuke},
        title = "{Mixing of metals during star cluster formation: statistics and implications for chemical tagging}",
      journal = {\mnras},
         year = 2018,
        month = dec,
       volume = {481},
       number = {4},
        pages = {5000-5013},
          doi = {10.1093/mnras/sty2625},
archivePrefix = {arXiv},
       eprint = {1807.01712},
 primaryClass = {astro-ph.GA},
       adsurl = {https://ui.adsabs.harvard.edu/abs/2018MNRAS.481.5000A}
}

@ARTICLE{galahdr4,
       author = {{Buder}, Sven and {Kos}, Janez and {Wang}, Xi Ella and {McKenzie}, Madeleine and {Howell}, Madeleine and {Martell}, Sarah and {Hayden}, Michael R. and {Zucker}, Daniel B. and {Nordlander}, Thomas and {Montet}, Benjamin and {Traven}, Gregor and {Bland-Hawthorn}, Joss and {de Silva}, Gayandhi M. and {Freeman}, Kenneth and {Lewis}, Geraint and {Lind}, Karin and {Sharma}, Sanjib and {Simpson}, Jeffrey D. and {Stello}, Dennis and {Zwitter}, Tomaz and {Amarsi}, Anish M. and {Armstrong}, Joseph J. and {Banks}, Kirsten and {Beavis}, Mark and {Beeson}, Kevin-Luke and {Chen}, Boquan and {Ciuc{\u{a}}}, Ioana and {da Costa}, Gary S. and {de Grijs}, Richard and {Martin}, Bailey and {Nataf}, David Moise and {Ness}, Melissa and {Rains}, Adam D. and {Scarr}, Tim and {Vogrin{\v{c}}i{\v{c}}}, Rok and {Wang}, Zixian Purmortal and {Wittenmyer}, Rob A. and {Xie}, Yi Anne and {The Galah Collaboration}},
        title = "{The GALAH survey: Data release 4}",
      journal = {\pasa},
         year = 2025,
        month = may,
       volume = {42},
          eid = {e051},
        pages = {e051},
          doi = {10.1017/pasa.2025.26},
archivePrefix = {arXiv},
       eprint = {2409.19858},
 primaryClass = {astro-ph.GA},
       adsurl = {https://ui.adsabs.harvard.edu/abs/2025PASA...42...51B}
}

@ARTICLE{castro_ginard2024,
       author = {{Castro-Ginard}, Alfred and {Penoyre}, Zephyr and {Casey}, Andrew R. and {Brown}, Anthony G.~A. and {Belokurov}, Vasily and {Cantat-Gaudin}, Tristan and {Drimmel}, Ronald and {Fouesneau}, Morgan and {Khanna}, Shourya and {Kurbatov}, Evgeny P. and {Price-Whelan}, Adrian M. and {Rix}, Hans-Walter and {Smart}, Richard L.},
        title = "{Gaia DR3 detectability of unresolved binary systems}",
      journal = {\aap},
         year = 2024,
        month = aug,
       volume = {688},
          eid = {A1},
        pages = {A1},
          doi = {10.1051/0004-6361/202450172},
archivePrefix = {arXiv},
       eprint = {2404.14127},
 primaryClass = {astro-ph.GA},
       adsurl = {https://ui.adsabs.harvard.edu/abs/2024A&A...688A...1C}
}

@ARTICLE{kounkel2021,
       author = {{Kounkel}, Marina and {Covey}, Kevin R. and {Stassun}, Keivan G. and {Price-Whelan}, Adrian M. and {Holtzman}, Jon and {Chojnowski}, Drew and {Longa-Pe{\~n}a}, Pen{\'e}lope and {Rom{\'a}n-Z{\'u}{\~n}iga}, Carlos G. and {Hernandez}, Jesus and {Serna}, Javier and {Badenes}, Carles and {De Lee}, Nathan and {Majewski}, Steven and {Stringfellow}, Guy S. and {Kratter}, Kaitlin M. and {Moe}, Maxwell and {Frinchaboy}, Peter M. and {Beaton}, Rachael L. and {Fern{\'a}ndez-Trincado}, Jos{\'e} G. and {Mahadevan}, Suvrath and {Minniti}, Dante and {Beers}, Timothy C. and {Schneider}, Donald P. and {Barba}, Rodolfo and {Brownstein}, Joel R. and {Garc{\'\i}a-Hern{\'a}ndez}, Domingo An{\'\i}bal and {Pan}, Kaike and {Bizyaev}, Dmitry},
        title = "{Double-lined Spectroscopic Binaries in the APOGEE DR16 and DR17 Data}",
      journal = {\aj},
         year = 2021,
        month = nov,
       volume = {162},
       number = {5},
          eid = {184},
        pages = {184},
          doi = {10.3847/1538-3881/ac1798},
archivePrefix = {arXiv},
       eprint = {2107.10860},
 primaryClass = {astro-ph.SR},
       adsurl = {https://ui.adsabs.harvard.edu/abs/2021AJ....162..184K}
}

@ARTICLE{geller2021,
       author = {{Geller}, Aaron M. and {Mathieu}, Robert D. and {Latham}, David W. and {Pollack}, Maxwell and {Torres}, Guillermo and {Leiner}, Emily M.},
        title = "{Stellar Radial Velocities in the Old Open Cluster M67 (NGC 2682). II. The Spectroscopic Binary Population}",
      journal = {\aj},
         year = 2021,
        month = apr,
       volume = {161},
       number = {4},
          eid = {190},
        pages = {190},
          doi = {10.3847/1538-3881/abdd23},
archivePrefix = {arXiv},
       eprint = {2101.07883},
 primaryClass = {astro-ph.SR},
       adsurl = {https://ui.adsabs.harvard.edu/abs/2021AJ....161..190G}
}

@INPROCEEDINGS{mathieu2000,
       author = {{Mathieu}, R.~D.},
        title = "{The WIYN Open Cluster Study}",
    booktitle = {Stellar Clusters and Associations: Convection, Rotation, and Dynamos},
         year = 2000,
       editor = {{Pallavicini}, R. and {Micela}, G. and {Sciortino}, S.},
       series = {Astronomical Society of the Pacific Conference Series},
       volume = {198},
        month = jan,
        pages = {517},
       adsurl = {https://ui.adsabs.harvard.edu/abs/2000ASPC..198..517M}
}

@INPROCEEDINGS{shroyer2020,
       author = {{Shroyer}, J.~E. and {Carlberg}, J.},
        title = "{Characterizing Infrared Radial Velocity Jitter in Red Giant Stars}",
    booktitle = {American Astronomical Society Meeting Abstracts \#235},
         year = 2020,
       series = {American Astronomical Society Meeting Abstracts},
       volume = {235},
        month = jan,
          eid = {110.20},
        pages = {110.20},
       adsurl = {https://ui.adsabs.harvard.edu/abs/2020AAS...23511020S}
}

@ARTICLE{eisenstein2011,
       author = {{Eisenstein}, Daniel J. and {Weinberg}, David H. and {Agol}, Eric and {Aihara}, Hiroaki and {Allende Prieto}, Carlos and {Anderson}, Scott F. and {Arns}, James A. and {Aubourg}, {\'E}ric and {Bailey}, Stephen and {Balbinot}, Eduardo and {Barkhouser}, Robert and {Beers}, Timothy C. and {Berlind}, Andreas A. and {Bickerton}, Steven J. and {Bizyaev}, Dmitry and {Blanton}, Michael R. and {Bochanski}, John J. and {Bolton}, Adam S. and {Bosman}, Casey T. and {Bovy}, Jo and {Brandt}, W.~N. and {Breslauer}, Ben and {Brewington}, Howard J. and {Brinkmann}, J. and {Brown}, Peter J. and {Brownstein}, Joel R. and {Burger}, Dan and {Busca}, Nicolas G. and {Campbell}, Heather and {Cargile}, Phillip A. and {Carithers}, William C. and {Carlberg}, Joleen K. and {Carr}, Michael A. and {Chang}, Liang and {Chen}, Yanmei and {Chiappini}, Cristina and {Comparat}, Johan and {Connolly}, Natalia and {Cortes}, Marina and {Croft}, Rupert A.~C. and {Cunha}, Katia and {da Costa}, Luiz N. and {Davenport}, James R.~A. and {Dawson}, Kyle and {De Lee}, Nathan and {Porto de Mello}, Gustavo F. and {de Simoni}, Fernando and {Dean}, Janice and {Dhital}, Saurav and {Ealet}, Anne and {Ebelke}, Garrett L. and {Edmondson}, Edward M. and {Eiting}, Jacob M. and {Escoffier}, Stephanie and {Esposito}, Massimiliano and {Evans}, Michael L. and {Fan}, Xiaohui and {Femen{\'\i}a Castell{\'a}}, Bruno and {Dutra Ferreira}, Leticia and {Fitzgerald}, Greg and {Fleming}, Scott W. and {Font-Ribera}, Andreu and {Ford}, Eric B. and {Frinchaboy}, Peter M. and {Garc{\'\i}a P{\'e}rez}, Ana Elia and {Gaudi}, B. Scott and {Ge}, Jian and {Ghezzi}, Luan and {Gillespie}, Bruce A. and {Gilmore}, G. and {Girardi}, L{\'e}o and {Gott}, J. Richard and {Gould}, Andrew and {Grebel}, Eva K. and {Gunn}, James E. and {Hamilton}, Jean-Christophe and {Harding}, Paul and {Harris}, David W. and {Hawley}, Suzanne L. and {Hearty}, Frederick R. and {Hennawi}, Joseph F. and {Gonz{\'a}lez Hern{\'a}ndez}, Jonay I. and {Ho}, Shirley and {Hogg}, David W. and {Holtzman}, Jon A. and {Honscheid}, Klaus and {Inada}, Naohisa and {Ivans}, Inese I. and {Jiang}, Linhua and {Jiang}, Peng and {Johnson}, Jennifer A. and {Jordan}, Cathy and {Jordan}, Wendell P. and {Kauffmann}, Guinevere and {Kazin}, Eyal and {Kirkby}, David and {Klaene}, Mark A. and {Knapp}, G.~R. and {Kneib}, Jean-Paul and {Kochanek}, C.~S. and {Koesterke}, Lars and {Kollmeier}, Juna A. and {Kron}, Richard G. and {Lampeitl}, Hubert and {Lang}, Dustin and {Lawler}, James E. and {Le Goff}, Jean-Marc and {Lee}, Brian L. and {Lee}, Young Sun and {Leisenring}, Jarron M. and {Lin}, Yen-Ting and {Liu}, Jian and {Long}, Daniel C. and {Loomis}, Craig P. and {Lucatello}, Sara and {Lundgren}, Britt and {Lupton}, Robert H. and {Ma}, Bo and {Ma}, Zhibo and {MacDonald}, Nicholas and {Mack}, Claude and {Mahadevan}, Suvrath and {Maia}, Marcio A.~G. and {Majewski}, Steven R. and {Makler}, Martin and {Malanushenko}, Elena and {Malanushenko}, Viktor and {Mandelbaum}, Rachel and {Maraston}, Claudia and {Margala}, Daniel and {Maseman}, Paul and {Masters}, Karen L. and {McBride}, Cameron K. and {McDonald}, Patrick and {McGreer}, Ian D. and {McMahon}, Richard G. and {Mena Requejo}, Olga and {M{\'e}nard}, Brice and {Miralda-Escud{\'e}}, Jordi and {Morrison}, Heather L. and {Mullally}, Fergal and {Muna}, Demitri and {Murayama}, Hitoshi and {Myers}, Adam D. and {Naugle}, Tracy and {Neto}, Angelo Fausti and {Nguyen}, Duy Cuong and {Nichol}, Robert C. and {Nidever}, David L. and {O'Connell}, Robert W. and {Ogando}, Ricardo L.~C. and {Olmstead}, Matthew D. and {Oravetz}, Daniel J. and {Padmanabhan}, Nikhil and {Paegert}, Martin and {Palanque-Delabrouille}, Nathalie and {Pan}, Kaike and {Pandey}, Parul and {Parejko}, John K. and {P{\^a}ris}, Isabelle and {Pellegrini}, Paulo and {Pepper}, Joshua and {Percival}, Will J. and {Petitjean}, Patrick and {Pfaffenberger}, Robert and {Pforr}, Janine and {Phleps}, Stefanie and {Pichon}, Christophe and {Pieri}, Matthew M. and {Prada}, Francisco and {Price-Whelan}, Adrian M. and {Raddick}, M. Jordan and {Ramos}, Beatriz H.~F. and {Reid}, I. Neill and {Reyle}, Celine and {Rich}, James and {Richards}, Gordon T. and {Rieke}, George H. and {Rieke}, Marcia J. and {Rix}, Hans-Walter and {Robin}, Annie C. and {Rocha-Pinto}, Helio J. and {Rockosi}, Constance M. and {Roe}, Natalie A. and {Rollinde}, Emmanuel and {Ross}, Ashley J. and {Ross}, Nicholas P. and {Rossetto}, Bruno and {S{\'a}nchez}, Ariel G. and {Santiago}, Basilio and {Sayres}, Conor and {Schiavon}, Ricardo and {Schlegel}, David J. and {Schlesinger}, Katharine J. and {Schmidt}, Sarah J. and {Schneider}, Donald P. and {Sellgren}, Kris and {Shelden}, Alaina and {Sheldon}, Erin and {Shetrone}, Matthew},
        title = "{SDSS-III: Massive Spectroscopic Surveys of the Distant Universe, the Milky Way, and Extra-Solar Planetary Systems}",
      journal = {\aj},
         year = 2011,
        month = sep,
       volume = {142},
       number = {3},
          eid = {72},
        pages = {72},
          doi = {10.1088/0004-6256/142/3/72},
archivePrefix = {arXiv},
       eprint = {1101.1529},
 primaryClass = {astro-ph.IM},
       adsurl = {https://ui.adsabs.harvard.edu/abs/2011AJ....142...72E}
}

@ARTICLE{sinha2024,
       author = {{Sinha}, Amaya and {Zasowski}, Gail and {Frinchaboy}, Peter and {Cunha}, Katia and {Souto}, Diogo and {Tayar}, Jamie and {Stassun}, Keivan},
        title = "{A Comprehensive Study of Open Cluster Chemical Homogeneity Using APOGEE and Milky Way Mapper Abundances}",
      journal = {\apj},
         year = 2024,
        month = nov,
       volume = {975},
       number = {1},
          eid = {89},
        pages = {89},
          doi = {10.3847/1538-4357/ad78e1},
archivePrefix = {arXiv},
       eprint = {2409.02360},
 primaryClass = {astro-ph.GA},
       adsurl = {https://ui.adsabs.harvard.edu/abs/2024ApJ...975...89S}
}

@ARTICLE{prsa2016,
       author = {{Pr{\v{s}}a}, Andrej and {Harmanec}, Petr and {Torres}, Guillermo and {Mamajek}, Eric and {Asplund}, Martin and {Capitaine}, Nicole and {Christensen-Dalsgaard}, J{\o}rgen and {Depagne}, {\'E}ric and {Haberreiter}, Margit and {Hekker}, Saskia and {Hilton}, James and {Kopp}, Greg and {Kostov}, Veselin and {Kurtz}, Donald W. and {Laskar}, Jacques and {Mason}, Brian D. and {Milone}, Eugene F. and {Montgomery}, Michele and {Richards}, Mercedes and {Schmutz}, Werner and {Schou}, Jesper and {Stewart}, Susan G.},
        title = "{Nominal Values for Selected Solar and Planetary Quantities: IAU 2015 Resolution B3}",
      journal = {\aj},
         year = 2016,
        month = aug,
       volume = {152},
       number = {2},
          eid = {41},
        pages = {41},
          doi = {10.3847/0004-6256/152/2/41},
archivePrefix = {arXiv},
       eprint = {1605.09788},
 primaryClass = {astro-ph.SR},
       adsurl = {https://ui.adsabs.harvard.edu/abs/2016AJ....152...41P}
}

@ARTICLE{meszaros2025,
       author = {{M{\'e}sz{\'a}ros}, Szabolcs and {Jofr{\'e}}, Paula and {Johnson}, Jennifer A. and {Bird}, Jonathan C. and {Bovy}, Jo and {Casey}, Andrew R. and {Chaname}, Julio and {Cunha}, Katia and {De Lee}, Nathan and {Frinchaboy}, Peter and {Guiglion}, Guillaume and {Heged{\H{u}}s}, Viola and {Ji}, Alex P. and {Kollmeier}, Juna A. and {Ness}, Melissa K. and {Otto}, Jonah and {Pinsonneault}, Marc H. and {Roman-Lopes}, Alexandre and {Saydjari}, Andrew and {Sinha}, Amaya and {Song}, Ying-Yi and {Stringfellow}, Guy S. and {Stassun}, Keivan G. and {Tayar}, Jamie and {Tkachenko}, Andrew and {Valentini}, Marica and {Way}, Zachary and {Weingrill}, J{\"o}rg},
        title = "{SDSS-V Milky Way Mapper (MWM): ASPCAP Stellar Parameters and Abundances in SDSS-V Data Release 19}",
      journal = {arXiv e-prints},
         year = 2025,
        month = jun,
          eid = {arXiv:2506.07845},
        pages = {arXiv:2506.07845},
          doi = {10.48550/arXiv.2506.07845},
archivePrefix = {arXiv},
       eprint = {2506.07845},
 primaryClass = {astro-ph.SR},
       adsurl = {https://ui.adsabs.harvard.edu/abs/2025arXiv250607845M}
}

@ARTICLE{Hawkins2016,
       author = {{Hawkins}, K. and {Jofr{\'e}}, P. and {Heiter}, U. and {Soubiran}, C. and {Blanco-Cuaresma}, S. and {Casagrande}, L. and {Gilmore}, G. and {Lind}, K. and {Magrini}, L. and {Masseron}, T. and {Pancino}, E. and {Randich}, S. and {Worley}, C.~C.},
        title = "{Gaia FGK benchmark stars: new candidates at low metallicities}",
      journal = {\aap},
         year = 2016,
        month = jul,
       volume = {592},
          eid = {A70},
        pages = {A70},
          doi = {10.1051/0004-6361/201628268},
archivePrefix = {arXiv},
       eprint = {1605.08229},
 primaryClass = {astro-ph.SR},
       adsurl = {https://ui.adsabs.harvard.edu/abs/2016A&A...592A..70H}
}

@ARTICLE{cg2020,
       author = {{Cantat-Gaudin}, T. and {Anders}, F. and {Castro-Ginard}, A. and {Jordi}, C. and {Romero-G{\'o}mez}, M. and {Soubiran}, C. and {Casamiquela}, L. and {Tarricq}, Y. and {Moitinho}, A. and {Vallenari}, A. and {Bragaglia}, A. and {Krone-Martins}, A. and {Kounkel}, M.},
        title = "{Painting a portrait of the Galactic disc with its stellar clusters}",
      journal = {\aap},
         year = 2020,
        month = aug,
       volume = {640},
          eid = {A1},
        pages = {A1},
          doi = {10.1051/0004-6361/202038192},
archivePrefix = {arXiv},
       eprint = {2004.07274},
 primaryClass = {astro-ph.GA},
       adsurl = {https://ui.adsabs.harvard.edu/abs/2020A&A...640A...1C}
}

@ARTICLE{dr3_nss,
       author = {{Gaia Collaboration} and {Arenou}, F. and {Babusiaux}, C. and {Barstow}, M.~A. and {Faigler}, S. and {Jorissen}, A. and {Kervella}, P. and {Mazeh}, T. and {Mowlavi}, N. and {Panuzzo}, P. and {Sahlmann}, J. and {Shahaf}, S. and {Sozzetti}, A. and {Bauchet}, N. and {Damerdji}, Y. and {Gavras}, P. and {Giacobbe}, P. and {Gosset}, E. and {Halbwachs}, J. -L. and {Holl}, B. and {Lattanzi}, M.~G. and {Leclerc}, N. and {Morel}, T. and {Pourbaix}, D. and {Re Fiorentin}, P. and {Sadowski}, G. and {S{\'e}gransan}, D. and {Siopis}, C. and {Teyssier}, D. and {Zwitter}, T. and {Planquart}, L. and {Brown}, A.~G.~A. and {Vallenari}, A. and {Prusti}, T. and {de Bruijne}, J.~H.~J. and {Biermann}, M. and {Creevey}, O.~L. and {Ducourant}, C. and {Evans}, D.~W. and {Eyer}, L. and {Guerra}, R. and {Hutton}, A. and {Jordi}, C. and {Klioner}, S.~A. and {Lammers}, U.~L. and {Lindegren}, L. and {Luri}, X. and {Mignard}, F. and {Panem}, C. and {Randich}, S. and {Sartoretti}, P. and {Soubiran}, C. and {Tanga}, P. and {Walton}, N.~A. and {Bailer-Jones}, C.~A.~L. and {Bastian}, U. and {Drimmel}, R. and {Jansen}, F. and {Katz}, D. and {van Leeuwen}, F. and {Bakker}, J. and {Cacciari}, C. and {Casta{\~n}eda}, J. and {De Angeli}, F. and {Fabricius}, C. and {Fouesneau}, M. and {Fr{\'e}mat}, Y. and {Galluccio}, L. and {Guerrier}, A. and {Heiter}, U. and {Masana}, E. and {Messineo}, R. and {Nicolas}, C. and {Nienartowicz}, K. and {Pailler}, F. and {Riclet}, F. and {Roux}, W. and {Seabroke}, G.~M. and {Sordo}, R. and {Th{\'e}venin}, F. and {Gracia-Abril}, G. and {Portell}, J. and {Altmann}, M. and {Andrae}, R. and {Audard}, M. and {Bellas-Velidis}, I. and {Benson}, K. and {Berthier}, J. and {Blomme}, R. and {Burgess}, P.~W. and {Busonero}, D. and {Busso}, G. and {C{\'a}novas}, H. and {Carry}, B. and {Cellino}, A. and {Cheek}, N. and {Clementini}, G. and {Davidson}, M. and {de Teodoro}, P. and {Nu{\~n}ez Campos}, M. and {Delchambre}, L. and {Dell'Oro}, A. and {Esquej}, P. and {Fern{\'a}ndez-Hern{\'a}ndez}, J. and {Fraile}, E. and {Garabato}, D. and {Garc{\'\i}a-Lario}, P. and {Haigron}, R. and {Hambly}, N.~C. and {Harrison}, D.~L. and {Hern{\'a}ndez}, J. and {Hestroffer}, D. and {Hodgkin}, S.~T. and {Jan{\ss}en}, K. and {Jevardat de Fombelle}, G. and {Jordan}, S. and {Krone-Martins}, A. and {Lanzafame}, A.~C. and {L{\"o}ffler}, W. and {Marchal}, O. and {Marrese}, P.~M. and {Moitinho}, A. and {Muinonen}, K. and {Osborne}, P. and {Pancino}, E. and {Pauwels}, T. and {Recio-Blanco}, A. and {Reyl{\'e}}, C. and {Riello}, M. and {Rimoldini}, L. and {Roegiers}, T. and {Rybizki}, J. and {Sarro}, L.~M. and {Smith}, M. and {Utrilla}, E. and {van Leeuwen}, M. and {Abbas}, U. and {{\'A}brah{\'a}m}, P. and {Abreu Aramburu}, A. and {Aerts}, C. and {Aguado}, J.~J. and {Ajaj}, M. and {Aldea-Montero}, F. and {Altavilla}, G. and {{\'A}lvarez}, M.~A. and {Alves}, J. and {Anders}, F. and {Anderson}, R.~I. and {Anglada Varela}, E. and {Antoja}, T. and {Baines}, D. and {Baker}, S.~G. and {Balaguer-N{\'u}{\~n}ez}, L. and {Balbinot}, E. and {Balog}, Z. and {Barache}, C. and {Barbato}, D. and {Barros}, M. and {Bartolom{\'e}}, S. and {Bassilana}, J. -L. and {Becciani}, U. and {Bellazzini}, M. and {Berihuete}, A. and {Bernet}, M. and {Bertone}, S. and {Bianchi}, L. and {Binnenfeld}, A. and {Blanco-Cuaresma}, S. and {Blazere}, A. and {Boch}, T. and {Bombrun}, A. and {Bossini}, D. and {Bouquillon}, S. and {Bragaglia}, A. and {Bramante}, L. and {Breedt}, E. and {Bressan}, A. and {Brouillet}, N. and {Brugaletta}, E. and {Bucciarelli}, B. and {Burlacu}, A. and {Butkevich}, A.~G. and {Buzzi}, R. and {Caffau}, E. and {Cancelliere}, R. and {Cantat-Gaudin}, T. and {Carballo}, R. and {Carlucci}, T. and {Carnerero}, M.~I. and {Carrasco}, J.~M. and {Casamiquela}, L. and {Castellani}, M. and {Castro-Ginard}, A. and {Chaoul}, L. and {Charlot}, P. and {Chemin}, L. and {Chiaramida}, V. and {Chiavassa}, A. and {Chornay}, N. and {Comoretto}, G.},
        title = "{Gaia Data Release 3. Stellar multiplicity, a teaser for the hidden treasure}",
      journal = {\aap},
         year = 2023,
        month = jun,
       volume = {674},
          eid = {A34},
        pages = {A34},
          doi = {10.1051/0004-6361/202243782},
archivePrefix = {arXiv},
       eprint = {2206.05595},
 primaryClass = {astro-ph.SR},
       adsurl = {https://ui.adsabs.harvard.edu/abs/2023A&A...674A..34G}
}

@ARTICLE{muellerhorn2025,
       author = {{M{\"u}ller-Horn}, Johanna and {G{\"o}ttgens}, Fabian and {Dreizler}, Stefan and {Kamann}, Sebastian and {Martens}, Sven and {Saracino}, Sara and {Ye}, Claire S.},
        title = "{Binary properties of the globular cluster 47 Tuc (NGC 104): A dearth of short-period binaries}",
      journal = {\aap},
         year = 2025,
        month = jan,
       volume = {693},
          eid = {A161},
        pages = {A161},
          doi = {10.1051/0004-6361/202450709},
archivePrefix = {arXiv},
       eprint = {2412.13189},
 primaryClass = {astro-ph.SR},
       adsurl = {https://ui.adsabs.harvard.edu/abs/2025A&A...693A.161M}
}

@ARTICLE{dr19,
       author = {{SDSS Collaboration} and {Adamane Pallathadka}, Gautham and {Aghakhanloo}, Mojgan and {Aird}, James and {Almeida}, Andr{\'e}s and {Amrita}, Singh and {Anders}, Friedrich and {Anderson}, Scott F. and {Arseneau}, Stefan and {Gonz{\'a}lez Avila}, Consuelo and {Aviram}, Shir and {Aydar}, Catarina and {Badenes}, Carles and {Barrera-Ballesteros}, Jorge K. and {Bauer}, Franz E. and {Behmard}, Aida and {Berg}, Michelle and {Besser}, F. and {Moni Bidin}, Christian and {Bizyaev}, Dmitry and {Blanc}, Guillermo and {Blanton}, Michael R. and {Bovy}, Jo and {Brandt}, William Nielsen and {Brownstein}, Joel R. and {Buchner}, Johannes and {Bulbul}, Esra and {Burchett}, Joseph N. and {Carigi}, Leticia and {Carlberg}, Joleen K. and {Casey}, Andrew R. and {Chakraborty}, Priyanka and {Chanam{\'e}}, Julio and {Chandra}, Vedant and {Chiappini}, Cristina and {Chilingarian}, Igor and {Comparat}, Johan and {Covey}, Kevin and {Crumpler}, Nicole and {Cunha}, Katia and {D'Onghia}, Elena and {Dai}, Xinyu and {Darling}, Jeremy and {Davis}, Megan and {De Lee}, Nathan and {Deacon}, Niall and {M{\'e}ndez Delgado}, Jos{\'e} Eduardo and {Demasi}, Sebastian and {Demianenko}, Mariia and {Demke}, Delvin and {Donor}, John and {Drory}, Niv and {Villa Durango}, Monica Alejandra and {Dwelly}, Tom and {Egorov}, Oleg and {Egorova}, Evgeniya and {El-Badry}, Kareem and {Eracleous}, Mike and {Fan}, Xiaohui and {Farr}, Emily and {Finkbeiner}, Douglas P. and {Fries}, Logan and {Frinchaboy}, Peter and {Gentile Fusillo}, Nicola Pietro and {Serrano F{\'e}lix}, Luis Daniel and {Gaensicke}, Boris and {Galligan}, Emma and {Garc{\'\i}a}, Pablo and {Gelfand}, Joseph and {Grabowski}, Katie and {Grebel}, Eva and {Green}, Paul J and {Greve}, Hannah and {Grier}, Catherine and {Griffith}, Emily and {Guetzoyan}, Paloma and {Gupta}, Pramod and {Hackshaw}, Zoe and {Hall}, Patrick B. and {Hawkins}, Keith and {Heged{\H{u}}s}, Viola and {Hekker}, Saskia and {Herbst}, T.~M. and {Hermes}, J.~J. and {Hern{\'a}ndez-Garc{\'\i}a}, Lorena and {Hiremath}, Pranavi and {Hogg}, David W and {Holtzman}, Jon and {Horne}, Keith and {Horta}, Danny and {Huang}, Yang and {Hutchinson}, Brian and {H{\"a}berle}, Maximilian and {Ibarra-Medel}, Hector Javier and {Ji}, Alexander P. and {Jofre}, Paula and {Johnson}, James W. and {Johnson}, Jennifer and {Johnston}, Evelyn J. and {Kaldor}, Mary and {Katkov}, Ivan and {Khalatyan}, Arman and {Khoperskov}, Sergey and {Klessen}, Ralf and {Kluge}, Matthias and {Koekemoer}, Anton M. and {Kollmeier}, Juna A. and {Kounkel}, Marina and {Kreckel}, Kathryn and {Krishnarao}, Dhanesh and {Krumpe}, Mirko and {Lacerna}, Ivan and {Laporte}, Chervin and {Lepine}, Sebastien and {Li}, Jing and {Liang}, Fu-Heng and {Limberg}, Guilherme and {Liu}, Xin and {Loebman}, Sarah and {Long}, Knox and {Lu}, Yuxi and {Lucey}, Madeline and {Lugo-Aranda}, Alejandra Z. and {Mart{\'\i}nez Martinez-Aldama}, Mary Loli and {McKinnon}, Kevin and {Medan}, Ilija and {Merloni}, Andrea and {Morrison}, Sean and {Myers}, Natalie and {M{\'e}sz{\'a}ros}, Szabolcs and {M{\"u}ller-Horn}, Johanna and {Nepal}, Samir and {Ness}, Melissa and {Nidever}, David and {Nitschelm}, Christian and {Oravetz}, Audrey and {Otto}, Jonah and {Pan}, Kaike and {P{\'e}rez Paolino}, Facundo and {Negrete Pe{\~n}aloza}, Castalia Alenka and {Pinsonneault}, Marc and {Taghizadeh Popp}, Manuchehr and {Price-Whelan}, Adrian and {Pulatova}, Nadiia and {Queiroz}, Anna Barbara and {Raddick}, Jordan and {Rankine}, Amy and {Rix}, Hans-Walter and {Rom{\'a}n-Z{\'u}{\~n}iga}, Carlos and {Fern{\'a}ndez Rosso}, Daniela and {Runnoe}, Jessie and {Mahmud Saad}, Serat and {Salvato}, Mara and {Sanchez}, Sebastian F. and {Sattler}, Natascha and {Saydjari}, Andrew and {Sayres}, Conor and {Schlaufman}, Kevin and {Schneider}, Donald P. and {Schwope}, Axel and {Seaton}, Lucas M. and {Seeburger}, Rhys and {Serna}, Javier and {Sharma}, Sanjib and {Shen}, Yue and {Sinha}, Amaya and {Sizemore}, Brian and {Sniegowska}, Marzena and {Song}, Yingyi and {Souto}, Diogo and {Stassun}, Keivan and {Steinmetz}, Matthias and {Stone}, Zachary and {Stone-Martinez}, Alexander and {Stringfellow}, Guy S. and {Mata S{\'a}nchez}, Aurora and {S{\'a}nchez-Gallego}, Jos{\'e} and {Tan}, Jonathan and {Tayar}, Jamie and {Thai}, Riley and {Thakar}, Ani and {Thibodeaux}, Pierre and {Ting}, Yuan-Sen and {Tkachenko}, Andrew and {Trakhtenbrot}, Benny and {Fernandez Trincado}, Jose G. and {Troup}, Nicholas and {Trump}, Jonathan R. and {Ulloa}, Natalie and {Van der Marel}, Roeland P. and {Vera}, Pablo and {Villanova}, Sandro and {Villase{\~n}or}, Jaime and {Wang}, Ji and {Way}, Zachary and {Weijmans}, Anne-Marie and {Wheeler}, Adam and {Wilson}, John C. and {Wofford}, Aida and {Wong}, Tony},
        title = "{The Nineteenth Data Release of the Sloan Digital Sky Survey}",
      journal = {arXiv e-prints},
         year = 2025,
        month = jul,
          eid = {arXiv:2507.07093},
        pages = {arXiv:2507.07093},
          doi = {10.48550/arXiv.2507.07093},
archivePrefix = {arXiv},
       eprint = {2507.07093},
 primaryClass = {astro-ph.GA},
       adsurl = {https://ui.adsabs.harvard.edu/abs/2025arXiv250707093S}
}

@ARTICLE{kollmeier2025,
       author = {{Kollmeier}, Juna A. and {Rix}, Hans-Walter and {Aerts}, Conny and {Aird}, James and {Alfaro}, Pablo Vera and {Almeida}, Andr{\'e}s and {Anderson}, Scott F. and {Jim{\'e}nez Arranz}, {\'O}scar and {Arseneau}, Stefan M. and {Assef}, Roberto and {Aviram}, Shir and {Aydar}, Catarina and {Badenes}, Carles and {Bandyopadhyay}, Avrajit and {Barger}, Kat and {Barkhouser}, Robert H. and {Bauer}, Franz E. and {Bender}, Chad and {Besser}, Felipe and {Bhattarai}, Binod and {Bilgi}, Pavaman and {Bird}, Jonathan and {Bizyaev}, Dmitry and {Blanc}, Guillermo A. and {Blanton}, Michael R. and {Bochanski}, John and {Bovy}, Jo and {Brandon}, Christopher and {Brandt}, William Nielsen and {Brownstein}, Joel R. and {Buchner}, Johannes and {Burchett}, Joseph N. and {Carlberg}, Joleen and {Casey}, Andrew R. and {Castaneda-Carlos}, Lesly and {Chakraborty}, Priyanka and {Chanam{\'e}}, Julio and {Chandra}, Vedant and {Cherinka}, Brian and {Chilingarian}, Igor and {Comparat}, Johan and {Cosens}, Maren and {Covey}, Kevin and {Crane}, Jeffrey D. and {Crumpler}, Nicole R. and {Cunha}, Katia and {Cunningham}, Tim and {Dai}, Xinyu and {Darling}, Jeremy and {Davidson}, Jr., James W. and {Davis}, Megan C. and {De Lee}, Nathan and {Deacon}, Niall and {M{\'e}ndez Delgado}, Jos{\'e} Eduardo and {Demasi}, Sebastian and {Demianenko}, Mariia and {Derwent}, Mark and {D'Onghia}, Elena and {Di Mille}, Francesco and {Dias}, Bruno and {Donor}, John and {Drory}, Niv and {Dwelly}, Tom and {Egorov}, Oleg and {Egorova}, Evgeniya and {El-Badry}, Kareem and {Engelman}, Mike and {Eracleous}, Mike and {Fan}, Xiaohui and {Farr}, Emily and {Fries}, Logan and {Frinchaboy}, Peter and {Froning}, Cynthia S. and {G{\"a}nsicke}, Boris T. and {Garc{\'\i}a}, Pablo and {Gelfand}, Joseph and {Gentile Fusillo}, Nicola Pietro and {Glover}, Simon and {Grabowski}, Katie and {Grebel}, Eva K. and {Green}, Paul J and {Grier}, Catherine and {Gupta}, Pramod and {Gray}, Aidan C. and {H{\"a}berle}, Maximilian and {Hall}, Patrick B. and {Hammond}, Randolph P. and {Hawkins}, Keith and {Harding}, Albert C. and {Heged{\H{u}}s}, Viola and {Herbst}, Tom and {Hermes}, J.~J. and {Rodr{\'\i}guez Hidalgo}, Paola and {Hilder}, Thomas and {Hogg}, David W and {Holtzman}, Jon A. and {Horta}, Danny and {Huang}, Yang and {Hwang}, Hsiang-Chih and {Ibarra-Medel}, Hector Javier and {Imig}, Julie and {Inight}, Keith and {Jana}, Arghajit and {Ji}, Alexander P. and {Jofre}, Paula and {Johns}, Matt and {Johnson}, Jennifer and {Johnson}, James W. and {Johnston}, Evelyn J. and {Jones}, Amy M and {Katkov}, Ivan and {Koekemoer}, Anton M. and {Kounkel}, Marina and {Kreckel}, Kathryn and {Krishnarao}, Dhanesh and {Krumpe}, Mirko and {Kumari}, Nimisha and {Kupfer}, Thomas and {Lacerna}, Ivan and {Laporte}, Chervin and {Lepine}, Sebastien and {Li}, Jing and {Liu}, Xin and {Loebman}, Sarah and {Long}, Knox and {Roman-Lopes}, Alexandre and {Lu}, Yuxi and {Majewski}, Steven Raymond and {Maoz}, Dan and {McKinnon}, Kevin A. and {Medan}, Ilija and {Merloni}, Andrea and {Minniti}, Dante and {Morrison}, Sean and {Myers}, Natalie and {M{\'e}sz{\'a}ros}, Szabolcs and {Nandra}, Kirpal and {Nayak}, Prasanta K. and {Ness}, Melissa K and {Nidever}, David L. and {O'Brien}, Thomas and {Oeur}, Micah and {Oravetz}, Audrey and {Oravetz}, Daniel and {Otto}, Jonah and {Adamane Pallathadka}, Gautham and {Palunas}, Povilas and {Pan}, Kaike and {Pappalardo}, Daniel and {Pandey}, Rakesh and {Negrete Pe{\~n}aloza}, Castalia Alenka and {Pinsonneault}, Marc H. and {Pogge}, Richard W. and {Taghizadeh Popp}, Manuchehr and {Price-Whelan}, Adrian M. and {Pulatova}, Nadiia and {Qiu}, Dan and {Ramirez}, Solange and {Rankine}, Amy and {Ricci}, Claudio and {Runnoe}, Jessie C. and {Sanchez}, Sebastian and {Salvato}, Mara and {Sattler}, Natascha and {Saydjari}, Andrew K. and {Sayres}, Conor and {Schlaufman}, Kevin C. and {Schneider}, Donald P. and {Schreiber}, Matthias R. and {Schwope}, Axel and {Serna}, Javier and {Shen}, Yue and {Sif{\'o}n}, Crist{\'o}bal and {Singh}, Amrita and {Sinha}, Amaya and {Smee}, Stephen and {Song}, Ying-Yi and {Souto}, Diogo and {Stassun}, Keivan G. and {Steinmetz}, Matthias and {Stone-Martinez}, Alexander and {Stringfellow}, Guy and {Stutz}, Amelia and {Jos{\'e}} and {S{\'a}} and {nchez-Gallego} and {Tan}, Jonathan C. and {Tayar}, Jamie and {Thai}, Riley and {Thakar}, Ani and {Ting}, Yuan-Sen and {Tkachenko}, Andrew and {Tovmasian}, Gagik and {Trakhtenbrot}, Benny and {Fern{\'a}ndez-Trincado}, Jos{\'e} G. and {Troup}, Nicholas and {Trump}, Jonathan and {Tuttle}, Sarah and {van der Marel}, Roeland P. and {Villanova}, Sandro},
        title = "{Sloan Digital Sky Survey-V: Pioneering Panoptic Spectroscopy}",
      journal = {arXiv e-prints},
         year = 2025,
        month = jul,
          eid = {arXiv:2507.06989},
        pages = {arXiv:2507.06989},
          doi = {10.48550/arXiv.2507.06989},
archivePrefix = {arXiv},
       eprint = {2507.06989},
 primaryClass = {astro-ph.IM},
       adsurl = {https://ui.adsabs.harvard.edu/abs/2025arXiv250706989K}
}

@ARTICLE{moe2019,
       author = {{Moe}, Maxwell and {Kratter}, Kaitlin M. and {Badenes}, Carles},
        title = "{The Close Binary Fraction of Solar-type Stars Is Strongly Anticorrelated with Metallicity}",
      journal = {\apj},
         year = 2019,
        month = apr,
       volume = {875},
       number = {1},
          eid = {61},
        pages = {61},
          doi = {10.3847/1538-4357/ab0d88},
archivePrefix = {arXiv},
       eprint = {1808.02116},
 primaryClass = {astro-ph.SR},
       adsurl = {https://ui.adsabs.harvard.edu/abs/2019ApJ...875...61M}
}

@ARTICLE{bidelman1951,
       author = {{Bidelman}, William P. and {Keenan}, Philip C.},
        title = "{The Ba II Stars.}",
      journal = {\apj},
         year = 1951,
        month = nov,
       volume = {114},
        pages = {473},
          doi = {10.1086/145488},
       adsurl = {https://ui.adsabs.harvard.edu/abs/1951ApJ...114..473B}
}

@ARTICLE{sandage1953,
       author = {{Sandage}, A.~R.},
        title = "{The color-magnitude diagram for the globular cluster M 3.}",
      journal = {\aj},
         year = 1953,
        month = jan,
       volume = {58},
        pages = {61-75},
          doi = {10.1086/106822},
       adsurl = {https://ui.adsabs.harvard.edu/abs/1953AJ.....58...61S}
}

@INPROCEEDINGS{offner2023,
       author = {{Offner}, S.~S.~R. and {Moe}, M. and {Kratter}, K.~M. and {Sadavoy}, S.~I. and {Jensen}, E.~L.~N. and {Tobin}, J.~J.},
        title = "{The Origin and Evolution of Multiple Star Systems}",
    booktitle = {Protostars and Planets VII},
         year = 2023,
       editor = {{Inutsuka}, S. and {Aikawa}, Y. and {Muto}, T. and {Tomida}, K. and {Tamura}, M.},
       series = {Astronomical Society of the Pacific Conference Series},
       volume = {534},
        month = jul,
        pages = {275},
          doi = {10.48550/arXiv.2203.10066},
archivePrefix = {arXiv},
       eprint = {2203.10066},
 primaryClass = {astro-ph.SR},
       adsurl = {https://ui.adsabs.harvard.edu/abs/2023ASPC..534..275O}
}

@Article{numpy,
 title         = {Array programming with {NumPy}},
 author        = {Charles R. Harris and K. Jarrod Millman and St{\'{e}}fan J.
                 van der Walt and Ralf Gommers and Pauli Virtanen and David
                 Cournapeau and Eric Wieser and Julian Taylor and Sebastian
                 Berg and Nathaniel J. Smith and Robert Kern and Matti Picus
                 and Stephan Hoyer and Marten H. van Kerkwijk and Matthew
                 Brett and Allan Haldane and Jaime Fern{\'{a}}ndez del
                 R{\'{i}}o and Mark Wiebe and Pearu Peterson and Pierre
                 G{\'{e}}rard-Marchant and Kevin Sheppard and Tyler Reddy and
                 Warren Weckesser and Hameer Abbasi and Christoph Gohlke and
                 Travis E. Oliphant},
 year          = {2020},
 month         = sep,
 journal       = {Nature},
 volume        = {585},
 number        = {7825},
 pages         = {357--362},
 doi           = {10.1038/s41586-020-2649-2},
 publisher     = {Springer Science and Business Media {LLC}},
 url           = {https://doi.org/10.1038/s41586-020-2649-2}
}

@Article{matplotlib,
  Author    = {Hunter, J. D.},
  Title     = {Matplotlib: A 2D graphics environment},
  Journal   = {Computing in Science \& Engineering},
  Volume    = {9},
  Number    = {3},
  Pages     = {90--95},
  publisher = {IEEE COMPUTER SOC},
  doi       = {10.1109/MCSE.2007.55},
  year      = 2007
}

@InProceedings{pandas_2,
  author    = { {W}es {M}c{K}inney },
  title     = { {D}ata {S}tructures for {S}tatistical {C}omputing in {P}ython },
  booktitle = { {P}roceedings of the 9th {P}ython in {S}cience {C}onference },
  pages     = { 56 - 61 },
  year      = { 2010 },
  editor    = { {S}t\'efan van der {W}alt and {J}arrod {M}illman },
  doi       = { 10.25080/Majora-92bf1922-00a }
}

@software{pandas_1,
    author       = {The pandas development team},
    title        = {pandas-dev/pandas: Pandas},
    month        = feb,
    year         = 2020,
    publisher    = {Zenodo},
    version      = {2.2.2},
    doi          = {10.5281/zenodo.3509134},
    url          = {https://doi.org/10.5281/zenodo.3509134}
}

@ARTICLE{scipy2020,
  author  = {Virtanen, Pauli and Gommers, Ralf and Oliphant, Travis E. and
            Haberland, Matt and Reddy, Tyler and Cournapeau, David and
            Burovski, Evgeni and Peterson, Pearu and Weckesser, Warren and
            Bright, Jonathan and {van der Walt}, St{\'e}fan J. and
            Brett, Matthew and Wilson, Joshua and Millman, K. Jarrod and
            Mayorov, Nikolay and Nelson, Andrew R. J. and Jones, Eric and
            Kern, Robert and Larson, Eric and Carey, C J and
            Polat, {\.I}lhan and Feng, Yu and Moore, Eric W. and
            {VanderPlas}, Jake and Laxalde, Denis and Perktold, Josef and
            Cimrman, Robert and Henriksen, Ian and Quintero, E. A. and
            Harris, Charles R. and Archibald, Anne M. and
            Ribeiro, Ant{\^o}nio H. and Pedregosa, Fabian and
            {van Mulbregt}, Paul and {SciPy 1.0 Contributors}},
  title   = {{{SciPy} 1.0: Fundamental Algorithms for Scientific
            Computing in Python}},
  journal = {Nature Methods},
  year    = {2020},
  volume  = {17},
  pages   = {261--272},
  adsurl  = {https://rdcu.be/b08Wh},
  doi     = {10.1038/s41592-019-0686-2},
}

@software{bacchus,
       author = {{Masseron}, Thomas and {Merle}, Thibault and {Hawkins}, Keith},
        title = "{BACCHUS: Brussels Automatic Code for Characterizing High accUracy Spectra}",
 howpublished = {Astrophysics Source Code Library, record ascl:1605.004},
         year = 2016,
        month = may,
          eid = {ascl:1605.004},
       adsurl = {https://ui.adsabs.harvard.edu/abs/2016ascl.soft05004M}
}

@ARTICLE{ness2021,
       author = {{Ness}, Melissa K. and {Wheeler}, Adam J. and {McKinnon}, Kevin and {Horta}, Danny and {Casey}, Andrew R. and {Cunningham}, Emily C. and {Price-Whelan}, Adrian M.},
        title = "{The Homogeneity of the Star-forming Environment of the Milky Way Disk over Time}",
      journal = {\apj},
         year = 2022,
        month = feb,
       volume = {926},
       number = {2},
          eid = {144},
        pages = {144},
          doi = {10.3847/1538-4357/ac4754},
archivePrefix = {arXiv},
       eprint = {2109.05722},
 primaryClass = {astro-ph.GA},
       adsurl = {https://ui.adsabs.harvard.edu/abs/2022ApJ...926..144N}
}

@ARTICLE{kepler,
       author = {{Borucki}, William J. and {Koch}, David and {Basri}, Gibor and {Batalha}, Natalie and {Brown}, Timothy and {Caldwell}, Douglas and {Caldwell}, John and {Christensen-Dalsgaard}, J{\o}rgen and {Cochran}, William D. and {DeVore}, Edna and {Dunham}, Edward W. and {Dupree}, Andrea K. and {Gautier}, Thomas N. and {Geary}, John C. and {Gilliland}, Ronald and {Gould}, Alan and {Howell}, Steve B. and {Jenkins}, Jon M. and {Kondo}, Yoji and {Latham}, David W. and {Marcy}, Geoffrey W. and {Meibom}, S{\o}ren and {Kjeldsen}, Hans and {Lissauer}, Jack J. and {Monet}, David G. and {Morrison}, David and {Sasselov}, Dimitar and {Tarter}, Jill and {Boss}, Alan and {Brownlee}, Don and {Owen}, Toby and {Buzasi}, Derek and {Charbonneau}, David and {Doyle}, Laurance and {Fortney}, Jonathan and {Ford}, Eric B. and {Holman}, Matthew J. and {Seager}, Sara and {Steffen}, Jason H. and {Welsh}, William F. and {Rowe}, Jason and {Anderson}, Howard and {Buchhave}, Lars and {Ciardi}, David and {Walkowicz}, Lucianne and {Sherry}, William and {Horch}, Elliott and {Isaacson}, Howard and {Everett}, Mark E. and {Fischer}, Debra and {Torres}, Guillermo and {Johnson}, John Asher and {Endl}, Michael and {MacQueen}, Phillip and {Bryson}, Stephen T. and {Dotson}, Jessie and {Haas}, Michael and {Kolodziejczak}, Jeffrey and {Van Cleve}, Jeffrey and {Chandrasekaran}, Hema and {Twicken}, Joseph D. and {Quintana}, Elisa V. and {Clarke}, Bruce D. and {Allen}, Christopher and {Li}, Jie and {Wu}, Haley and {Tenenbaum}, Peter and {Verner}, Ekaterina and {Bruhweiler}, Frederick and {Barnes}, Jason and {Prsa}, Andrej},
        title = "{Kepler Planet-Detection Mission: Introduction and First Results}",
      journal = {Science},
         year = 2010,
        month = feb,
       volume = {327},
       number = {5968},
        pages = {977},
          doi = {10.1126/science.1185402},
       adsurl = {https://ui.adsabs.harvard.edu/abs/2010Sci...327..977B}
}

@ARTICLE{K2,
       author = {{Howell}, Steve B. and {Sobeck}, Charlie and {Haas}, Michael and {Still}, Martin and {Barclay}, Thomas and {Mullally}, Fergal and {Troeltzsch}, John and {Aigrain}, Suzanne and {Bryson}, Stephen T. and {Caldwell}, Doug and {Chaplin}, William J. and {Cochran}, William D. and {Huber}, Daniel and {Marcy}, Geoffrey W. and {Miglio}, Andrea and {Najita}, Joan R. and {Smith}, Marcie and {Twicken}, J.~D. and {Fortney}, Jonathan J.},
        title = "{The K2 Mission: Characterization and Early Results}",
      journal = {\pasp},
         year = 2014,
        month = apr,
       volume = {126},
       number = {938},
        pages = {398},
          doi = {10.1086/676406},
archivePrefix = {arXiv},
       eprint = {1402.5163},
 primaryClass = {astro-ph.IM},
       adsurl = {https://ui.adsabs.harvard.edu/abs/2014PASP..126..398H}
}

@ARTICLE{mist3,
       author = {{Paxton}, Bill and {Bildsten}, Lars and {Dotter}, Aaron and {Herwig}, Falk and {Lesaffre}, Pierre and {Timmes}, Frank},
        title = "{Modules for Experiments in Stellar Astrophysics (MESA)}",
      journal = {\apjs},
         year = 2011,
        month = jan,
       volume = {192},
       number = {1},
          eid = {3},
        pages = {3},
          doi = {10.1088/0067-0049/192/1/3},
archivePrefix = {arXiv},
       eprint = {1009.1622},
 primaryClass = {astro-ph.SR},
       adsurl = {https://ui.adsabs.harvard.edu/abs/2011ApJS..192....3P}
}

@ARTICLE{mist2,
       author = {{Choi}, Jieun and {Dotter}, Aaron and {Conroy}, Charlie and {Cantiello}, Matteo and {Paxton}, Bill and {Johnson}, Benjamin D.},
        title = "{Mesa Isochrones and Stellar Tracks (MIST). I. Solar-scaled Models}",
      journal = {\apj},
         year = 2016,
        month = jun,
       volume = {823},
       number = {2},
          eid = {102},
        pages = {102},
          doi = {10.3847/0004-637X/823/2/102},
archivePrefix = {arXiv},
       eprint = {1604.08592},
 primaryClass = {astro-ph.SR},
       adsurl = {https://ui.adsabs.harvard.edu/abs/2016ApJ...823..102C}
}

@ARTICLE{mist1,
       author = {{Dotter}, Aaron},
        title = "{MESA Isochrones and Stellar Tracks (MIST) 0: Methods for the Construction of Stellar Isochrones}",
      journal = {\apjs},
         year = 2016,
        month = jan,
       volume = {222},
       number = {1},
          eid = {8},
        pages = {8},
          doi = {10.3847/0067-0049/222/1/8},
archivePrefix = {arXiv},
       eprint = {1601.05144},
 primaryClass = {astro-ph.SR},
       adsurl = {https://ui.adsabs.harvard.edu/abs/2016ApJS..222....8D}
}

@ARTICLE{lada,
       author = {{Lada}, Charles J. and {Lada}, Elizabeth A.},
        title = "{Embedded Clusters in Molecular Clouds}",
      journal = {\araa},
         year = 2003,
        month = jan,
       volume = {41},
        pages = {57-115},
          doi = {10.1146/annurev.astro.41.011802.094844},
archivePrefix = {arXiv},
       eprint = {astro-ph/0301540},
 primaryClass = {astro-ph},
       adsurl = {https://ui.adsabs.harvard.edu/abs/2003ARA&A..41...57L}
}

@ARTICLE{galpy,
       author = {{Bovy}, Jo},
        title = "{galpy: A python Library for Galactic Dynamics}",
      journal = {\apjs},
         year = 2015,
        month = feb,
       volume = {216},
       number = {2},
          eid = {29},
        pages = {29},
          doi = {10.1088/0067-0049/216/2/29},
archivePrefix = {arXiv},
       eprint = {1412.3451},
 primaryClass = {astro-ph.GA},
       adsurl = {https://ui.adsabs.harvard.edu/abs/2015ApJS..216...29B}
}

@INPROCEEDINGS{fps,
       author = {{Pogge}, Richard W. and {Derwent}, Mark A. and {O'Brien}, Thomas P. and {Jurgenson}, Colby A. and {Pappalardo}, Daniel and {Engelman}, Michael and {Brandon}, Christopher and {Brady}, Julia and {Clawson}, Nicholas and {Shover}, Jon and {Mason}, Jerry and {Kneib}, Jean-Paul and {Araujo}, Ricardo and {Bouri}, Mohamed and {Kronig}, Luzius and {Grossen}, Lo{\"\i}c. and {Gillet}, Denis and {Macktoobian}, Matin and {Tuttle}, Sarah E. and {Farr}, Emily and {S{\'a}nchez-Gallego}, Jos{\'e} and {Sayres}, Conor},
        title = "{A robotic Focal Plane System (FPS) for the Sloan Digital Sky Survey V}",
    booktitle = {Ground-based and Airborne Instrumentation for Astronomy VIII},
         year = 2020,
       editor = {{Evans}, Christopher J. and {Bryant}, Julia J. and {Motohara}, Kentaro},
       series = {Society of Photo-Optical Instrumentation Engineers (SPIE) Conference Series},
       volume = {11447},
        month = dec,
          eid = {1144781},
        pages = {1144781},
          doi = {10.1117/12.2561113},
       adsurl = {https://ui.adsabs.harvard.edu/abs/2020SPIE11447E..81P}
}

@INPROCEEDINGS{connector,
       author = {{Barkhouser}, Robert H. and {Smee}, Stephen A. and {Hammond}, Randolph P. and {Harding}, Albert C. and {Gray}, Aidan C. and {Ramirez}, Solange and {Wachter}, Stefanie and {Kollmeier}, Juna and {Downey}, John and {Eriksen}, Jamey E. and {Wilson}, John C.},
        title = "{Design of the new SDSS 2.5m telescope wide field corrector for SDSS-V}",
    booktitle = {Ground-based and Airborne Telescopes IX},
         year = 2022,
       editor = {{Marshall}, Heather K. and {Spyromilio}, Jason and {Usuda}, Tomonori},
       series = {Society of Photo-Optical Instrumentation Engineers (SPIE) Conference Series},
       volume = {12182},
        month = aug,
          eid = {121823O},
        pages = {121823O},
          doi = {10.1117/12.2630655},
       adsurl = {https://ui.adsabs.harvard.edu/abs/2022SPIE12182E..3OB}
}

@ARTICLE{dr18,
       author = {{Almeida}, Andr{\'e}s and {Anderson}, Scott F. and {Argudo-Fern{\'a}ndez}, Maria and {Badenes}, Carles and {Barger}, Kat and {Barrera-Ballesteros}, Jorge K. and {Bender}, Chad F. and {Benitez}, Erika and {Besser}, Felipe and {Bird}, Jonathan C. and {Bizyaev}, Dmitry and {Blanton}, Michael R. and {Bochanski}, John and {Bovy}, Jo and {Brandt}, William Nielsen and {Brownstein}, Joel R. and {Buchner}, Johannes and {Bulbul}, Esra and {Burchett}, Joseph N. and {Cano D{\'\i}az}, Mariana and {Carlberg}, Joleen K. and {Casey}, Andrew R. and {Chandra}, Vedant and {Cherinka}, Brian and {Chiappini}, Cristina and {Coker}, Abigail A. and {Comparat}, Johan and {Conroy}, Charlie and {Contardo}, Gabriella and {Cortes}, Arlin and {Covey}, Kevin and {Crane}, Jeffrey D. and {Cunha}, Katia and {Dabbieri}, Collin and {Davidson}, James W. and {Davis}, Megan C. and {de Andrade Queiroz}, Anna Barbara and {De Lee}, Nathan and {M{\'e}ndez Delgado}, Jos{\'e} Eduardo and {Demasi}, Sebastian and {Di Mille}, Francesco and {Donor}, John and {Dow}, Peter and {Dwelly}, Tom and {Eracleous}, Mike and {Eriksen}, Jamey and {Fan}, Xiaohui and {Farr}, Emily and {Frederick}, Sara and {Fries}, Logan and {Frinchaboy}, Peter and {G{\"a}nsicke}, Boris T. and {Ge}, Junqiang and {Gonz{\'a}lez {\'A}vila}, Consuelo and {Grabowski}, Katie and {Grier}, Catherine and {Guiglion}, Guillaume and {Gupta}, Pramod and {Hall}, Patrick and {Hawkins}, Keith and {Hayes}, Christian R. and {Hermes}, J.~J. and {Hern{\'a}ndez-Garc{\'\i}a}, Lorena and {Hogg}, David W. and {Holtzman}, Jon A. and {Ibarra-Medel}, Hector Javier and {Ji}, Alexander and {Jofre}, Paula and {Johnson}, Jennifer A. and {Jones}, Amy M. and {Kinemuchi}, Karen and {Kluge}, Matthias and {Koekemoer}, Anton and {Kollmeier}, Juna A. and {Kounkel}, Marina and {Krishnarao}, Dhanesh and {Krumpe}, Mirko and {Lacerna}, Ivan and {Lago}, Paulo Jakson Assuncao and {Laporte}, Chervin and {Liu}, Chao and {Liu}, Ang and {Liu}, Xin and {Lopes}, Alexandre Roman and {Macktoobian}, Matin and {Majewski}, Steven R. and {Malanushenko}, Viktor and {Maoz}, Dan and {Masseron}, Thomas and {Masters}, Karen L. and {Matijevic}, Gal and {McBride}, Aidan and {Medan}, Ilija and {Merloni}, Andrea and {Morrison}, Sean and {Myers}, Natalie and {M{\'e}sz{\'a}ros}, Szabolcs and {Negrete}, C. Alenka and {Nidever}, David L. and {Nitschelm}, Christian and {Oravetz}, Daniel and {Oravetz}, Audrey and {Pan}, Kaike and {Peng}, Yingjie and {Pinsonneault}, Marc H. and {Pogge}, Rick and {Qiu}, Dan and {Ramirez}, Solange V. and {Rix}, Hans-Walter and {Fern{\'a}ndez Rosso}, Daniela and {Runnoe}, Jessie and {Salvato}, Mara and {Sanchez}, Sebastian F. and {Santana}, Felipe A. and {Saydjari}, Andrew and {Sayres}, Conor and {Schlaufman}, Kevin C. and {Schneider}, Donald P. and {Schwope}, Axel and {Serna}, Javier and {Shen}, Yue and {Sobeck}, Jennifer and {Song}, Ying-Yi and {Souto}, Diogo and {Spoo}, Taylor and {Stassun}, Keivan G. and {Steinmetz}, Matthias and {Straumit}, Ilya and {Stringfellow}, Guy and {S{\'a}nchez-Gallego}, Jos{\'e} and {Taghizadeh-Popp}, Manuchehr and {Tayar}, Jamie and {Thakar}, Ani and {Tissera}, Patricia B. and {Tkachenko}, Andrew and {Hernandez Toledo}, Hector and {Trakhtenbrot}, Benny and {Fern{\'a}ndez-Trincado}, Jos{\'e} G. and {Troup}, Nicholas and {Trump}, Jonathan R. and {Tuttle}, Sarah and {Ulloa}, Natalie and {Vazquez-Mata}, Jose Antonio and {Vera Alfaro}, Pablo and {Villanova}, Sandro and {Wachter}, Stefanie and {Weijmans}, Anne-Marie and {Wheeler}, Adam and {Wilson}, John and {Wojno}, Leigh and {Wolf}, Julien and {Xue}, Xiang-Xiang and {Ybarra}, Jason E. and {Zari}, Eleonora and {Zasowski}, Gail},
        title = "{The Eighteenth Data Release of the Sloan Digital Sky Surveys: Targeting and First Spectra from SDSS-V}",
      journal = {\apjs},
         year = 2023,
        month = aug,
       volume = {267},
       number = {2},
          eid = {44},
        pages = {44},
          doi = {10.3847/1538-4365/acda98},
archivePrefix = {arXiv},
       eprint = {2301.07688},
 primaryClass = {astro-ph.GA},
       adsurl = {https://ui.adsabs.harvard.edu/abs/2023ApJS..267...44A}
}

@ARTICLE{sdssv,
       author = {{Kollmeier}, Juna A. and {Zasowski}, Gail and {Rix}, Hans-Walter and {Johns}, Matt and {Anderson}, Scott F. and {Drory}, Niv and {Johnson}, Jennifer A. and {Pogge}, Richard W. and {Bird}, Jonathan C. and {Blanc}, Guillermo A. and {Brownstein}, Joel R. and {Crane}, Jeffrey D. and {De Lee}, Nathan M. and {Klaene}, Mark A. and {Kreckel}, Kathryn and {MacDonald}, Nick and {Merloni}, Andrea and {Ness}, Melissa K. and {O'Brien}, Thomas and {Sanchez-Gallego}, Jose R. and {Sayres}, Conor C. and {Shen}, Yue and {Thakar}, Ani R. and {Tkachenko}, Andrew and {Aerts}, Conny and {Blanton}, Michael R. and {Eisenstein}, Daniel J. and {Holtzman}, Jon A. and {Maoz}, Dan and {Nandra}, Kirpal and {Rockosi}, Constance and {Weinberg}, David H. and {Bovy}, Jo and {Casey}, Andrew R. and {Chaname}, Julio and {Clerc}, Nicolas and {Conroy}, Charlie and {Eracleous}, Michael and {G{\"a}nsicke}, Boris T. and {Hekker}, Saskia and {Horne}, Keith and {Kauffmann}, Jens and {McQuinn}, Kristen B.~W. and {Pellegrini}, Eric W. and {Schinnerer}, Eva and {Schlafly}, Edward F. and {Schwope}, Axel D. and {Seibert}, Mark and {Teske}, Johanna K. and {van Saders}, Jennifer L.},
        title = "{SDSS-V: Pioneering Panoptic Spectroscopy}",
      journal = {arXiv e-prints},
         year = 2017,
        month = nov,
          eid = {arXiv:1711.03234},
        pages = {arXiv:1711.03234},
          doi = {10.48550/arXiv.1711.03234},
archivePrefix = {arXiv},
       eprint = {1711.03234},
 primaryClass = {astro-ph.GA},
       adsurl = {https://ui.adsabs.harvard.edu/abs/2017arXiv171103234K}
}

@ARTICLE{manea2023,
       author = {{Manea}, Catherine and {Hawkins}, Keith and {Ness}, Melissa K. and {Buder}, Sven and {Martell}, Sarah L. and {Zucker}, Daniel B.},
        title = "{Chemical Doppelgangers in GALAH DR3: The Distinguishing Power of Neutron-capture Elements among Milky Way Disk Stars}",
      journal = {\apj},
         year = 2024,
        month = sep,
       volume = {972},
       number = {1},
          eid = {69},
        pages = {69},
          doi = {10.3847/1538-4357/ad58d9},
archivePrefix = {arXiv},
       eprint = {2310.15257},
 primaryClass = {astro-ph.SR},
       adsurl = {https://ui.adsabs.harvard.edu/abs/2024ApJ...972...69M}
}

@ARTICLE{cheng2021,
       author = {{Cheng}, Chloe M. and {Price-Jones}, Natalie and {Bovy}, Jo},
        title = "{Testing the chemical homogeneity of chemically tagged dissolved birth clusters}",
      journal = {\mnras},
         year = 2021,
        month = oct,
       volume = {506},
       number = {4},
        pages = {5573-5588},
          doi = {10.1093/mnras/stab2106},
archivePrefix = {arXiv},
       eprint = {2010.09721},
 primaryClass = {astro-ph.GA},
       adsurl = {https://ui.adsabs.harvard.edu/abs/2021MNRAS.506.5573C}
}

@ARTICLE{badenes2018,
       author = {{Badenes}, Carles and {Mazzola}, Christine and {Thompson}, Todd A. and {Covey}, Kevin and {Freeman}, Peter E. and {Walker}, Matthew G. and {Moe}, Maxwell and {Troup}, Nicholas and {Nidever}, David and {Allende Prieto}, Carlos and {Andrews}, Brett and {Barb{\'a}}, Rodolfo H. and {Beers}, Timothy C. and {Bovy}, Jo and {Carlberg}, Joleen K. and {De Lee}, Nathan and {Johnson}, Jennifer and {Lewis}, Hannah and {Majewski}, Steven R. and {Pinsonneault}, Marc and {Sobeck}, Jennifer and {Stassun}, Keivan G. and {Stringfellow}, Guy S. and {Zasowski}, Gail},
        title = "{Stellar Multiplicity Meets Stellar Evolution and Metallicity: The APOGEE View}",
      journal = {\apj},
         year = 2018,
        month = feb,
       volume = {854},
       number = {2},
          eid = {147},
        pages = {147},
          doi = {10.3847/1538-4357/aaa765},
archivePrefix = {arXiv},
       eprint = {1711.00660},
 primaryClass = {astro-ph.SR},
       adsurl = {https://ui.adsabs.harvard.edu/abs/2018ApJ...854..147B}
}

@ARTICLE{galpy1,
       author = {{Bovy}, Jo},
        title = "{galpy: A python Library for Galactic Dynamics}",
      journal = {\apjs},
         year = 2015,
        month = feb,
       volume = {216},
       number = {2},
          eid = {29},
        pages = {29},
          doi = {10.1088/0067-0049/216/2/29},
archivePrefix = {arXiv},
       eprint = {1412.3451},
 primaryClass = {astro-ph.GA},
       adsurl = {https://ui.adsabs.harvard.edu/abs/2015ApJS..216...29B}
}

% Don't change these lines
\bsp	% typesetting comment
\label{lastpage}
\end{document}